%% file: main.tex
\documentclass{article}

\usepackage[preprint]{neurips_2025}

\usepackage[utf8]{inputenc} 
\usepackage[T1]{fontenc}    
\usepackage{hyperref}       
\usepackage{url}            
\usepackage{booktabs}       
\usepackage{amsfonts}       
\usepackage{nicefrac}       
\usepackage{microtype}      
\usepackage{xcolor}         

\input{preamble}
\usepackage{wrapfig}

\usepackage{multirow}
\usepackage{colortbl}
\usepackage{xcolor}
\usepackage{amssymb}
\definecolor{xl}{HTML}{B0E3E6}
\definecolor{long}{HTML}{FAD9D5}
\definecolor{think}{HTML}{DAE8FC}
\definecolor{chat}{HTML}{D0CEE2}
\definecolor{nvidiaGreen}{RGB}{118,185,0}
\definecolor{qwenPurple}{RGB}{116,81,207}
\definecolor{closedGray}{RGB}{90,90,110}

\title{~\includegraphics[height=5pt,trim=0cm 4cm 0 10cm]{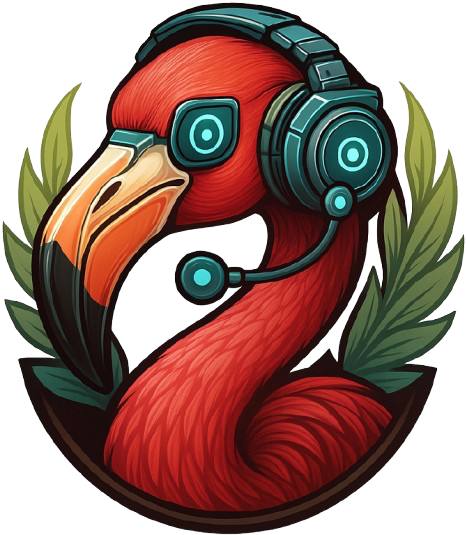}Audio Flamingo 3: Advancing Audio Intelligence with Fully Open Large Audio Language Models}

\usepackage{pifont} 

\author{%
  \textbf{Arushi Goel}\textsuperscript{\ding{72}1}, \textbf{Sreyan Ghosh}\textsuperscript{\ding{72}12}, \textbf{Jaehyeon Kim}\textsuperscript{1}, \textbf{Sonal Kumar}\textsuperscript{2}, \textbf{Zhifeng Kong}\textsuperscript{1}, \textbf{Sang-gil Lee}\textsuperscript{1}, \\
  \textbf{Chao-Han Huck Yang}\textsuperscript{\textbf{1}}, \textbf{Ramani Duraiswami}\textsuperscript{\textbf{2}}, \textbf{Dinesh Manocha}\textsuperscript{\textbf{2}}, \textbf{Rafael Valle}\textsuperscript{\textbf{1}}, \textbf{Bryan Catanzaro}\textsuperscript{\textbf{1}} \\ \\
  NVIDIA, USA\textsuperscript{1}, University of Maryland, College Park, USA\textsuperscript{2} \\ \\
  \textsuperscript{\ding{72}}Equal contribution. Alphabetically ordered.\\ \\ Correspondence: arushig@nvidia.com, sreyang@umd.edu \\ \\
  \textbf{Project:} \url{https://research.nvidia.com/labs/adlr/AF3/}
}
\begin{document}

\maketitle
\vspace{-4mm}
\begin{abstract}
We present Audio Flamingo 3 (AF3), a \textit{fully open} state-of-the-art (SOTA) large audio-language model that advances reasoning and understanding across speech, sound, and music. AF3 introduces: (i) AF-Whisper, a unified audio encoder trained using a novel strategy for joint representation learning across all 3 modalities of speech, sound, and music; (ii) flexible, on-demand thinking, allowing the model to do chain-of-thought-type reasoning before answering; (iii) multi-turn, multi-audio chat; (iv) long audio understanding and reasoning (including speech) up to 10 minutes; and (v) voice-to-voice interaction. To enable these capabilities, we propose several large-scale training datasets curated using novel strategies, including AudioSkills-XL, LongAudio-XL, AF-Think, and AF-Chat, and train AF3 with a novel five-stage curriculum-based training strategy. Trained on only open-source audio data, AF3 achieves new SOTA results on over 20+ (long) audio understanding and reasoning benchmarks, surpassing both open-weight and closed-source models trained on much larger datasets.

\vspace{-3mm}
\end{abstract}

\section{Introduction} \label{sec:introduction}
\input{intro}
\vspace{-1mm}

\section{Related Work}
\label{sec:related_work}

\input{related_work}

\section{Methodology}\label{sec:approach}
\input{approach}

\vspace{-2mm}
\section{Experiments}
\label{sec:experiments}
\vspace{-1mm}
\input{experiments}

 \section{Conclusion, Limitations and Future Work}
\label{sec:discussion}
\vspace{-1mm}
\input{discussion}

\clearpage
\newpage
\bibliography{references}\label{sec:references}
\bibliographystyle{abbrv}

\newpage
\appendix
\section*{Appendix}
\label{sec:appendix}

\input{appendix}

\end{document}

%% file: preamble.tex
\newcommand{\cmark}{{\color{green}\checkmark}}
\newcommand{\xmark}{{\color{red}\texttimes}}

\usepackage{amsmath,amsfonts}
\usepackage{microtype}
\usepackage{graphicx}
\usepackage{subcaption}
\usepackage{booktabs} 
\usepackage{colortbl}
\usepackage{xcolor}

\usepackage{pgfplots,pgfplotstable}
\pgfplotsset{compat=1.18}
\usepgfplotslibrary{polar}
\usepackage{multirow}
\usepackage{adjustbox}  

\usepackage{hyperref}
\usepackage{url}
\usepackage{bbding}
\usepackage{pifont}
\usepackage{wasysym}
\usepackage{amssymb}
\usepackage[capitalize,noabbrev]{cleveref}

\hypersetup{
    pdfborderstyle={/S/U/W 1}, 
    linkbordercolor=red,       
    citebordercolor=green,     
    filebordercolor=magenta,   
   urlbordercolor=cyan        
}

\usepackage{enumitem}

\usepackage{interval}
\usepackage{tabularx}
\usepackage{xspace}
\usepackage{mathtools}  
\usepackage{siunitx}    

\usepackage{placeins}
\makeatletter
\renewcommand{\paragraph}{%
  \@startsection{paragraph}{4}%
  {\z@}{0.5em}{-5em}%
  {\normalfont\normalsize\bfseries}%
}

\makeatother
\usepackage{algorithm2e}
\usepackage{amsmath}



%% file: intro.tex
\begin{wrapfigure}{r}{0.43\textwidth}
     \centering
     \begin{subfigure}[b]{0.43\textwidth}
         \centering
         \includegraphics[trim= 0.4cm 0.5cm 0cm 1.5cm, width=\textwidth]{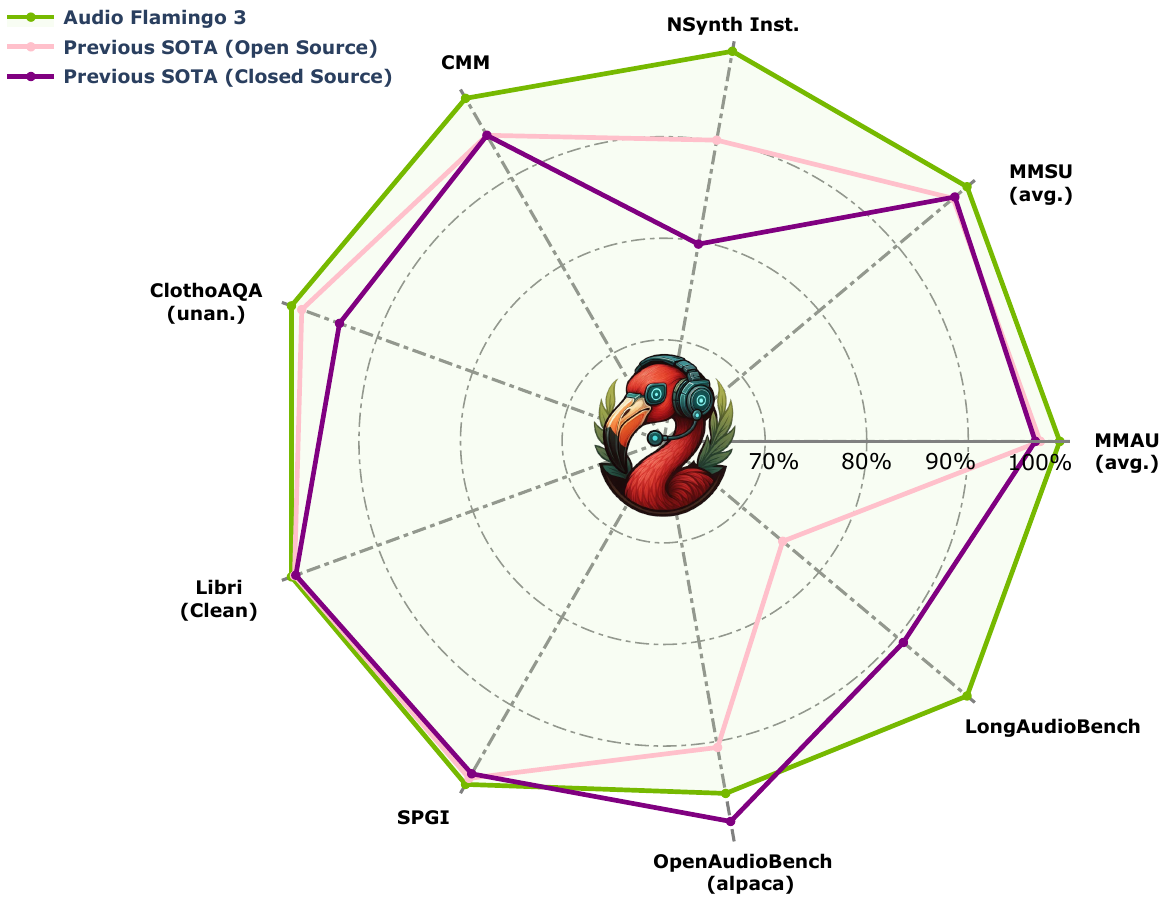}
     \end{subfigure}
        \caption{\small AF3 vs. prior SOTA LALMs (values normalized and WER=100-WER). AF3 outperforms most open-source/weights (e.g., Qwen2.5-Omni) and closed (e.g., Gemini 2.5 Pro) LALMs while being \textit{fully open.}}
        \label{fig:af3_radial}
        \vspace{-2.5em}
\end{wrapfigure}  

Audio—including speech, sounds, and music—is central to human perception and interaction. It enables us to understand our surroundings, engage in conversations, express emotions, interpret videos, and enjoy music. For AI systems to approach artificial general intelligence (AGI)~\citep{morris2024position}, they must similarly develop the ability to comprehend and reason over diverse audio signals. While Large Language Models (LLMs) excel at language-based reasoning, their audio comprehension remains limited — both in accessibility and capability~\cite{hurst2024gpt,touvron2023llama}. Extending LLMs to process and reason over audio is essential for building truly context-aware, intelligent agents.

Audio-Language Models (ALMs) extend the capabilities of LMs to the auditory domain. Early works such as CLAP~\cite{elizalde2022clap} align audio and text in a shared embedding space, enabling them with tasks like retrieval~\citep{oncescu2021audio}. More recently, the emergence of Large ALMs (LALMs)—decoder-only language models augmented with audio understanding~\cite{chu2023qwenaudio,chu2024qwenaudio2,team2023gemini}—has unlocked powerful capabilities, including open-ended audio question-answering (AQA) that demands both reasoning and world knowledge~\citep{sakshi2024mmau}. These capabilities have further enabled tasks like audio analysis~\cite{drossos2020clotho,kim2019audiocaps}, conversational assistants~\cite{daniel2018toward}, etc.

\begin{table}[t]
\centering
\renewcommand{\arraystretch}{1.3}
\resizebox{\textwidth}{!}{%
\begin{tabular}{lccc ccc cc ccc ccc}
\toprule
\textbf{Models} & \multicolumn{3}{c}{\textbf{Audio Understanding}} & \multicolumn{2}{c}{\textbf{Voice}} & \multicolumn{2}{c}{\textbf{Multi-turn Chat}} & \multicolumn{3}{c}{\textbf{Long Audio ($>$30 secs)}} & \multicolumn{3}{c}{\textbf{Open-Source}} \\
\cmidrule(lr){2-4} \cmidrule(lr){5-6} \cmidrule(lr){7-8} \cmidrule(lr){9-11} \cmidrule(lr){12-14}
& Sound & Music & Speech & In & Out* & Single A & Multiple A & Speech & Sound & Music & Model & Data & Code \\
\midrule

LTU         & \cmark & \cmark & \xmark & \xmark & \xmark & \xmark & \xmark & \xmark & \xmark & \xmark & \cmark & \cmark & \cmark \\
LTU-AS         & \cmark & \cmark & \cmark & \xmark & \xmark & \xmark & \xmark & \xmark & \xmark & \xmark & \cmark & \cmark & \cmark \\
GAMA         & \cmark & \cmark & \xmark & \xmark & \xmark & \xmark & \xmark & \xmark & \xmark & \xmark & \cmark & \cmark & \cmark \\
SALMONN         & \cmark & \cmark & \cmark & \xmark & \xmark & \xmark & \xmark & \xmark & \xmark & \xmark & \cmark & \cmark & \cmark \\
MuLLaMa         & \xmark & \cmark & \xmark & \xmark & \xmark & \xmark & \xmark & \xmark & \xmark & \xmark & \cmark & \cmark & \cmark \\
Phi-4-mm         & \cmark & \cmark & \cmark & \xmark & \xmark & \xmark & \xmark & \cmark & \cmark & \cmark & \cmark & \xmark & \xmark\\
Qwen-Audio         & \cmark & \cmark & \cmark & \cmark & \cmark & \cmark & \xmark & \xmark & \xmark & \xmark & \cmark & \xmark & \xmark\\
Qwen2-Audio        & \cmark & \cmark & \cmark & \cmark & \cmark & \cmark & \xmark & \xmark & \xmark & \xmark & \cmark & \xmark & \xmark\\
Qwen2.5-Omni        & \cmark & \cmark & \cmark & \cmark & \cmark & \cmark & \xmark & \cmark & \cmark & \cmark & \cmark & \xmark & \xmark\\
GPT-4o Audio       & \cmark & \cmark & \cmark & \cmark & \cmark & \cmark & \cmark & \cmark & \cmark & \cmark & \xmark & \xmark & \xmark\\
Gemini 2.0 / 2.5       & \cmark & \cmark & \cmark & \cmark & \cmark & \cmark & \xmark & \cmark & \cmark & \cmark & \xmark & \xmark & \xmark\\
Audio Flamingo     & \cmark & \cmark & \xmark & \xmark & \xmark & \cmark & \xmark & \xmark & \xmark & \xmark & \cmark & \cmark & \cmark\\
Audio Flamingo 2   & \cmark & \cmark & \xmark & \xmark & \xmark & \xmark & \xmark & \xmark & \cmark & \cmark & \cmark & \cmark & \cmark \\
\rowcolor{gray!10}
\textbf{Audio Flamingo 3} & \cmark & \cmark & \cmark & \cmark & \cmark & \cmark & \cmark & \cmark & \cmark & \cmark & \cmark & \cmark & \cmark\\
\bottomrule
\end{tabular}
}
\vspace{2mm}
\caption{\small Comparison of various LALMs in terms of capabilities and openness. AF3 stands out as the most capable and open model to date, achieving SOTA results across benchmarks (A in Chat stands for Audio). *\textit{Voice-out is powered by our novel streaming TTS implementation, which is also applicable to other LALMs.}}
\label{tab:audio_models}
\vspace{-7mm}
\end{table}

However, existing models still fall short in key areas critical to AGI, such as expert-level reasoning~\cite{morris2024position,sakshi2024mmau}, multi-turn and multi-audio dialogue~\cite{goel2024audio}, and long audio understanding~\cite{kong2025audioflamingo2}. We identify two core limitations: (i) most LALMs are trained primarily on short audio for recognition tasks rather than ones that require deliberate reasoning; and (ii) in turn, they lack exposure to the skill sets required for complex tasks. Additionally, most LALMs that support all three modalities of speech, sound, and music are closed-source: while some have publicly released models
weights~\cite{chu2023qwenaudio,chu2024qwenaudio2,abouelenin2025phi}, they offer limited to no information about their data, code,
or recipes (more details in \cref{tab:audio_models}).

{\noindent \textbf{Main Contributions.}} To address these issues, we introduce \textbf{Audio Flamingo 3 (AF3)}, a \textit{fully open-source}\footnote{By \textit{fully open}, we mean that the model’s weights, training data, and code will be publicly released, with full transparency about the training methodology. Due to the licensing and scope of the training data used in the work, all releases will be under a research-only license.} LALM with state-of-the-art performance in audio understanding and reasoning across 20+ benchmarks. In addition, AF3 brings several novel capabilities, including multi-turn, multi-audio chat, on-demand thinking, voice-to-voice interaction, and long-context audio reasoning (up to 10 minutes). We propose three core innovations to enable these capabilities: (i) \textbf{Data}: We focus on curating high-quality data at scale and propose (a) \textit{AudioSkills-XL}: a large-scale dataset of 8M diverse AQA pairs, (b) \textit{LongAudio-XL}: large-scale dataset of 1.25M diverse audio QA pairs for long audio reasoning; (c) \textit{AF-Chat}: a multi-turn multi-audio chat dataset curated using a novel algorithm with 75k instances and (d) \textit{AF-Think}: a dataset with 250k+ AQA pairs with short length prefixes to encourage CoT-type reasoning before arriving at the answer (ii) \textbf{AF-Whisper}: We train AF-Whisper, a unified audio encoder pretrained using a novel strategy on large-scale audio-caption pairs, capable of learning general-purpose representations across speech, sounds, and music; and (iii) \textbf{Learning Curriculum}: We train AF3 with a five-stage curriculum-based training strategy that progressively increases context length and task complexity. In summary, our main contributions are:

\begin{itemize}[leftmargin=*, labelsep=1em]
\vspace{-2mm}
\setlength\parskip{0em}
    \item We introduce \textbf{Audio Flamingo 3 (AF3)}, the most open and capable foundational LALM to date. AF3 introduces key capabilities including: (i) long-context audio QA (extending beyond sounds as in~\citep{kong2025audioflamingo2} and including speech), and (ii) flexible, on-demand thinking, enabling the model to generate concise, CoT-style reasoning steps when prompted. AF3 achieves state-of-the-art performance on 20+ audio understanding and reasoning benchmarks.
    \item We also present \textbf{AF3-Chat}, a fine-tuned variant of AF3 designed for multi-turn, multi-audio chat and voice-to-voice interaction.
    \item We propose novelties in data curation, audio encoder representation learning, and training strategies. Being fully open, we release our code, training recipes, and 4 new datasets to promote research in this space.
\end{itemize}

%% file: related_work.tex
\textbf{Audio Language Models.} The rapid progress of LLMs has catalyzed the development of multimodal LLMs (MLLMs) capable of understanding and reasoning across diverse data modalities, including audio. Within this space, ALMs specifically target reasoning over auditory inputs such as speech, sounds, and music. ALMs typically follow two main architectural paradigms: (i) \textit{Encoder-only ALMs}, which learn a joint embedding space for audio and text, enabling tasks like cross-modal retrieval. Representative models include CLAP~\citep{elizalde2022clap}, Wav2CLIP~\citep{wu2021wav2clip}, and AudioCLIP~\citep{guzhov2021audioclip}.
(ii) \textit{Encoder-decoder ALMs}, also referred to as LALMs, which use decoder-only LLMs augmented with an audio encoder. Notable examples include LTU~\citep{gong2023ltu}, LTU-AS~\citep{gong2023ltu-as}, SALMONN~\citep{tang2023salmonn}, Pengi~\citep{deshmukh2023pengi}, Audio Flamingo~\citep{kong2024audioflamingo}, Audio Flamingo 2~\citep{kong2025audioflamingo2}, AudioGPT~\citep{huang2023audiogpt}, GAMA~\citep{ghosh2024gama}, Qwen-Audio~\citep{chu2023qwenaudio}, and Qwen2-Audio~\citep{chu2024qwenaudio2}. These LALMs have significantly improved performance on core audio understanding tasks such as automatic speech recognition (ASR)~\citep{radford2022whisper}, audio captioning~\citep{kim2019audiocaps}, and acoustic scene classification~\citep{chen2022beats}. More importantly, they have enabled new capabilities such as open-ended audio question answering, which requires complex reasoning and external world knowledge.

Despite these advancements, current LALMs fall short in supporting various capabilities, including multi-turn, multi-audio chat, long-context audio comprehension, etc. Moreover, most LALMs are limited to specific audio types, lacking the ability to unify understanding across speech, sounds, and music. Finally, the most advanced LALMs remain only partially open, releasing model checkpoints without accompanying training code or data. This lack of transparency limits reproducibility and impedes scientific progress by obscuring the development process. 

\textbf{Reasoning and Long-Context Understanding.} Recent progress in LLMs has increasingly emphasized long-context understanding. In the vision-language space, substantial strides have been made in modeling long videos~\cite{chen2024longvila}. In the audio domain, AF2 marked the first step toward long-context audio comprehension, though it is limited to sounds and music.

Parallel efforts have aimed to enhance reasoning in LLMs and MLLMs through improved reasoning datasets~\cite{sakshi2024mmau,weck2024muchomusic}, advancements in multimodal perception~\cite{xu2024llava,team2023gemini}, and emerging paradigms like chain-of-thought (CoT) prompting~\citep{ma2025audio}, which encourages models to "think before answering." In developing AF3, we combine these advances—integrating controlled reasoning supervision, long-context training, and modality diversity—to equip the model with strong reasoning capabilities and long-context comprehension, including speech.

%% file: approach.tex


\subsection{Audio Flamingo 3 Architecture}
\label{subsec:af3_arch}

\begin{figure}[!t]
    \centering
    \includegraphics[width=\linewidth]{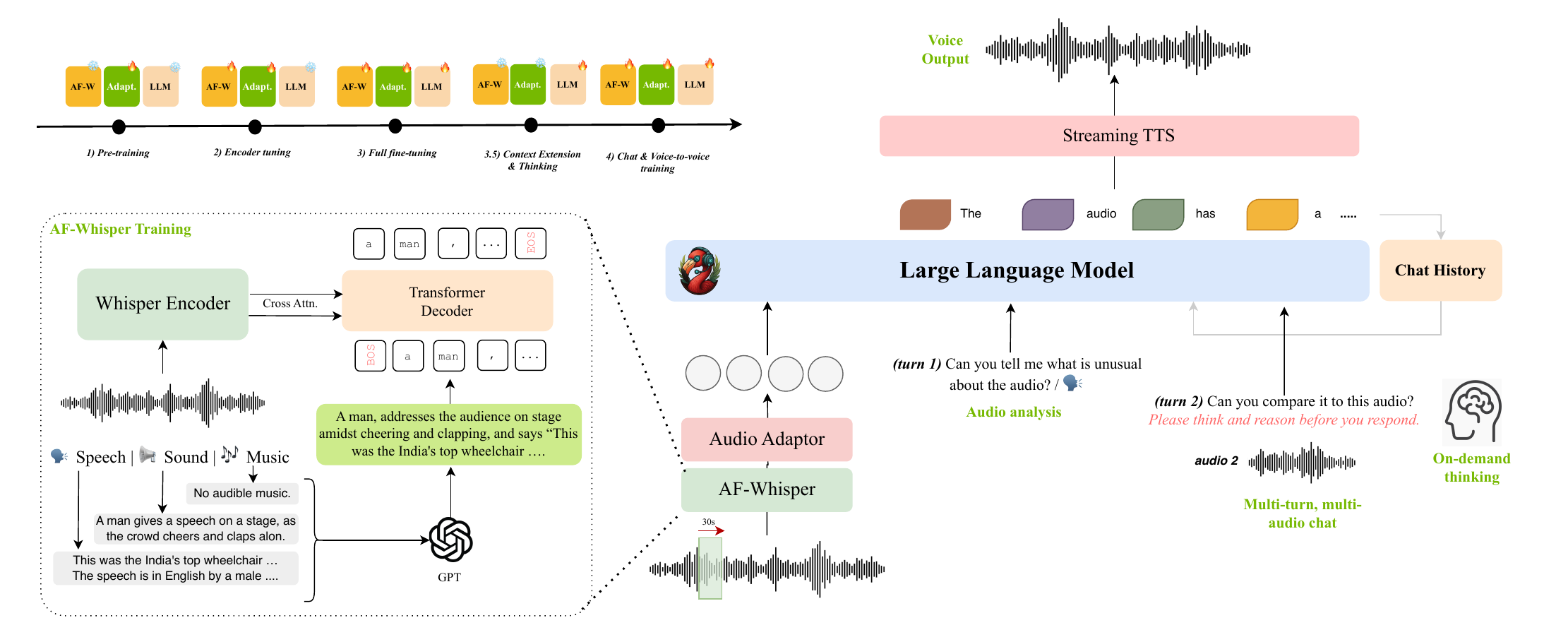}
    \caption{\small Overview of Audio Flamingo 3, AF-Whisper training, and five-stage curriculum training.}
    \label{fig.af3_method}
    \vspace{-2mm}
\end{figure}

In this section, we discuss our proposed architecture for Audio Flamingo 3 as shown in \cref{fig.af3_method}. AF3 consists of i) AF-Whisper: an audio encoder with sliding window feature extraction, ii) audio projector, iii) an LLM, and iv) a streaming TTS. We provide details of each component below.

{\noindent \textbf{AF-Whisper Audio Encoder.}}
Prior work in audio representation learning typically treats speech, sounds, and music as separate modalities, and LALMs often rely on distinct encoders for each~\cite{tang2023salmonn,ghosh2024gama}. Using separate encoders for LALMs increases model complexity, introduces frame-rate mismatches, and can lead to training instability. To address this, we propose AF-Whisper, a unified audio encoder trained with a simple yet effective representation learning strategy to model all three audio types.

As illustrated in \cref{fig.af3_method}, we start with the pre-trained Whisper large-v3 encoder~\cite{radford2022whisper}, attach it to a standard Transformer decoder, and train using the audio captioning task with the next-token-prediction objective. To achieve this, we generate a natural language caption for each audio, describing its speech, sound, and music content. First, we pool several datasets and then prompt GPT-4.1 to generate the audio caption. For prompting, we use available metadata for each sample, which includes transcripts, ambient sound descriptions, and music attributes. For samples lacking any of the 3 metadata, we synthesize it using AF2~\citep{kong2025audioflamingo2} or Whisper-Large-v3 ASR~\citep{radford2022whisper}. All datasets used for training are detailed in Section~\ref{sec.af_whisper_datasets}. We choose Whisper as the backbone due to its existing speech understanding capabilities and its dense, high-resolution audio features, which are more informative than those from models like CLAP~\citep{elizalde2022clap}. We connect it with a Transformer decoder using cross-attention (similar to RECAP\cite{liu2023recap} and AF2~\citep{kong2025audioflamingo2}) with 24 layers, 8 attention heads, and 1024 hidden size.

{\noindent \textbf{Feature Extraction.}} Given an audio input $A$, we first resample it to 16kHz mono. The raw waveform is then transformed into a 128-channel mel-spectrogram using a window size of 25ms and a hop size of 10ms. This mel-spectrogram is processed by AF-Whisper, producing hidden representations, denoted as $h_a = f_a(A)$, where $h_a \in \mathbb{R}^{N \times d}$. As shown in \cref{fig.af3_method}, each audio is processed in 30-second chunks of non-overlapping sliding windows, and $N$ or the temporal resolution depends on the length of the audio and the maximum number of sliding windows (which varies according to the stage of training). AF-Whisper produces audio features at a frame rate of 50Hz, and we further apply a pooling layer with a stride of two similar to \cite{chu2024qwenaudio2}. $d$ denotes the hidden dimension, which is 1280.

\textbf{Audio Adaptor.} To align the audio modality with the text embedding space of the LLM, we introduce audio adaptor layers, denoted by $\mathcal{A}(.)$. Specifically, the encoded hidden representations $h_a$ from AF-Whisper are passed through these adaptor layers to produce embeddings: $a = \mathcal{A}(h_a)$. These resulting embeddings serve as prompts to the LLM, alongside the textual instruction.

\textbf{Large Language Model (LLM).} We employ Qwen-2.5-7B~\citep{yang2024qwen2} as our backbone, a decoder-only causal LLM with 7B parameters, 36 hidden layers, and 16 attention heads.

\textbf{Streaming TTS.} To enable voice-to-voice interaction, we employ a TTS module for streaming speech generation, supporting streaming inputs and outputs. Our TTS module employs a decoder-only transformer architecture: it predicts the subsequent audio token conditioned on incoming subword text tokens from the LLM and the history of previously generated audio tokens. Similar streaming TTS techniques have been explored with LLMs~\citep{xie2024mini} (for voice-out on LLM outputs), but not in the context of LALMs (which we define as models designed to perceive and reason over diverse audio inputs). Since not a core novelty of our work, we provide more details, including training and architecture, in Appendix~\ref{sec.af3_voice}.

\vspace{-3mm}
\section{Audio Flamingo 3 Training Data}
\vspace{-1mm}

We present detailed statistics for all datasets used to train AF3 in Table~\ref{tab:sft_datasets_app}. AF3 has a total of 5 stages of training, where each stage employs a unique combination of datasets with unique weights (number of passes over that dataset for that particular stage). For Stages 1 and 2, we use open-source, recognition-focused foundational datasets converted to QA format. In the following sub-sections, we introduce our four novel skill-focused and unique datasets, each accompanied by custom data curation strategies, used in Stages 3, 3.5, and 4, which form a core contribution of this work.

\vspace{-1mm}
\subsection{AudioSkills-XL: Expanding AudioSkills with Reasoning-Focused QAs}
\label{subsec:audioskills}
\vspace{-1mm}

Audio QA pairs derived from foundational benchmarks focused on recognition tasks (e.g., ASR, acoustic event classification) are insufficient for training models in expert-level reasoning~\cite{sakshi2024mmau}. Therefore, in Stage 3 fine-tuning, we prioritize the development of reasoning and problem-solving abilities by curating large-scale, high-quality Audio QA data. Inspired by AF2, we limit this stage to short audio clips ($\leq$30s) and defer long audio reasoning to later stages. We expand the AudioSkills dataset~\cite{kong2025audioflamingo2} by 4.5M new Audio QA pairs (majorly multiple-choice questions (MCQ)-based) to create AudioSkills-XL, a high-quality corpus containing 8M Audio QA pairs, using two strategies: 

\noindent (1) \textbf{We expand coverage of existing reasoning skills and introduce new ones using additional audio sources, increasing the dataset by 3.5M QA pairs}: (a) For sounds, we incorporate data from YouTube8M and synthetic sources. (b) For music, we include Music4All~\cite{santana2020music4all} and the Million Song Dataset~\cite{Bertin-Mahieux2011}. For YouTube8M, we adapt captions from AudioSetCaps~\cite{bai2024audiosetcapsnipsws} and generate QA using GPT-4.1 with general reasoning prompts from AF2. Additionally, we introduce new reasoning skills and design corresponding prompts to support them. For music, we generate data for novel skills (as AudioSkills was focused more on sounds; details in Table~\ref{tab:af_skills_tab}) and go beyond captions - we leverage metadata such as song titles, artist names, album names, etc (see Fig.~\ref{fig:music_qa_example} for full list) to generate more complex, reasoning-focused QAs. We also use this metadata to generate rich music captions for Stage 1 and 2 pre-training (see Fig.~\ref{fig:music_qa_example}), demonstrating how text-based knowledge can enhance audio understanding, particularly in knowledge-driven domains like music. This method can be seen as synthetic knowledge generation, where we leverage text-based knowledge to enrich audio understanding and enable models to acquire domain-specific knowledge from unlabeled audios in the wild. Our analysis shows that LLMs like GPT-4.1 hold substantial world knowledge about music, and that metadata improves QA quality significantly.

\noindent (2) \textbf{We augment AudioSkills with 1M speech QA samples} using YouTube8M~\citep{abu2016youtube}, LibriSpeech~\citep{panayotov2015librispeech} (read speech), GigaSpeech~\citep{chen2021gigaspeech} (conversational), and VoxCeleb2~\citep{chung2018voxceleb2} (interviews). From YouTube8M, we introduce a new task: Speech-in-Sound QA, where the model must reason over both speech content and ambient sounds to understand complex auditory scenes. To create these QAs, we create Speech-in-Sound-Caps, a new dataset with $\approx$2M speech-aware auditory scene captions from YouTube8M. To curate this, we first filter the dataset for English speech (using AF2) and transcribe the spoken content with Whisper-Large-v3. We then generate two types of descriptions: one capturing sound events and another summarizing speech characteristics such as tone, emotion, and pitch (both using AF2 and custom prompts; see Appendix~\ref{fig:speech_sound_prompt}). Finally, we prompt GPT-4.1 to synthesize a speech-aware scene caption. These captions significantly improve the quality of final audio captions (compared to only using sound information) by providing a more holistic representation of the audio. For LibriSpeech and GigaSpeech, we concatenate shorter segments into clips of 15–30 seconds, selecting information-dense segments filtered by prompting an LLM. To move beyond basic spoken content understanding common in most current datasets~\citep{zhao2023librisqa}, we design five distinct types of speech QA that require diverse reasoning skills (explained in the next subsection). 
\vspace{-2mm}
\subsection{LongAudio-XL: Expanding LongAudio with Long Speech QA}
\label{subsec:longaudio}
\vspace{-1mm}

To our knowledge, Long Speech QA (i.e., audio $\ge$ 30 seconds) has not been explored in prior work, despite its relevance to real-world applications such as long-form conversation understanding, meeting summarization, and narrative comprehension. To bridge this gap, we extend the existing LongAudio dataset~\citep{kong2025audioflamingo2} (focused on sounds and music) by incorporating over 1M reasoning-focused QA examples from long-form speech (30s-10min). We curate audios from diverse sources including: \textit{Single-speaker speech}: LibriSpeech (audiobooks)~\citep{panayotov2015librispeech}, EuroParl~\citep{koehn2005europarl}, VoxPopuli (parliamentary debates)~\citep{wang2021voxpopuli} and \textit{Multi-speaker conversations}: Spotify Podcasts~\citep{clifton-etal-2020-100000}, Switchboard~\citep{godfrey1992switchboard}, Fisher (dyadic calls)~\citep{cieri2004fisher}, MELD~\citep{poria2018meld}, DailyTalk~\citep{lee2023dailytalk}, MMDialog (natural dialogues)~\citep{feng2022mmdialog}. We merge consecutive short segments in chronological order to construct longer, coherent audios. We construct QAs across a wide range of skills, as illustrated in \Cref{fig:audioskills}:

\begin{figure*}[!t]
    \centering
    \includegraphics[width=\linewidth]{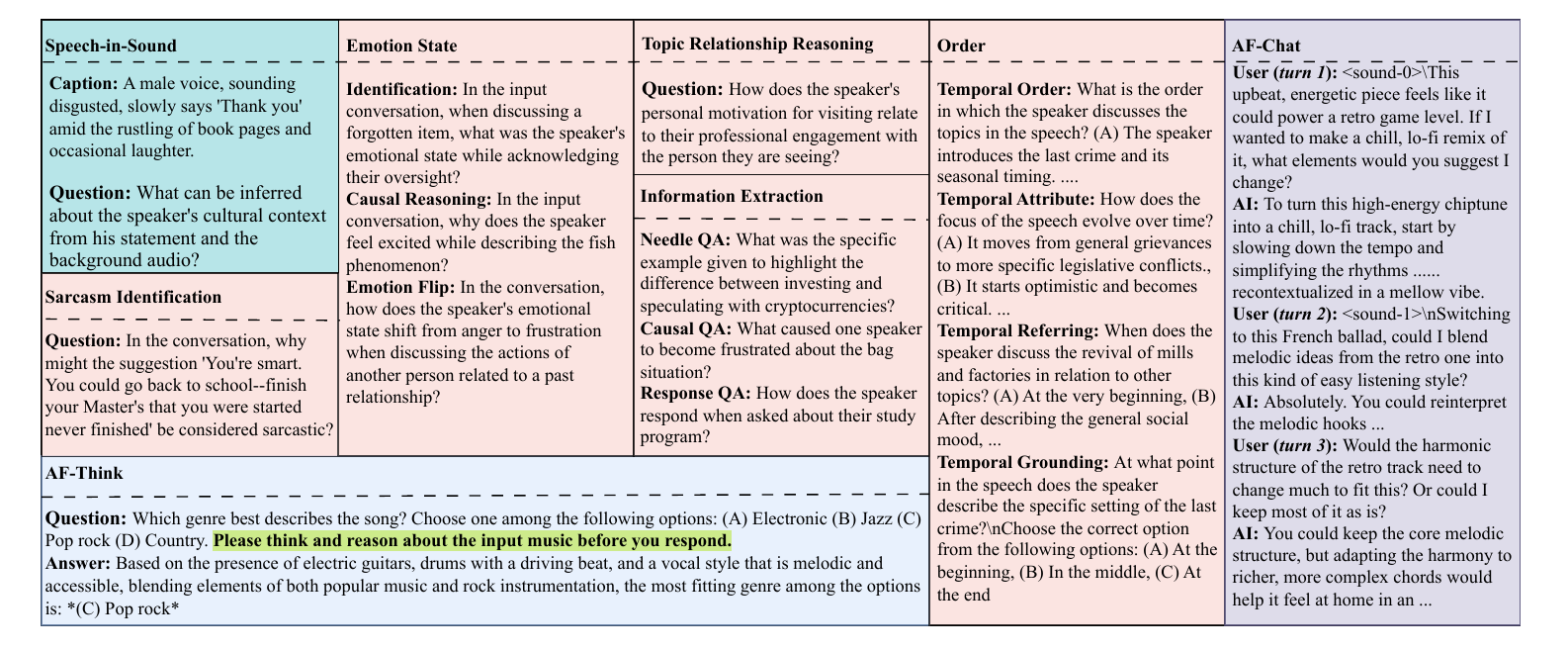}
    \caption{\small Examples from \colorbox{xl}{AudioSkill-XL}, \colorbox{long}{LongAudio-XL}, \colorbox{think}{AF-Think}, and \colorbox{chat}{AF-Chat.} We include additional examples in Appendix~\ref{sec.audioskills_app} and ~\ref{sec.longaudio_app}, featuring novel music reasoning QAs mentioned in detail in Section~\ref{subsubsec:music_know}.}
    \label{fig:audioskills}
    \vspace{-5mm}
\end{figure*}

\begin{enumerate}
\vspace{-3mm}
\setlength\parskip{0em}
    \item \textbf{Sarcasm Identification:} Inferring sarcasm by analyzing content, tone, and emotional cues.
    \item \textbf{Emotional State Reasoning:} \textbf{\textit{i) Identification:}} Determine the speaker's emotion at a specific utterance. \textbf{\textit{ii) Causal Reasoning:}} Identify the reason behind a speaker's emotional state using conversational context. \textbf{\textit{iii) Emotion Flip:}} Explain shifts in a speaker’s emotional state during the conversation.
    \item \textbf{Topic Relationship Reasoning:} Understand how two ideas or topics are related within the overall discourse.
    \item \textbf{Information Extraction (IE):} \textbf{\textit{i) Needle QA:}} Targeted QA on specific utterances or parts of the speech (e.g., entity or fact extraction, general knowledge linkage). \textbf{\textit{ii) Causal QA:}} Identify causes for a particular utterance in context. \textbf{\textit{iii) Response QA:}} Extract how one speaker responds to another’s statement. \textbf{\textit{iv) Topic QA:}} Identify the main topic of the speech or conversation.
    \item \textbf{Summarization:} Generate a concise summary of the speech content.
    \item \textbf{Order:} \textbf{\textit{i) Temporal Order:}} Understanding the sequential order of topics in the speech; \textbf{\textit{ii) Temporal Attribute:}} Understanding how topics change over time; \textbf{\textit{iii) Temporal Referring:}} Resolve references to specific time points (e.g., "at the end") \textbf{\textit{iv) Temporal Grounding:}} Identify when in the audio a specific topic was discussed.
\end{enumerate}    

\vspace{-2mm}
\subsection{AF-Think: Towards flexible, on-demand reasoning}
\vspace{-1mm}

Recent studies show that making an LLM “think”, similar to chain-of-thought (CoT) prompting~\citep{wei2022chain}, can improve reasoning performance in LLMs~\cite{guo2025deepseek}, especially for complex tasks like coding and math (e.g., DeepSeek-R1, OpenAI-o1). Visual MLLMs have also benefited from this paradigm~\cite{xu2024llava,wang2025multimodal}. In the audio domain, early attempts such as Audio-CoT~\cite{ma2025audio}, Audio-Reasoner~\cite{xie2025audio}, and R1-AQA~\cite{li2025reinforcement} have explored CoT-style reasoning, but often yield limited gains and involve complex or inefficient training procedures. Moreover, consistent with findings in~\cite{li2025reinforcement}, we observe that deep, explicit thinking does not always improve performance in audio understanding tasks.

In AF3, we adopt a lightweight thinking mechanism with two key modifications: (i) We create AF-Think, a dataset of 250k MCQ-based QAs with short, controlled thought preceding the answer. This additional thinking serve as a prefix to the answer and are limited to an average of approximately 40 words, providing concise yet effective context for audio QA (example in \Cref{fig:audioskills}). (ii) Instead of explicitly post-training for CoT, we add a special suffix to QA prompts (highlighted in \Cref{fig:audioskills}). We include AF-Think in the Stage 3.5 training mixture, upweighted relative to standard QA data. This allows AF3 to think only when prompted, offering \textit{flexible, on-demand additional reasoning}. 

To generate AF-Think, we first sample a subset of multiple-choice reasoning QAs from AudioSkills-XL and LongAudio-XL (originally with just the correct option as the answer). Next, we prompt Gemini 2.0 Flash with the input audio, the question, and the answer to generate short thinking prefixes. We found Gemini to hallucinate less and generate more accurate reasoning when guided by the ground-truth answer, rather than producing CoT from scratch. We restrict this process to only high-quality datasets and filter out noisy instances.

\vspace{-2mm}
\subsection{AF-Chat: Multi-turn Multi-audio Chat Data}
\vspace{-1mm}

While single-turn single-audio QA training equips LALMs to reason over individual audio inputs, enabling free-form, multi-turn, multi-audio conversations requires a dedicated chat alignment tuning stage, akin to the instruction-tuning phases used for LLMs~\cite{zhou2023lima}. Chat becomes significantly more complex when multiple audio inputs must be integrated across turns, requiring the model to track context, reason over relationships between past and current inputs, and generate coherent follow-ups. Despite its importance and chat being the most used application of LLMs, this capability remains underexplored in LALMs primarily due to the absence of open, high-quality training data.

To address this gap, we introduce \textit{AF-Chat}, a high-quality fine-tuning dataset consisting of 75k multi-turn, multi-audio chat instances. On average, each dialogue includes 4.6 audio clips and 6.2 dialogue turns, with a range of 2–8 audio clips and 2–10 turns. To construct this dataset, we draw from Speech-in-Sound Caps (for speech and sounds), and Music4All and MSD (for music). We follow a two-step curation process: First, for each seed audio, we identify its top 8 most semantically similar and dissimilar clips using a combination of captions, NV-Embed-v2~\cite{lee2024nv} embeddings, and FAISS-based clustering~\citep{douze2024faiss} (details in Appendix~\ref{subsec:clustering}). For every dialogue, we restrict the audios to this pool. This targeted clustering yields significantly higher-quality dialogues than random audio selection by ensuring each instance is grounded in a diverse yet semantically coherent audio pool.

Next, we prompt GPT-4.1 using carefully designed expert exemplars (Fig.~\ref{fig:music_chat} and ~\ref{fig:chat_sound}) to generate natural, multi-turn chat sessions under the following constraints: (i) the model may choose any subset of the similar/dissimilar audios (up to 10 turns), prioritizing conversation quality; (ii) not all turns require a new audio—follow-up and clarification questions are encouraged; and (iii) later turns may refer back to earlier audios or responses to simulate real conversational grounding. The design of AF-Chat is informed by extensive internal human studies to reflect how users naturally interact with audio-language models. As a result, it provides rich, diverse supervision for aligning LALMs to handle complex, contextual, and naturalistic audio conversations. Finally, we select 200 high-quality samples for the test set, known as AF-Chat-test, and ensure that the audios in these instances have audio clips that were not seen during training.

\vspace{-2mm}
\section{Audio Flamingo 3 Training Strategy}
\vspace{-1mm}

AF3 is trained using a five-stage strategy designed to progressively enhance its capabilities by increasing audio context length, improving data quality, and diversifying tasks. A full list of datasets used at each stage is provided in Appendix~\ref{tab:sft_datasets_app}. 

\textit{\textbf{Stage 1: Alignment pre-training.}} For this stage, we train only the audio adaptor layers while keeping the audio encoder and LLM frozen. This step aligns encoder representations with the language model. \textit{\textbf{Stage 2: Encoder Tuning.}} The main purpose of this stage is to adapt AF-Whisper to diverse datasets and broaden and improve its audio understanding capabilities. We fine-tune both the audio encoder and adaptor while keeping the LLM frozen. In both Stages 1 and 2, the audio context length is limited to 30 seconds, and training uses recognition-focused datasets (e.g., classification, captioning, and ASR). \textit{\textbf{Stage 3: Full Fine-Tuning.}} The primary purpose of this stage is to emphasize reasoning and skill acquisition by the LALM. As mentioned earlier, since skill-specific data is easy to scale on short audios, we still stick to short audios in this stage and use high-quality foundational and QA datasets and our proposed AudioSkills-XL. However, we increase the audio context length up to 2.5 minutes now to accommodate the moderately long audios in AudioSkills. The resulting model at the end of Stage 3.5 is referred to as AF3. \textit{\textbf{Stage 3.5: Context Extension and Thinking.}} This stage focuses on extending context length and encouraging CoT-style reasoning. In addition to the Stage 3 data mixture, we incorporate LongAudio-XL and AF-Think. We adopt LoRA-based training~\citep{hu2022lora}—similar to LTU and GAMA—by freezing the model's original weights and training LoRA adapters for the LLM. This approach allows end-users to flexibly enhance the model’s reasoning and long-context understanding capabilities on demand. \textit{\textbf{Stage 4: Chat and Voice Fine-Tuning.}} This stage focuses on enabling multi-turn, interactive, and voice-based dialogue. We fine-tune the entire model on our proposed AF-Chat dataset to equip AF3 with conversational audio understanding and response generation capabilities. The resulting model at the end of Stage 4 is referred to as AF3-Chat.

%% file: experiments.tex
\begin{table}[!ht]
\centering
\caption{\small Comparison of AF3 with other LALMs on various benchmarks (WER ↓ (Word Error Rate), ACC ↑ (Accuracy), and GPT4o ↑ (GPT evaluation)). We report scores for only the top-performing prior LALM. \textcolor{nvidiaGreen}{+Think} refers to AF3 with additional thinking. We highlight \textcolor{closedGray}{closed source}, \textcolor{qwenPurple}{open weights}, and \textcolor{nvidiaGreen}{open source} models.}
\vspace{2mm}
\resizebox{\linewidth}{!}{%
\begin{tabular}{llccc}
\toprule
\textbf{Task} & \textbf{Dataset} & \textbf{Prior SOTA} & \textbf{Metrics} & \textbf{Results} \\
\midrule

\multirow{41}{*}{\shortstack[c]{\textbf{Audio} \\ \textbf{Understanding} \\ \textbf{and} \\\textbf{Reasoning}}} 
& \multirow{3}{*}{\shortstack[l]{\textbf{MMAU-v05.15.25 (test)} \\ \textit{Sound | Music | Speech | Avg} }}
& \textcolor{qwenPurple}{Qwen2.5-O} & \multirow{3}{*}{ACC ↑} & \textbf{76.77} | 67.33 | \textbf{68.90} |  71.00 \\
& & \textcolor{nvidiaGreen}{Audio Flamingo 3} & & 75.83 | \textbf{74.47} | 66.97 | \textbf{72.42}  \\
& & \textcolor{nvidiaGreen}{+Think} & & \textcolor{black!50}{75.27} | \textcolor{black!50}{74.60} | \textcolor{black!50}{69.60} | \textcolor{black!50}{73.16}  \\\cmidrule{2-5}

& \multirow{3}{*}{\shortstack[l]{\textbf{MMAU-v05.15.25 (test-mini)} \\ \textit{Sound | Music | Speech | Avg} }}
& \textcolor{qwenPurple}{Qwen2.5-O} & \multirow{3}{*}{ACC ↑} & 78.10 | 65.90 | \textbf{70.60} |  71.50 \\
& & \textcolor{nvidiaGreen}{Audio Flamingo 3} & & \textbf{79.58} | \textbf{73.95} | 66.37 | \textbf{73.30}  \\
& & \textcolor{nvidiaGreen}{+Think} & & \textcolor{black!50}{79.88} | \textcolor{black!50}{76.55} | \textcolor{black!50}{66.37} | \textcolor{black!50}{74.26}  \\ \cmidrule{2-5}

& \multirow{3}{*}{\textbf{MMAR}} & \textcolor{qwenPurple}{Qwen2.5-O} & \multirow{3}{*}{ACC ↑} & 56.7 \\
& & \textcolor{nvidiaGreen}{Audio Flamingo 3} & & \textbf{58.5} \\ 
& & \textcolor{nvidiaGreen}{+Think} & & \textcolor{black!50}{60.1}  \\ \cmidrule{2-5}

& \multirow{3}{*}{\textbf{MMSU}} & \textcolor{closedGray}{Gemini-1.5-Pro} & \multirow{3}{*}{ACC ↑} & 60.7 \\
& & \textcolor{nvidiaGreen}{Audio Flamingo 3} & & \textbf{61.4} \\ 
& & \textcolor{nvidiaGreen}{+Think} & & \textcolor{black!50}{62.3}  \\ \cmidrule{2-5}

& \multirow{2}{*}{\shortstack[l]{\textbf{ClothoAQA} \\ \textit{unanimous | non-binary} }} & \textcolor{qwenPurple}{Qwen2.5-O} | \textcolor{qwenPurple}{Qwen2.5-O}& \multirow{2}{*}{ACC ↑} & 89.2 | 52.6  \\
& & \textcolor{nvidiaGreen}{Audio Flamingo 3} & & \textbf{91.1} | \textbf{56.2} \\ \cmidrule{2-5}

& \multirow{2}{*}{\shortstack[l]{\textbf{Audio Captioning} \\ \textit{Clotho-v2 | AudioCaps} }} & \textcolor{nvidiaGreen}{Audio Flamingo 2} | \textcolor{nvidiaGreen}{Audio Flamingo 2} &\multirow{2}{*}{CIDEr ↑} & 0.46 | 0.58  \\
& & \textcolor{nvidiaGreen}{Audio Flamingo 3} & & \textbf{0.50} | \textbf{0.70} \\ \cmidrule{2-5}

& \multirow{2}{*}{\shortstack[l]{\textbf{Audio Entailment} \\ \textit{Clotho | AudioCaps} }} & \textcolor{nvidiaGreen}{Audio Flamingo 2} | \textcolor{nvidiaGreen}{Audio Flamingo 2} & \multirow{2}{*}{ACC ↑} & 92.5 | 93.3  \\
& & \textcolor{nvidiaGreen}{Audio Flamingo 3} & & \textbf{93.3} |  \textbf{95.0} \\ \cmidrule{2-5}

& \multirow{2}{*}{\textbf{IEMOCAP}} & \textcolor{qwenPurple}{Qwen2-A-Inst} & \multirow{2}{*}{ACC ↑} & 59.2 \\
& & \textcolor{nvidiaGreen}{Audio Flamingo 3} & & \textbf{63.8} \\\cmidrule{2-5}

& \multirow{2}{*}{\textbf{CochlScene}} & \textcolor{nvidiaGreen}{Pengi} & \multirow{2}{*}{ACC ↑} & 91.6 \\
& & \textcolor{nvidiaGreen}{Audio Flamingo 3} & & \textbf{93.2} \\\cmidrule{2-5}

& \multirow{2}{*}{\textbf{NonSpeech7k}}
   & \textcolor{nvidiaGreen}{Audio Flamingo 2} & \multirow{2}{*}{ACC ↑} & 84.3 \\
& & \textcolor{nvidiaGreen}{Audio Flamingo 3} & & \textbf{85.9}  \\ \cmidrule{2-5}

& \multirow{2}{*}{\textbf{CMM Hallucination}} & \textcolor{closedGray}{Gemini 2.5 Pro} & \multirow{2}{*}{ACC ↑} & 82.0 \\
& & \textcolor{nvidiaGreen}{Audio Flamingo 3} & & \textbf{86.5} \\\cmidrule{2-5}
& \multirow{2}{*}{\textbf{CompA-R-\textit{test}}} & \textcolor{nvidiaGreen}{Audio Flamingo 2} &\multirow{2}{*}{ACC ↑}  & 96.4 \\
& & \textcolor{nvidiaGreen}{Audio Flamingo 3} & & \textbf{98.0}   \\
\cmidrule{2-5}

& \multirow{2}{*}{\textbf{MusicAVQA}} & \textcolor{qwenPurple}{Qwen2.5-O} & \multirow{2}{*}{ACC ↑} & 73.4 \\
& & \textcolor{nvidiaGreen}{Audio Flamingo 3} & & \textbf{76.7}  \\ \cmidrule{2-5}

& \multirow{2}{*}{\shortstack[l]{\textbf{NSynth} \\ \textit{Source | Instrument} }} & \textcolor{nvidiaGreen}{Pengi} | \textcolor{qwenPurple}{Qwen-A} & \multirow{2}{*}{ACC ↑} & 62.0 | 78.8 \\
& & \textcolor{nvidiaGreen}{Audio Flamingo 3} & & \textbf{65.5} | \textbf{78.9} \\ \cmidrule{2-5}

& \multirow{2}{*}{\shortstack[l]{\textbf{Music Instruct} \\ \textit{Long} }} & \textcolor{nvidiaGreen}{Audio Flamingo 2} & \multirow{2}{*}{ACC ↑} & 90.2 \\
& & \textcolor{nvidiaGreen}{Audio Flamingo 3} & &  \textbf{92.7} \\ \cmidrule{2-5}

& \multirow{3}{*}{\textbf{MuchoMusic}} & \textcolor{qwenPurple}{Qwen2-A-Inst} & \multirow{3}{*}{ACC ↑} & 46.2 \\
& & \textcolor{nvidiaGreen}{Audio Flamingo 3} & & \textbf{47.4}   \\ 
& & \textcolor{nvidiaGreen}{+Think} & & \textcolor{black!50}{47.6}   \\

\cmidrule{2-5}

& \multirow{2}{*}{\textbf{LibriSQA}} & \textcolor{closedGray}{Gemini 2.5 Pro} & \multirow{2}{*}{GPT4o ↑} & 8.7 \\
& & \textcolor{nvidiaGreen}{Audio Flamingo 3} & & \textbf{8.7}   \\ 

\cmidrule{2-5}
& \multirow{2}{*}{\textbf{LongAudioBench}}
    & \textcolor{closedGray}{Gemini 2.5 Pro} & \multirow{2}{*}{GPT4o ↑} & 60.4 \\
& & \textcolor{nvidiaGreen}{Audio Flamingo 3} & & \textbf{68.6}   \\ 
& \multirow{2}{*}{\textbf{+Speech (ours)}}
    & \textcolor{closedGray}{Gemini 2.5 Pro}  & \multirow{2}{*}{GPT4o ↑} & 66.2 \\
& & \textcolor{nvidiaGreen}{Audio Flamingo 3} & &  \textbf{72.9}  \\ 
\midrule

\multirow{12}{*}{\shortstack[c]{\textbf{Automatic} \\ \textbf{Speech} \\ \textbf{Recognition} \\\textbf{(ASR)}}} 
& \multirow{2}{*}{\shortstack[l]{\textbf{LibriSpeech (en)} \\ \textit{test-clean | test-other}}}

& \textcolor{qwenPurple}{Phi-4-mm} | \textcolor{qwenPurple}{Qwen2.5-O} & \multirow{2}{*}{WER ↓} &  1.67 | 3.4 \\ 
& & \textcolor{nvidiaGreen}{Audio Flamingo 3} & &  \textbf{1.57} | \textbf{3.13} \\ 
\cmidrule{2-5}

& \multirow{2}{*}{\textbf{SPGISpeech (en)}} 
& \textcolor{qwenPurple}{Qwen2-A-Inst} & \multirow{2}{*}{WER ↓}& 3.0 \\
& & \textcolor{nvidiaGreen}{Audio Flamingo 3} & & \textbf{1.86} \\ \cmidrule{2-5}

& \multirow{2}{*}{\textbf{TEDLIUM (en)}} & \textcolor{qwenPurple}{Phi-4-mm} & \multirow{2}{*}{WER ↓} & \textbf{2.9} \\

& & \textcolor{nvidiaGreen}{Audio Flamingo 3} & & 3.5 \\ \cmidrule{2-5}

& \multirow{2}{*}{\textbf{GigaSpeech (en)}} & \textcolor{qwenPurple}{Phi-4-mm} & \multirow{2}{*}{WER ↓} & \textbf{9.78} \\
& & \textcolor{nvidiaGreen}{Audio Flamingo 3} & &  10.27 \\ \cmidrule{2-5}

& \multirow{2}{*}{\textbf{Common Voice 15 (en)}} & \textcolor{qwenPurple}{Phi-4-mm} & \multirow{2}{*}{WER ↓} & 7.61 \\
& & \textcolor{nvidiaGreen}{Audio Flamingo 3} & & \textbf{7.4} \\ \cmidrule{2-5}

& \multirow{2}{*}{\textbf{VoxPopuli (en)}} & \textcolor{qwenPurple}{Phi-4-mm} & \multirow{2}{*}{WER ↓} & 5.91 \\
& & \textcolor{nvidiaGreen}{Audio Flamingo 3} & & \textbf{5.55} \\

\bottomrule
\end{tabular}%
}
\vspace{-4mm}
\label{tab:main_results}
\end{table}

\noindent \textbf{Experimental Setup.} We train AF3 on 128 NVIDIA A100 GPUs, each with 80GB of memory. Details about batch size, learning rates, and optimizers for each stage of training are in \cref{sec.af3_training_details}. 

\noindent \textbf{Baselines.} We evaluate our model against recent SOTA LALMs, including GAMA~\citep{ghosh2024gama}, Audio Flamingo~\citep{kong2024audioflamingo}, Audio Flamingo 2~\citep{kong2025audioflamingo2}, Qwen-A(udio)~\citep{chu2023qwenaudio}, Qwen2-A(udio)~\citep{chu2024qwenaudio2}, Qwen2-A(udio)-(Inst)ruct, Qwen2.5-O(mni)~\citep{xu2025qwen2}, R1-AQA~\cite{li2025reinforcement}, Pengi~\citep{deshmukh2023pengi}, Phi-4-mm~\citep{abouelenin2025phi}, Baichun Audio~\citep{li2025baichuan}, Step-Audio-Chat~\citep{huang2025step}, LTU~\citep{gong2023ltu}, LTU-AS~\citep{gong2023ltu-as}, SALMONN~\citep{tang2023salmonn}, AudioGPT~\citep{huang2023audiogpt}, and Gemini (2.0 Flash, 1.5 Pro, 2.5 Flash and 2.5 Pro)~\citep{team2023gemini} (note we do not evaluate Gemini on ASR benchmarks due to low rate limits), as well as GPT-4o-audio~\citep{hurst2024gpt}. For LongAudioBench, for models that do not support longer audio, we follow the cascaded approach for evaluation proposed by~\citep{kong2025audioflamingo2}. For Table~\ref{tab:voice_results}, we only compare against open LALMs. All results reported in the tables correspond to the best-performing model. Evaluation for voice-to-voice capabilities is beyond our scope.

\noindent \textbf{Evaluation Datasets.} We evaluate AF3 on a variety of tasks and benchmarks, including \textit{audio classification} (CochlScene~\citep{jeong2022cochlscene}, NSynth (Source and Instrument)~\citep{engel2017neural}, NonSpeech7k~\citep{rashid2023nonspeech7k}, IEMOCAP~\citep{busso2008iemocap}), \textit{audio QA} (ClothoAQA~\citep{lipping2022clotho}, MusicAVQA~\citep{li2022learning}, Music Instruct~\citep{deng2023musilingo}, LibriSQA~\citep{zhao2023librisqa}), \textit{reasoning-focused audio QA} (MMAU~\citep{sakshi2024mmau} (v05.15.25), MuchoMusic (perceptual version)~\citep{zang2025you,weck2024muchomusic}, MMAR~\citep{ma2025mmarchallengingbenchmarkdeep}, MMSU~\citep{wang2025mmsu}, CompA-R-test~\citep{ghoshcompa}, Audio Entailment~\citep{deshmukh2025audio}), \textit{multimodal hallucination detection} (CMM~\citep{leng2024curse}), \textit{audio captioning} (Clotho-v2~\citep{drossos2020clotho}, AudioCaps~\citep{kim2019audiocaps}), \textit{ASR} (Librispeech (clean and other)~\citep{panayotov2015librispeech}, SPGISpeech~\citep{o2021spgispeech}, TEDLIUM~\citep{rousseau2012ted,hernandez2018ted}, GigaSpeech (Large)~\citep{chen2021gigaspeech}, Common Voice 15~\citep{commonvoice:2020} and Voxpopuli~\citep{wang2021voxpopuli}) and \textit{long audio captioning and QA} (LongAudioBench -- which we augment with 2.5k human-annotated long-speech QA instances). For evaluating chat capabilities, we conduct a human study of model outputs on AF-Chat-test (more details in \cref{sec.afchat}) and compare only with Qwen2-Audio. Each annotator is asked to rate the response of the model for every turn on a scale of 1-5 for factuality, usefulness, and depth. We report results averaged across all instances across all turns. Furthermore, we evaluate the voice-text capabilities of our AF3-Chat model on two datasets, OpenAudioBench~\citep{li2025baichuan} and VoiceBench~\citep{chen2024voicebench}. These benchmarks consist of voice queries (synthetically generated speech from text queries) and assess aspects such as instruction following, question answering, trivia knowledge, and reasoning. Finally, we evaluate our speech generation module using zero-shot TTS evaluation on the English subset of the SEED benchmark~\citep{anastassiou2024seed}. All baseline results reported in this work are based on our own evaluations; we did not rely on results from prior literature (in some cases, we were unable to reproduce the numbers as originally reported). To calculate accuracy, we use either exact string matching with the ground truth or CLAP-based retrieval following~\citep{deshmukh2023pengi}, implemented with open-source AF-CLAP~\citep{kong2025audioflamingo2}. For MCQ, AF3 typically outputs only the selected option. In cases where the model provides more verbose or open-ended responses (e.g., with thinking mode), we apply multiple regex patterns to extract the chosen option.

\begin{table}[t]
\centering
\caption{\small Comparison of AF3 with open LALMs on AF-Chat, voice-text and TTS benchmarks. WER ↓ (Word Error Rate), SIM ↑ (Similarity), and GPT4o ↑ (GPT evaluation) indicate that lower or higher is better.}
\vspace{2mm}
\resizebox{\linewidth}{!}{
\begin{tabular}{lllcc}
\toprule
\textbf{Task} & \textbf{Dataset} & \textbf{Model} & \textbf{Metrics} & \textbf{Results} \\
\midrule

\multirow{2}{*}{\textbf{Multi-audio chat}}
& \multirow{2}{*}{\shortstack[l]{\textbf{AF-Chat-test} \\ \textit{Factuality | Usefulness | Depth}}} & \textcolor{qwenPurple}{Qwen2.5-O} & \multirow{2}{*}{GPT4o ↑} & 2.4 | 2.7 | 3.2 \\
& & \textcolor{nvidiaGreen}{AF3-Chat} & & \textbf{3.6} | \textbf{3.4} | \textbf{3.9} \\ \midrule

\multirow{6}{*}{\textbf{Voice-Text}}
& \multirow{3}{*}{\shortstack[l]{\textbf{OpenAudioBench} \\ \textit{alpaca-eval | llama-questions |} \\ \textit{trivia-qa }}} & \textcolor{qwenPurple}{Qwen2-A-Inst} & \multirow{3}{*}{GPT4o ↑} & 57.19 | 69.67 | 40.30  \\

& & \textcolor{qwenPurple}{Qwen2.5-O}  & & 72.76 | 75.33 | \textbf{57.06} \\

& & \textcolor{nvidiaGreen}{AF3-Chat} & & \textbf{76.26} | \textbf{80.33} | 53.05   \\ \cmidrule{2-5}

& \multirow{3}{*}{\shortstack[l]{\textbf{VoiceBench} \\ \textit{AlpacaEval | AdvBench |} \\ \textit{OpenBookQA | Commoneval}}}
    & \textcolor{qwenPurple}{Qwen2-A-Inst} & \multirow{3}{*}{GPT4o ↑} & 3.69 | 98.85 | 49.01 | 3.40 \\

& & \textcolor{qwenPurple}{Qwen2.5-O}  & & \textbf{4.33} | \textbf{99.62} | \textbf{79.12} | \textbf{3.84} \\

& & \textcolor{nvidiaGreen}{AF3-Chat} & & 4.19 | 98.26 | 66.81 | 3.40  \\ \midrule

\multirow{2}{*}{\textbf{Speech Generation}}
& \multirow{2}{*}{\shortstack[l]{\textbf{SEED (test-en)} \\ \textit{Content Cons. | Speaker Sim. | Inf. Time }}} & \textcolor{qwenPurple}{Qwen2.5-O} & \multirow{2}{*}{WER ↓ | SIM ↑ | Time ↓} & 2.72 |  \textbf{0.63} | 14.62s (1.26s) \\
& & \textcolor{nvidiaGreen}{AF3-Chat} & & \textbf{2.02} | 0.61 | \textbf{ 5.94s (0.02s)} \\ 

\bottomrule
\end{tabular}}
\label{tab:voice_results}
\end{table}

\subsection{Audio Understanding and Reasoning Evaluation}

\noindent \textbf{AF3 is the strongest and fully open-source LALM. }\Cref{tab:main_results} shows AF3 outperforming previous SOTA open-weight and closed-source models across a wide range of audio understanding and reasoning benchmarks. AF3 sets new highs on MMAU (72.42) (note for Qwen2.5-Omni on MMAU we report the ``parsed score'' for fair evaluation), ClothoAQA (91.1), Clotho Entailment (92.9), and CMM Hallucination (86.7). On tasks like NSynth and MusicInstruct, it shows significant gains, highlighting strong sound and music understanding. For LongAudioBench (sound and speech), AF3 outperforms Gemini 2.5 Pro by a wide margin, demonstrating its strength in long-context reasoning. We also evaluate AF3 with thinking prompts (+Think) on reasoning-heavy benchmarks like MMAU and MuchoMusic, observing a performance boost. Although the thinking mode is activated after Stage 3.5 only when using our specific thinking prompt, the checkpoint remains usable without it. We report average scores of 73.16 and 74.26 on MMAU-test and MMAU-test-mini, respectively. Additionally, AF3 achieves state-of-the-art ASR results on LibriSpeech, SPGISpeech, and VoxPopuli—even compared to dedicated ASR models—despite not being trained on large-scale ASR datasets like many open-weight models. We illustrate a demo of AF3's capabilities in Fig.~\ref{fig:superman}.

\subsection{Chat and TTS Evaluation}

\textbf{Multi-turn multi-audio chat evaluation. } On \textit{AF-Chat-test} AF3-Chat shows a relative improvement of 30\% over Qwen2.5-Omni, thereby showing the capability of effectively handling extended dialog turns, allowing for deeper contextual reasoning and more accurate references to multiple audio inputs.

\textbf{Voice-Text and Speech Generation Evaluation. }  \Cref{tab:voice_results} evaluates AF3-Chat on two key tasks: voice-to-text and text-to-speech generation. In the voice-to-text setting (spoken QA), AF3-Chat achieves strong gains across all of OpenAudioBench, surpassing Qwen2.5-Omni. On VoiceBench, which tests spoken QA robustness across AdvBench, CommonEval, and OpenBookQA, AF3-Chat performs comparably to Qwen2.5-Omni and Qwen2-Audio Chat. For TTS (evaluated on SEED test-en), AF3-Chat shows improved performance with a lower WER of 2.02 (vs. 2.72 for Qwen2.5-Omni) and a speaker similarity score of 0.61, closely matching Qwen2.5’s 0.63.

Furthermore, AF3-Chat exhibits significant advantages in generation speed. For a 10-second audio generation on an A100 GPU, AF3-Chat’s text-to-audio token generation is 5.94 seconds with an additional 0.02 seconds for waveform synthesis. In comparison, the Talker model of Qwen2.5-Omni requires 14.62 seconds for token generation and an additional 1.26 seconds for waveform synthesis. This efficiency allows our streaming text-to-speech to achieve a time-to-first-token of 0.15 seconds and an inter-token latency of 0.06 seconds (both including waveform synthesis), producing a 10-second audio clip in 6.68 seconds.

\subsection{Ablation Studies}
In this section, we ablate our key components (using just 10\% of the training data) to support the paper's main claims. 

\noindent \textbf{Evaluating AF-Whisper as a Unified Encoder.} \Cref{tab:ablation} compares AF3 trained with our unified AF-Whisper encoder against a dual-encoder setup using CLAP for sounds/music and Whisper-v3 for speech~\citep{elizalde2022clap,radford2022whisper}. AF-Whisper outperforms the dual-encoder model under the same data budget, demonstrating its effectiveness as a single encoder for sound, music, and speech.

\noindent \textbf{AudioSkills-XL: A Key Dataset for Performance Gains.}: To measure the impact of AudioSkills-XL, we ablate it from Stage 3 of training and compare results to the full setup. As shown in \cref{tab:ablation}, removing AudioSkills-XL causes a significant performance drop—particularly on MMAU—underscoring its role in improving generalization and robustness. These findings highlight the value of large-scale, skill-targeted audio QA data for fine-tuning multi-modal models.

\begin{table}[ht]
\centering
\caption{\small Comparison of AF3 w/ 10\% data, w/o AF-Whisper and w/o AudioSkills-XL.}
\resizebox{\linewidth}{!}{
\begin{tabular}{lcccccccc}
\toprule
Model &{\textbf{MMAU-Sound}} 
& { \textbf{MMAU-Music}} 
& { \textbf{MMAU-Speech}} 
&{ \textbf{Librispeech-clean}} 
& {\textbf{Librispeech-other}}  \\
&{ACC ↑} 
& {ACC ↑} 
&{ACC ↑} 
& {WER ↓} 
&{WER ↓}  \\
\midrule
w/ 10\% data       & 66.7 & 65.9 & 57.4 & 2.0 & 4.1 \\
+ w/o AF-Whisper   & 63.7 & 68.3 & 45.2 & 3.7 & 7.2   \\ \midrule
w/o AudioSkills-XL      & 56.1 & 42.1 & 14.3 & 1.6 & 3.6   \\ \midrule
Audio Flamingo 3   & \textbf{75.8} & \textbf{74.4} & \textbf{66.9} & \textbf{1.5} & \textbf{3.1}   \\
\bottomrule
\end{tabular}}
\label{tab:ablation}

\end{table}

%% file: discussion.tex
In this paper, we introduce Audio Flamingo 3, the most capable and open LALM. Our model leverages a custom Whisper, novel data curation techniques, and a 5-stage curriculum learning strategy. Audio Flamingo 3 not only achieves SOTA performance in audio understanding and reasoning but also introduces capabilities, including multi-turn multi-audio chat, on-demand thinking, and voice chat. We detail our practices, including architecture, training, inference, and the evaluation pipeline, and open-source two large datasets.
For future work, we aim to address current limitations, including: (1) mitigating the need for a cascaded system for voice chat, (2) making AF3 multi-lingual, and (3) reducing dependency on closed-source models for synthetic data.

%% file: appendix.tex
\definecolor{xl}{HTML}{B0E3E6}
\definecolor{long}{HTML}{FAD9D5}
\definecolor{think}{HTML}{DAE8FC}
\definecolor{chat}{HTML}{D0CEE2}
\definecolor{nvidiaGreen}{RGB}{118,185,0}
\definecolor{qwenPurple}{RGB}{116,81,207}
\definecolor{closedGray}{RGB}{90,90,110}
\renewcommand*{\thesection}{\Alph{section}}

\section{AF-Whisper}

\subsection{Training Details}
\label{sec.af_whisper_training}

We train AF-Whisper on 128 NVIDIA A100 80GB GPUs. During training, we use an effective batch size of 1024, the AdamW optimizer (learning rate = \(10^{-4}\), weight decay = 0.1), and train using fp16 precision. We train for 5 epochs on the complete dataset and sample instances randomly from the entire pool for each batch.

\subsection{Training Datasets}
\label{sec.af_whisper_datasets}

Table~\ref{tab:clap_datasets} lists the datasets used to train AF-Whisper. For each dataset, we follow the same process outlined in Section 3 of the main paper: generating transcripts, spoken language characteristics, and audio captions. When available, we incorporate gold-standard metadata for these elements (for e.g., transcripts for LibriSpeech or captions for AudioCaps). GPT-4.1 is prompted to produce the final caption using a format similar to Fig.~\ref{fig:transcript_prompt}, with a modified exemplar. For extracting spoken language characteristics using AF2, we use the following prompt: ``There is a human speaking in the audio. Describe in detail the characteristics of the spoken utterance, including pitch, emotion, mood, speed, and other speech dynamics.''
\vspace{-3mm}
\begin{table}[!h]
    \centering
    \caption{\small Statistics of audio-caption datasets used for AF-Whisper training.}
    \resizebox{0.5\columnwidth}{!}{
    \begin{tabular}{cc}
    \toprule
        Dataset & \#Audio-Text Pairs \\ \midrule
        GigaSpeech (L)~\citep{chen2021gigaspeech}& 2,266,371\\
        Speech-in-Sound Captions$^*$~\cite{abu2016youtube} & 1,999,959  \\
        SPGISpeech~\citep{o2021spgispeech} & 1,966,109\\
        Sound-VECaps~\cite{yuan2025sound} & 1,657,029  \\
        Million Songs Dataset~\cite{bertin2011million} & 1,169,997 \\
        Common Voice 15~\cite{commonvoice:2020} & 1,109,689\\
        MiraData~\cite{ju2024miradata} & 748,320 \\
        Action2sound$^*$~\cite{chen2024action2sound} & 306,602 \\
        NSynth~\cite{engel2017neural} & 289,205 \\
        LibriSpeech~\citep{panayotov2015librispeech} &  281,241\\
        Freesound~\cite{fonseca2017freesound} & 256,695 \\
        AudioSet Strong$^*$~\cite{hershey2021benefit} & 216,622 \\
        VGGSound~\cite{chen2020vggsound} & 185,161 \\
        VoxPopuli (en)~\citep{wang2021voxpopuli} & 177,019\\
        FMA~\cite{defferrard2016fma} & 106,412 \\
        Video Recap~\cite{islam2024video} & 64,627 \\
        CochlScene~\cite{jeong2022cochlscene} & 60,855 \\
        Music4All~\citep{santana2020music4all} & 109269 \\
        Switchboard~\citep{godfrey1992switchboard} & 76,652 \\
        FSD50k~\cite{fonseca2021fsd50k} & 40,966 \\
        MACS~\cite{morato2021diversity} & 31,675 \\
        BBC{\footnote{\url{https://sound-effects.bbcrewind.co.uk/}}} & 31,201 \\
        MagnaTagATune~\cite{law2010evaluation} & 25,863 \\
        SoundDescs~\cite{koepke2022audio} & 23,085 \\
        Clotho~\cite{drossos2020clotho} & 19,195 \\
        TAU-Urban~\cite{mesaros2018multi} & 14,400 \\
        MusicCaps~\cite{agostinelli2023musiclm} & 5,479 \\
        WavText5K~\cite{deshmukh2022audio} & 4,347 \\
        SONICS~\cite{rahman2024sonics} & 1,602 \\
        SoundBible{\footnote{\url{https://soundbible.com/}}}  & 935 \\
        MUSDB18~\cite{rafii2017musdb18} & 276 \\
        Medleydb-Pitch~\cite{bittner2014medleydb} & 103 \\
    \midrule
    \textbf{Total} & 13,246,961 \\
    \bottomrule
    \end{tabular}}
    \label{tab:clap_datasets}
\end{table}

\section{AudioSkills-XL}
\label{sec.audioskills_app}

Table~\ref{tab:af_skills_tab} provides all details, including statistics and references to prompts we used for generating AudioSkills-XL.

\begin{table*}[h]
\centering
\caption{\small Detailed statistics of AudioSkills-XL, categorized into individual reasoning types, together with details on open-source datasets, additional meta-data, and prompts used for QA generation. * indicates that these types are further categorized into skills, and we elaborate on this in Section~\ref{subsec:skill_wise}. Rows \textbf{not} grayed out are the contributions of this paper. Speech QA types are the same as LongAudio-XL and explained in Section 4.2, with examples in Figure 3 and more examples in Appendix~\ref{sec.longaudio_app}.}
\resizebox{\linewidth}{!}{
\begin{tabular}{lllll}
\toprule
\textbf{Question Type} & \textbf{Size} & \textbf{Datasets Used} & \textbf{Meta-Data Used} & \textbf{Prompt Reference} \\
\midrule
\textcolor{black!50}{Temporal} & \textcolor{black!50}{188K} & \textcolor{black!50}{Table 14 in~\citep{kong2025audioflamingo2}} & \textcolor{black!50}{Table 14 in~\citep{kong2025audioflamingo2}} & \textcolor{black!50}{Table 14 in~\citep{kong2025audioflamingo2}} \\
+ ours & 350K & Synthetic Data & - & pythonic \\
\textcolor{black!50}{Attribute Identification} & \textcolor{black!50}{201K} & \textcolor{black!50}{Table 14 in~\citep{kong2025audioflamingo2}} & \textcolor{black!50}{Table 14 in~\citep{kong2025audioflamingo2}} & \textcolor{black!50}{Table 14 in~\citep{kong2025audioflamingo2}} \\
\textcolor{black!50}{Counting} & \textcolor{black!50}{50K} & \textcolor{black!50}{Table 14 in~\citep{kong2025audioflamingo2}} & \textcolor{black!50}{Table 14 in~\citep{kong2025audioflamingo2}} & \textcolor{black!50}{Table 14 in~\citep{kong2025audioflamingo2}} \\
\textcolor{black!50}{Contextual Sound Event Reasoning} & \textcolor{black!50}{982K} & \textcolor{black!50}{Table 14 in~\citep{kong2025audioflamingo2}} & \textcolor{black!50}{Table 14 in~\citep{kong2025audioflamingo2}} & \textcolor{black!50}{Table 14 in~\citep{kong2025audioflamingo2}} \\
\textcolor{black!50}{Contextual Speech Event Reasoning} & \textcolor{black!50}{1,272K} & \textcolor{black!50}{Table 14 in~\citep{kong2025audioflamingo2}} & \textcolor{black!50}{Table 14 in~\citep{kong2025audioflamingo2}} & \textcolor{black!50}{Table 14 in~\citep{kong2025audioflamingo2}} \\
\textcolor{black!50}{Information Extraction} & \textcolor{black!50}{858K} & \textcolor{black!50}{Table 14 in~\citep{kong2025audioflamingo2}} & \textcolor{black!50}{Table 14 in~\citep{kong2025audioflamingo2}} & \textcolor{black!50}{Table 14 in~\citep{kong2025audioflamingo2}} \\
\textcolor{black!50}{General Reasoning} & \textcolor{black!50}{704K} & \textcolor{black!50}{Table 14 in~\citep{kong2025audioflamingo2}} & \textcolor{black!50}{Table 14 in~\citep{kong2025audioflamingo2}} & \textcolor{black!50}{Table 14 in~\citep{kong2025audioflamingo2}} \\
+ ours (only sound) & 300K & YouTube8M & caption & Fig.~\ref{fig:yes_no_qa} \\
Sound Reasoning* (ours) & 300K & YouTube8M & caption & Fig.~\ref{fig:sound_3_prompt},~\ref{fig:sound_2_prompt},~\ref{fig:sound_1_prompt} \\
Music Knowledge* (ours) & 1,000K & MusicBench, Music4All, MSD & captions, dataset-specific meta-data & Fig.~\ref{fig:music4all_knowledge},~\ref{fig:msd_knowledge} \\
Music Reasoning* (ours) & 1,000K & MusicBench, Music4All, MSD & captions, dataset-specific meta-data & Fig.~\ref{fig:msd_reasoning},~\ref{fig:music4all_openqa_prompt},~\ref{fig:music4all_reasoning_qa_prompt} \\
Speech-in-Sound QA (ours) & 1,739K & Speech-in-Sound Caps (YouTube8M) & Caption, Transcripts, Speech Characteristics & Fig.~\ref{fig:speech_sound_prompt}~\ref{fig:transcript_prompt} \\
Speech QA* (ours) & 200K & LibriSpeech, GigaSpeech, VoxCeleb2 & Transcripts & Fig.~\ref{fig:reason_behind_prompt},~\ref{fig:order_prompt},~\ref{fig:detail_prompt},~\ref{fig:respond_prompt} \\
\bottomrule
\end{tabular}}
\label{tab:af_skills_tab}
\end{table*}

\vspace{-2mm}
\subsection{Skill-Wise Breakdown}
\label{subsec:skill_wise}

\subsubsection{Music Reasoning}

\noindent \textbf{Genre and Style:} Focuses on the model's ability to infer musical genre or stylistic influences by analyzing instrumentation, arrangement, and production characteristics.

\noindent \textbf{Mood and Expression:} Focuses on how well the model interprets the emotional tone or affective content conveyed by the music, such as melancholy, uplifting, or aggressive moods.

\noindent \textbf{Temporal Relations Between Elements:} Focuses on the model’s understanding of structural evolution within the music over time, including transitions in energy, tempo, or instrumentation across different sections.

\noindent \textbf{Functional Context:} Focuses on the model to link the music with real-world settings or usage contexts (e.g., movie scenes, events), requiring understanding of appropriateness and intent.

\noindent \textbf{Lyrics:} Focuses on interpretation of lyrical themes and content where applicable, often demanding a blend of semantic understanding and musical context awareness.

\noindent \textbf{Historical and Cultural Context:} Focuses on whether the model can connect musical elements to their broader cultural or historical origins (e.g., jazz fusion, protest music), relying on external world knowledge.

\noindent \textbf{Music Texture:} Focuses on knowledge of the audio's timbral and sonic character by evaluating aspects such as the layering of instruments, vocal texture, and overall audio quality. This skill captures how dense, sparse, smooth, or gritty a piece sounds, requiring models to interpret descriptive attributes and production characteristics.

\noindent \textbf{Melody:} Focuses on understanding the primary musical contour or thematic tune in the audio. Melody-based QAs evaluate recognition of pitch movement, vocal/instrumental phrasing, and stylistic traits such as ornamentation or melodic structure, encouraging indirect inference over simple labeling.

\noindent \textbf{Rhythm and Tempo:} Focuses on the temporal structure of the music, including pulse, beat, speed, and time signature. These questions test whether the model can identify rhythmic complexity, tempo changes, and groove characteristics that define a track’s pacing or drive.

\noindent \textbf{Harmony and Chords:} Focuses on the models' ability to reason about harmonic progressions and chordal structures that shape the emotional and tonal qualities of the audio. This includes interpreting transitions, key relationships, and compositional patterns in harmony using indirect reasoning from musical cues.

\noindent \textbf{General Complex Reasoning QA:} Evaluates the model's ability to perform multi-dimensional inference on short music segments by combining musical knowledge, perceptual cues, and contextual understanding. These questions are grounded in rich musical attributes, such as dynamics, structure, genre fusion, narrative cues, emotional evolution, and historical style, and require the model to synthesize diverse information to arrive at the correct answer. This category tests higher-order music comprehension across expressive, structural, technical, and cultural dimensions, aiming to emulate how humans make sense of music beyond surface-level tagging.

\vspace{-2mm}
\subsubsection{Music Knowledge}
\label{subsubsec:music_know}

\noindent \textbf{Instrumentation:} Focuses on the model's ability to recognize the instruments used in the music and how their timbre, arrangement, or presence contributes to the overall sound and suitability for various contexts.

\noindent \textbf{Performance:} Focuses on understanding of the vocal or instrumental delivery, including vocal tone, articulation, expression, or the presence of unique performance techniques.

\noindent \textbf{Sound Texture:} Focuses on the density and layering of sound, such as sparse vs. rich textures, acoustic vs. electronic timbres, and how these contribute to the sonic identity of the piece.

\noindent \textbf{Metre and Rhythm:} Focuses on the temporal structure of the piece, including rhythmic patterns, tempo consistency or variation, and the use of syncopation or groove, which are essential for identifying genre or compositional style.

\noindent \textbf{Melody:} Focuses on how the model interprets the musical contour and phrasing of the primary tune, including vocal stylings, tonal range, and melodic progression.

\noindent \textbf{Dynamics and Expression:} Focuses on the model's sensitivity to dynamic shifts (e.g., soft to loud passages), expressive techniques, and emotional delivery throughout the performance.

\noindent \textbf{Harmony:} Focuses on the model’s ability to recognize chord progressions, harmonic structure, and tonal relationships, which contribute to the music's emotional or stylistic impact.

\begin{figure*}
    \centering
    \includegraphics[width=\linewidth]{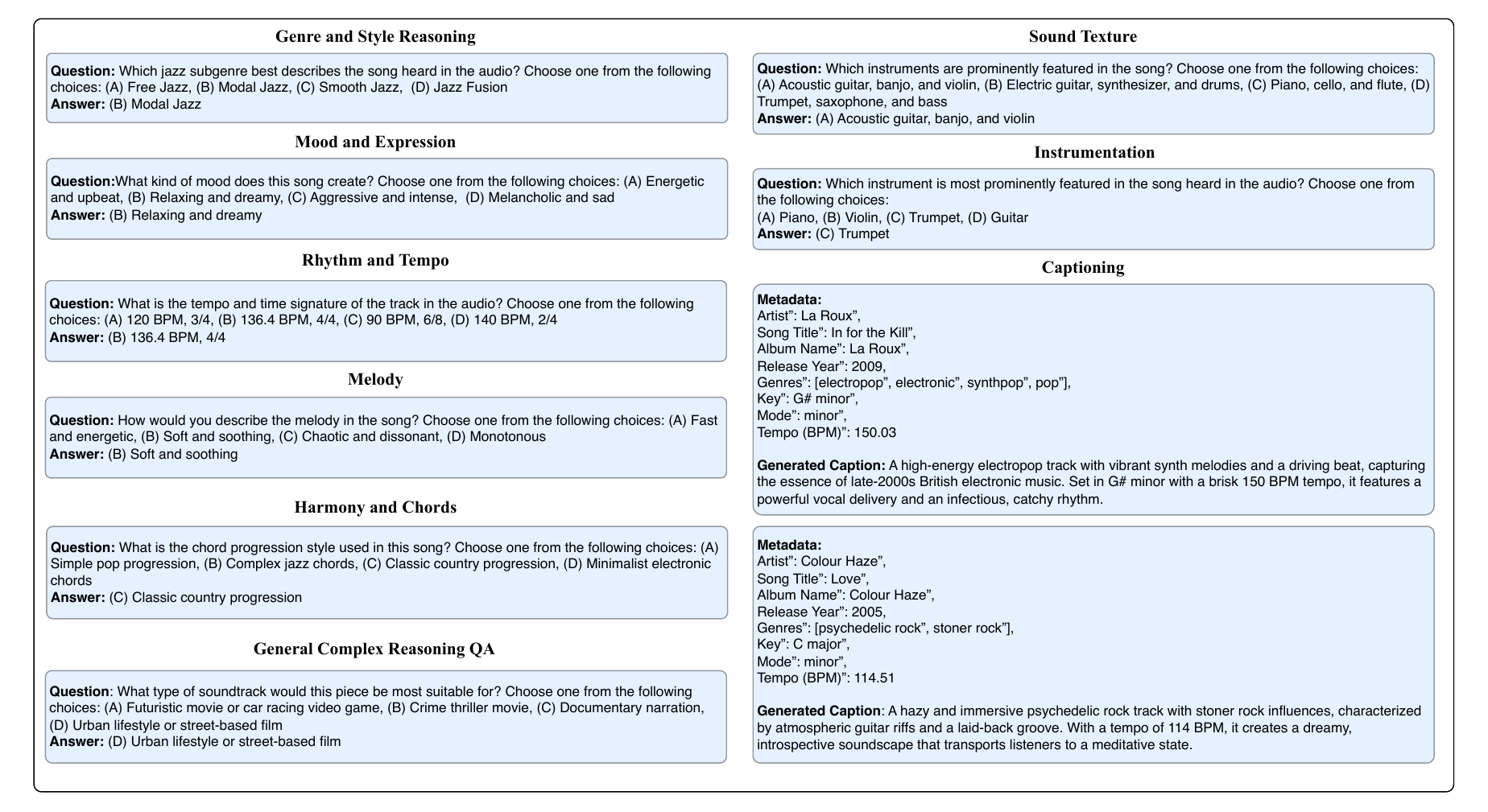}
    \caption{\small Examples of \textbf{Music Reasoning and Knowledge Questions} from AudioSkills-XL. Additionally, we also illustrate examples of music captions generated for audios in Music4All by prompting GPT-4.1 with metadata obtained from the dataset.}
    \label{fig:music_qa_example}
    \vspace{-5mm}
\end{figure*}

\subsubsection{Sound Reasoning}

\noindent \textbf{Speech-in-Sound QA:}  Focuses on reasoning over spoken content in addition to ambient sounds or music to answer complex questions about the input audio, including scene interpretation, action reasoning, etc.

\noindent \textbf{Eco-Acoustic Sounds QA:} Focuses on the model’s ability to interpret natural environmental conditions based on ambient audio cues. This includes reasoning over weather phenomena such as thunderstorms, snowfall, or rain using non-speech acoustic indicators like wind, water, or animal sounds.

\noindent \textbf{Acoustic Scene Reasoning:} Evaluates the model’s capability to infer real-world environments from ambient and structural sound patterns. These include background music, reverberation, crowd noise, and electronic elements, enabling scene classification (e.g., arcade, mall, theater) from complex audio mixes.

\noindent \textbf{Sound-Based Event Reasoning:} Focuses on identifying and reasoning over specific audio features or events, such as musical motifs, instrument timbres, or recurring sonic patterns, to infer event types or characteristic actions.

\begin{figure*}[h]
    \centering
    \includegraphics[width=0.9\linewidth]{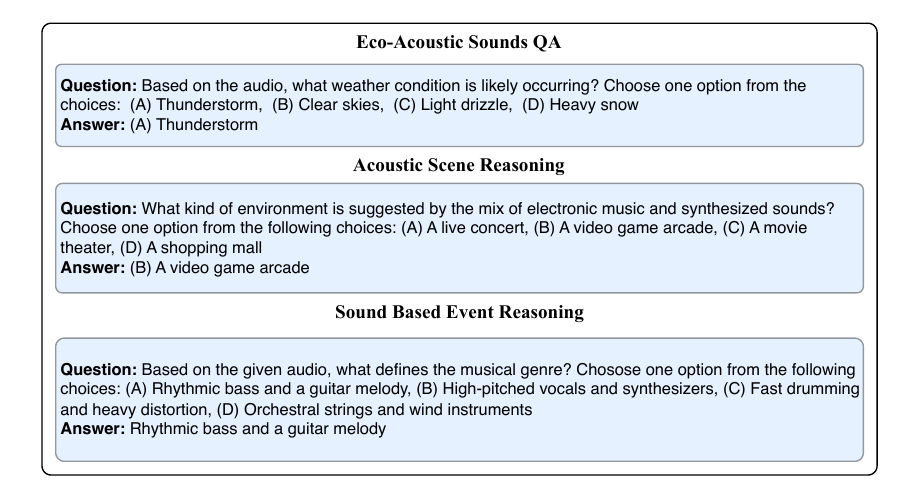}
    \vspace{-2mm}
    \caption{\small Examples of \textbf{Sound Reasoning QA}, together with the metadata used for generating them.}
    \label{fig:sound_qa_example}
    \vspace{-3mm}
\end{figure*}

\begin{figure*}[h]
    \centering
    \includegraphics[width=0.8\linewidth]{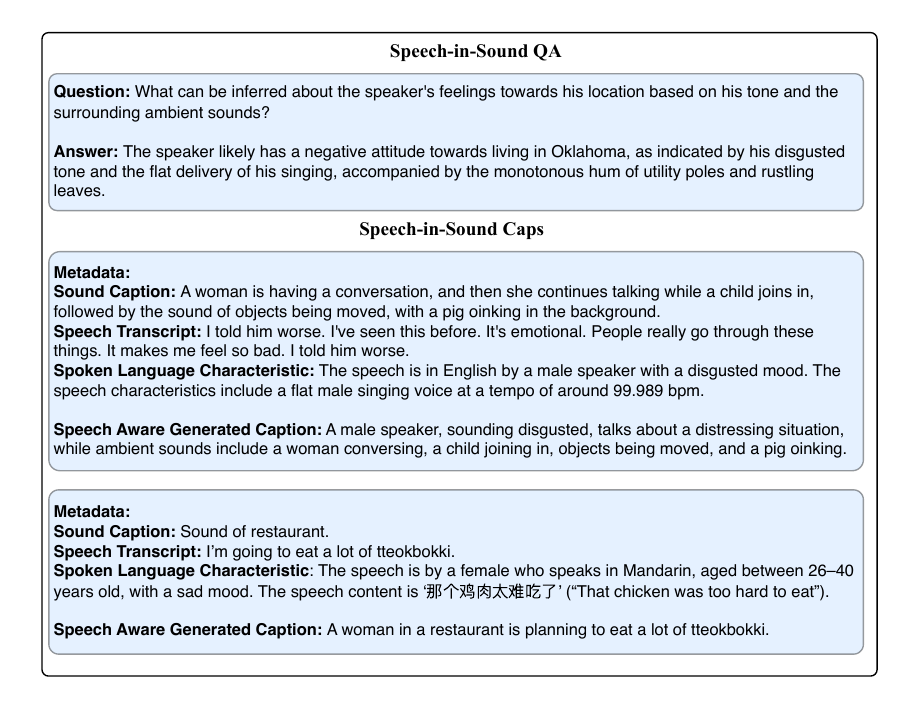}
    \caption{\small Examples of \textbf{Speech-in-Sound Caps and QA}, together with the metadata used for generating them.}
    \label{fig:speech_sound_example}
\end{figure*}

\begin{figure}[h]
    \centering
    \includegraphics[width=\linewidth]{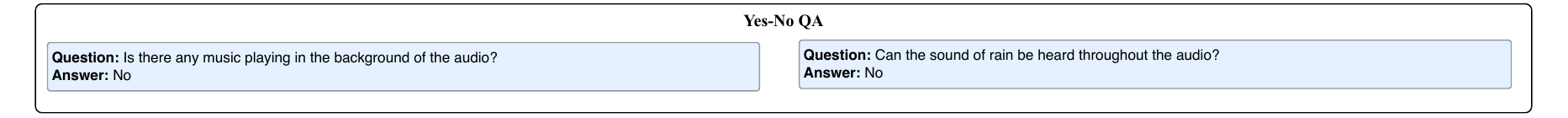}
    \caption{\small Examples of \textbf{general audio QA} generated as part of AudioSkills. We generate this as we find models struggle to say a ``No'' while responding to questions.}
    \label{fig:yesno_qa_example}
\end{figure}

\pagebreak

\section{LongAudio-XL}
\label{sec.longaudio_app}
Tables~\ref{tab:longaudio_data_1} and~\ref{tab:longaudio_data_2} present detailed skill-wise statistics for LongAudio-XL, including the source datasets and the minimum, maximum, and average durations of the audio samples.

\begin{figure*}[h]
    \centering
    \includegraphics[width=0.8\linewidth]{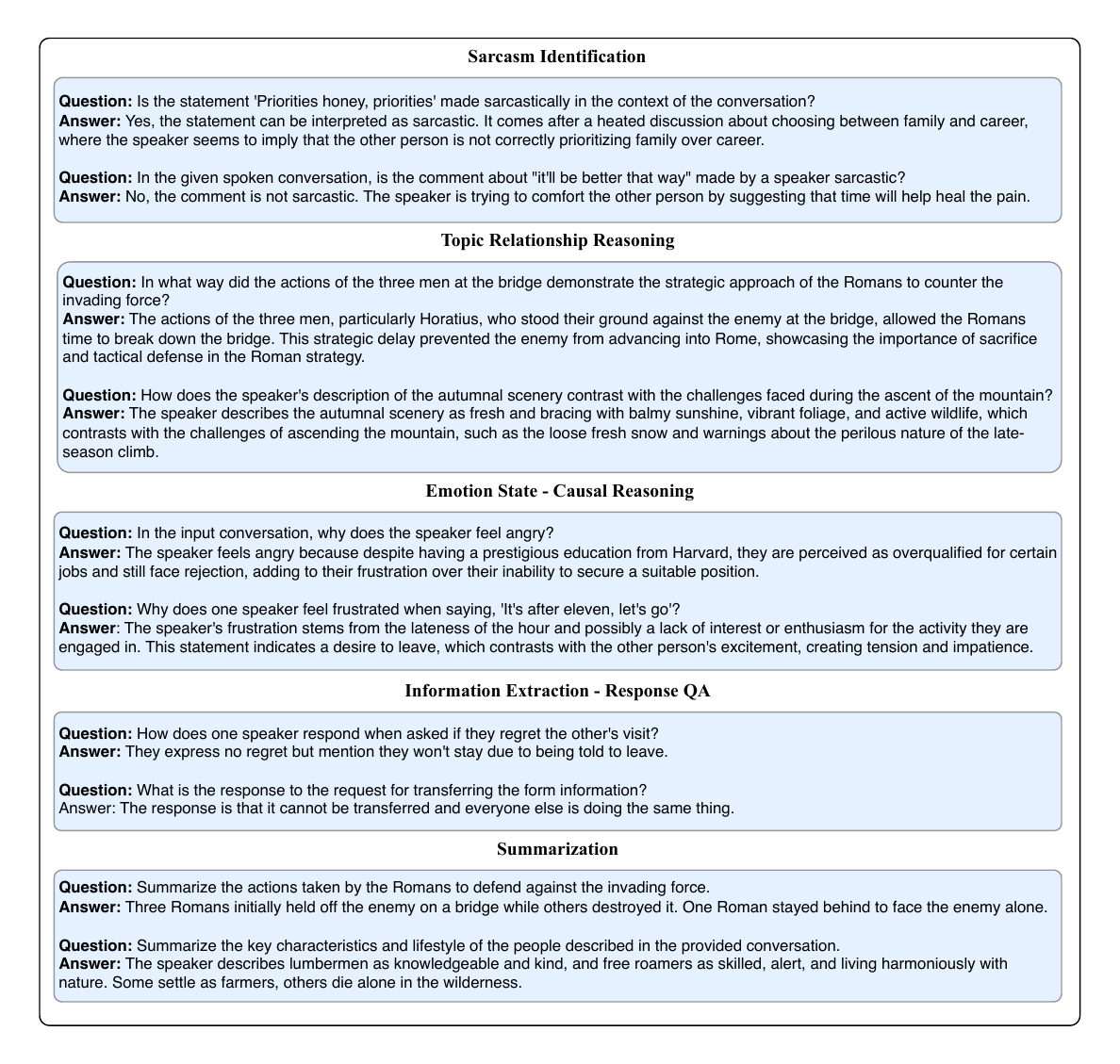}
    \caption{\small Examples of \textbf{LongAudio-XL}.}
    \label{fig:long_audio_eample}
    \vspace{-2em}
\end{figure*}

Below, we also show some examples form LongAudio-Xl in Fig~\ref{fig:long_audio_eample}

\section{AF-Think}
\label{sec.afthink}

Table~\ref{tab:af_think_tab} provides all details, including statistics and prompts for generating AF-Think.

\begin{table*}[h]
\centering
\caption{\small Detailed statistics of AF-Think. Most speech QA examples in this benchmark involve reasoning about ambient sounds in addition to spoken content. As our analysis shows, this added requirement increases task complexity, necessitating deeper inference to answer questions accurately.}
\resizebox{\linewidth}{!}{
\begin{tabular}{lllll}
\toprule
\textbf{Modality Type} & \textbf{Size} & \textbf{Datasets Used} & \textbf{Meta-Data Used} & \textbf{Prompt Reference} \\
\midrule
Speech & 100K & Speech-in-Sound QA, LongAudio-XL & transcripts, generated QAs & Fig.~\ref{fig:cot_emotion_shift},~\ref{fig:missed_prompt} \\
Sound only & 50K & AudioSkills-XL (AudioSet-SL, Youtube8M) & captions, QAs, dataset-specific meta-data & Fig.~\ref{fig:cot_sound},~\ref{fig:cot_ambient} \\
Music & 100K & AudioSkills-XL (Music4All, MSD) & captions, QAs, dataset-specific meta-data & Fig.~\ref{fig:cot_music_2},~\ref{fig:cot_music} \\
\bottomrule
\end{tabular}}
\label{tab:af_think_tab}
\end{table*}

Below, we also provide several examples from AF-Think in Fig.~\ref{fig:examples_think}:

\begin{figure*}[h]
    \centering
    \includegraphics[width=\linewidth]{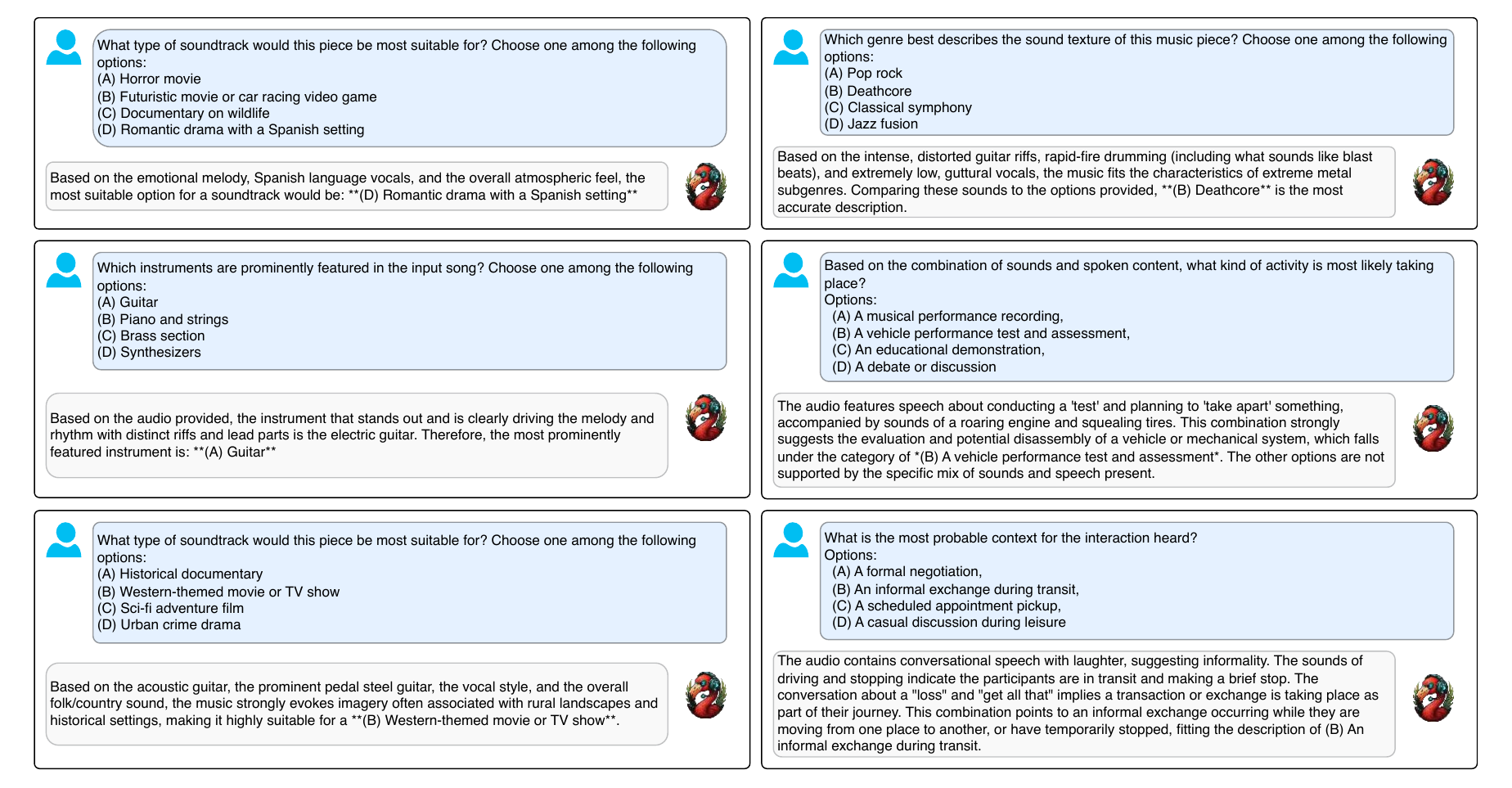}
    \caption{\small Examples of \textbf{AF-Think}, for music, speech and sounds.}
    \label{fig:examples_think}
\end{figure*}

\section{AF-Chat}
\label{sec.afchat}

Table~\ref{tab:af_chat_tab} provides all details, including statistics and prompts for generating AF-Chat.

\begin{table*}[h]
\centering
\caption{\small Detailed statistics of AF-Chat.}
\resizebox{\linewidth}{!}{
\begin{tabular}{lllll}
\toprule
\textbf{Modality Type} & \textbf{Size} & \textbf{Datasets Used} & \textbf{Meta-Data Used} & \textbf{Prompt Reference} \\
\midrule
Sound \& Speech & 35K & YouTube8M & captions, transcripts, speech characteristics & Fig.~\ref{fig:chat_sound} \\
Music & 40K & Music4All, MSD & captions, dataset-specific meta-data & Fig.~\ref{fig:music_chat} \\
\bottomrule
\end{tabular}}
\label{tab:af_chat_tab}
\end{table*}

Below, in Fig.~\ref{fig:chat_sound_example} and ~\ref{fig:chat_music_example} we also provide 2 examples from AF-Chat, one for each sound and music.

\begin{figure*}
    \centering
    \includegraphics[width=\linewidth]{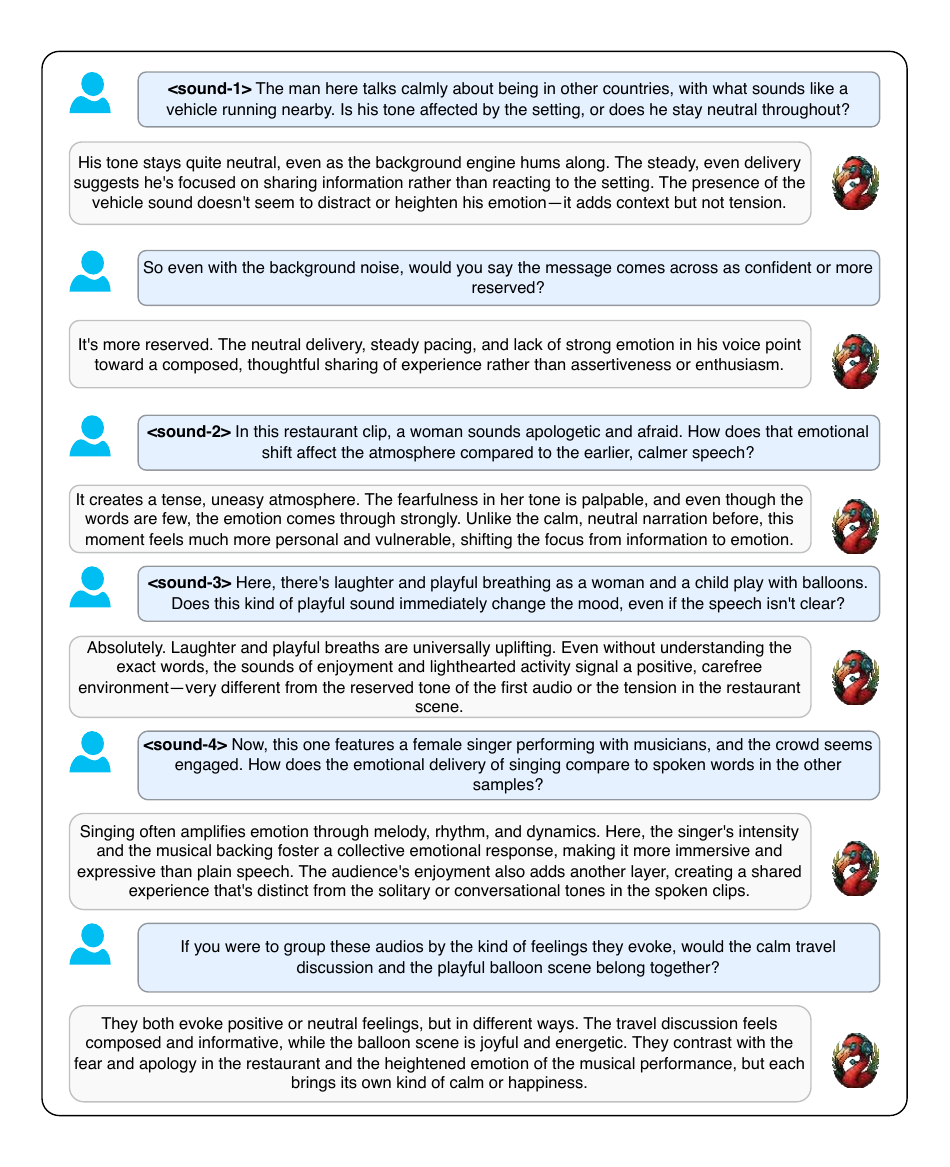}
    \caption{\small Example of AF-Chat for sound and speech.}
    \label{fig:chat_sound_example}
\end{figure*}

\begin{figure*}
    \centering
    \includegraphics[width=\linewidth]{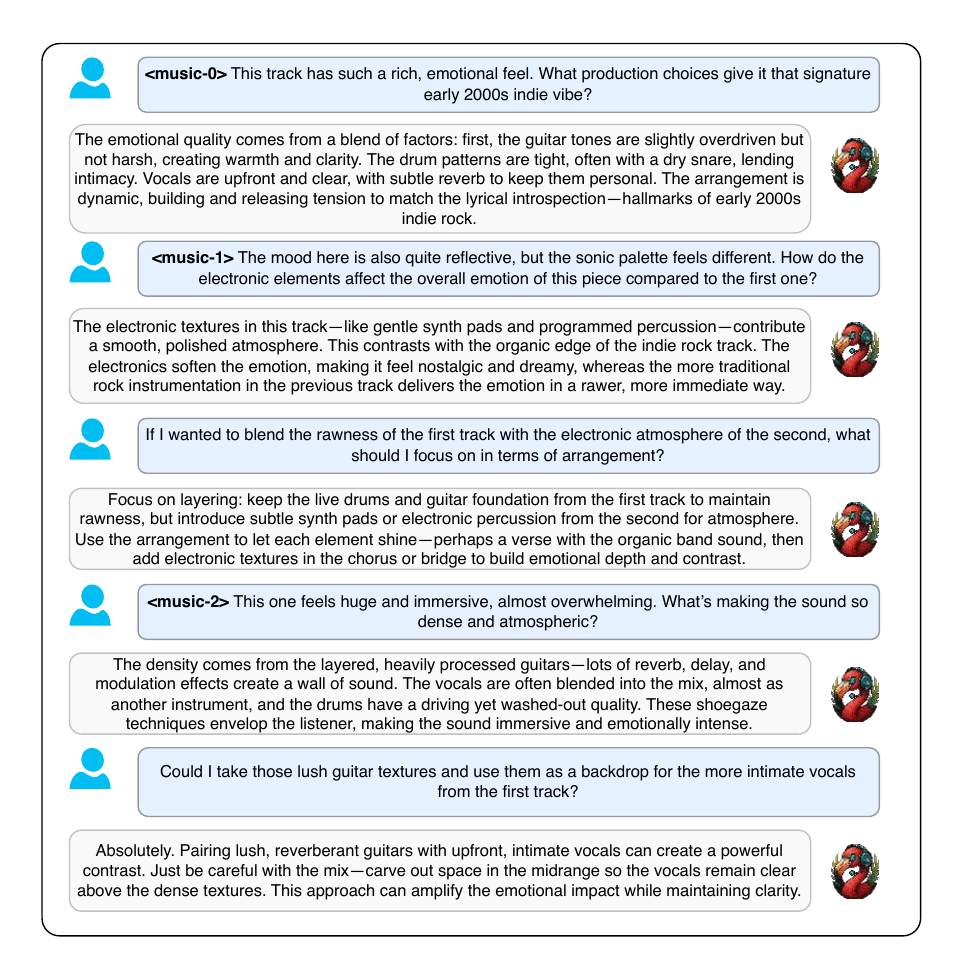}
    \caption{\small Example of AF-Chat for music.}
    \label{fig:chat_music_example}
\end{figure*}

\subsection{Human Study for AF-Chat-test}
\label{subsec:afchat_human}

The human verification process has been approved by our institution’s Institutional Review Board (IRB). For the human study, we hire 4 Ph.D. students proficient in audio research, including music. For each instance in each test-set dialogue, the students were asked to rate the output of the model on a scale of 1-5 across Factuality (how correct the response is), Usefulness (how useful the response is with respect to the context of the conversation), and Depth (how detailed the response is). For reference, we, the authors of the paper, provide responses scored 1-5 across the 3 aspects. The final score provided in Table 3 is an average of scores across all instances.

\subsection{Clustering for constructing AF-Chat}
\label{subsec:clustering}

To construct high-quality multi-turn, multi-audio dialogues for AF-Chat, we implement a targeted clustering strategy that ensures each dialogue is grounded in a semantically diverse but coherent audio context. Rather than sampling audio clips at random, which often leads to incoherent or loosely connected conversations, we curate each dialogue from a controlled pool of semantically related audio samples.

Specifically, for each seed audio, we retrieve its top 8 most semantically similar and top 8 most dissimilar clips from the dataset. Similarity is computed using captions, NV-Embed-v2 embeddings of the captions, and FAISS-based similarity search~\citep{douze2024faiss} of the embeddings.

For speech and environmental sounds, we use clips from Speech-in-Sound Caps. For music, we source from Music4All and the Million Song Dataset (MSD). Once the 16-candidate pool is formed (8 similar + 8 dissimilar), we restrict the dialogue construction process to this subset. GPT-4.1 is then prompted to construct multi-turn conversations (up to 10 turns) using any combination of these audio clips. This ensures:

\begin{enumerate}
    \item Topical consistency across turns using similar clips.
    \item Diversity and contrast through the inclusion of dissimilar audio.
    \item Clear referential structure, as questions may depend on or refer back to earlier audio.
\end{enumerate}

\begin{figure}[h]
    \centering
    \includegraphics[width=\linewidth]{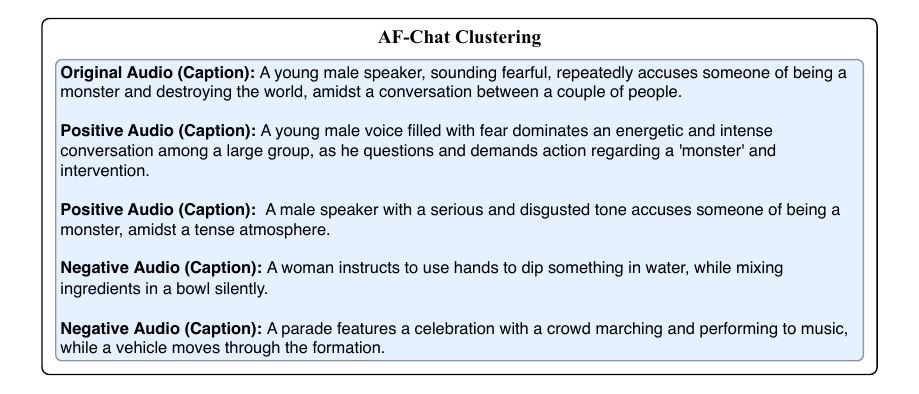}
    \caption{\small Examples of \textbf{audio clusters} obtained after clustering (Section~\ref{subsec:afchat_human}), used for constructing AF-Chat.}
    \label{fig:af_clustering_example}
\end{figure}

Our clustering strategy was informed by a preliminary human study (participant details similar to Section~\ref{subsec:afchat_human}), where participants engaged in multi-audio, multi-turn conversations with an LALM, focused on tasks such as sound design and music information retrieval. We observed that participants naturally gravitated toward using either highly similar or strongly contrasting audio clips within a dialogue. This behavioral insight motivated our use of similar and dissimilar audio clustering.

Empirically, this approach produced dialogues that were more natural, coherent, and diverse compared to those built from randomly selected audio pools. Moreover, AF3-Chat, when trained on this clustered dataset, outperformed the variant trained on randomly selected audio clips, both in terms of response relevance and conversational depth.

\begin{table*}[h]
\centering
\caption{\small Detailed skill-wise and dataset-wise statistics of LongAudio-XL.}
\begin{tabular}{lccccc}
\toprule
\textbf{QA Type} & \textbf{Dataset} & \textbf{\#Instances} & \textbf{Min Dur.(s)} & \textbf{Max Dur.(s)} & \textbf{Avg. Dur.(s)}\\
\midrule
Order & VoxPopuli~\citep{wang2021voxpopuli} & 16,926 & 1.87 & 294.80 & 89.55 \\
       & LibriSpeech~\cite{panayotov2015librispeech} & 2,340 & 16.02 & 147.59 & 82.01 \\
       & MELD~\cite{poria2018meld} & 4,135 & 1.06 & 108.01 & 30.42 \\
       & IEMOCAP~\citep{busso2008iemocap} & 599 & 82.00 & 542.00 & 272.45 \\
       & EuroParl~\citep{koehn2005europarl} & 11,885 & 2.59 & 176.14 & 69.34 \\
       & Fisher~\citep{cieri2004fisher} & 25,962 & 33.34 & 240.00 & 136.84 \\
       & Switchboard~\citep{godfrey1992switchboard} & 2,702 & 22.81 & 148.96 & 87.38 \\
       & MultiDialog~\citep{mesaros2018multi} & 27,927 & 1.31 & 499.33 & 135.10 \\
       & VoxCeleb2~\citep{chung2018voxceleb2} & 12,855 & 8.00 & 1273.60 & 71.12 \\
\midrule
Emotion Ident. & IEMOCAP~\cite{busso2008iemocap} & 300 & 82.00 & 542.00 & 272.22 \\
           & MELD~\cite{poria2018meld} & 1,847 & 1.78 & 108.01 & 33.20 \\
\midrule
Emotion Causal Reason. & IEMOCAP~\cite{busso2008iemocap} & 300 & 82.00 & 542.00 & 272.22 \\
          & MELD~\cite{poria2018meld} & 1,850 & 1.57 & 108.01 & 33.13 \\
\midrule
Emotion Flip Reason. & IEMOCAP~\cite{busso2008iemocap} & 299 & 82.00 & 542.00 & 272.62 \\
            & MELD~\cite{poria2018meld} & 1,807 & 1.64 & 108.01 & 33.57 \\
\midrule
Topic Relation. Reason. & VoxPopuli~\cite{wang2021voxpopuli} & 13,651 & 3.58 & 240.44 & 97.14 \\
            & LibriSpeech~\cite{panayotov2015librispeech} & 1,165 & 16.02 & 147.59 & 82.11 \\
            & MELD~\cite{poria2018meld} & 1,518 & 1.89 & 108.01 & 34.45 \\
            & IEMOCAP~\cite{busso2008iemocap} & 188 & 82.00 & 542.00 & 270.99 \\
            & EuroParl~\cite{koehn2005europarl} & 9,381 & 7.97 & 176.14 & 70.14 \\
            & Fisher~\cite{cieri2004fisher} & 20,453 & 33.34 & 240.00 & 136.10 \\
            & Switchboard~\cite{godfrey1992switchboard} & 998 & 24.58 & 148.96 & 90.05 \\
            & MultiDialog~\cite{mesaros2018multi} & 14,906 & 5.11 & 499.33 & 135.35 \\
            & DailyTalk~\cite{lee2023dailytalk} & 3,141 & 8.05 & 103.66 & 35.32 \\
            & VoxCeleb2~\cite{chung2018voxceleb2} & 5,414 & 8.51 & 1193.60 & 78.96 \\
\midrule
Sarcasm Ident. & IEMOCAP~\cite{busso2008iemocap} & 299 & 82.00 & 542.00 & 271.58 \\
     & MELD~\cite{poria2018meld} & 1,958 & 1.10 & 108.01 & 31.82 \\
\midrule
Summarization & VoxPopuli~\cite{wang2021voxpopuli} & 13,913 & 2.12 & 294.80 & 91.38 \\
         & LibriSpeech~\cite{panayotov2015librispeech} & 1,057 & 16.02 & 147.59 & 83.15 \\
         & MELD~\cite{poria2018meld} & 2,803 & 1.84 & 108.01 & 32.92 \\
         & IEMOCAP~\cite{busso2008iemocap} & 300 & 82.00 & 542.00 & 272.22 \\
         & EuroParl~\cite{koehn2005europarl} & 8,905 & 6.62 & 176.14 & 70.03 \\
         & Fisher~\cite{cieri2004fisher} & 15,500 & 0.33 & 240.00 & 135.60 \\
         & Switchboard~\cite{godfrey1992switchboard} & 1,346 & 24.58 & 148.96 & 87.60 \\
         & MultiDialog~\cite{mesaros2018multi} & 20,838 & 1.93 & 499.33 & 135.73 \\
         & DailyTalk~\cite{lee2023dailytalk} & 7,218 & 8.05 & 103.66 & 31.42 \\
         & VoxCeleb2~\cite{chung2018voxceleb2} & 5,894 & 7.94 & 1193.60 & 70.87 \\
         & Spotify Podcasts~\cite{clifton-etal-2020-100000} & 103920 & 0.06 & 18206.44 & 2002.99 \\

\midrule
Needle QA (IE) & DailyTalk~\cite{lee2023dailytalk}   & 13,563  & 5.72  & 103.66 & 31.12 \\
& EuroParl~\cite{koehn2005europarl}    & 18,426  & 6.57  & 176.14 & 70.10 \\
& Fisher~\cite{cieri2004fisher}      & 37,779  & 18.59 & 240.00 & 135.99 \\
& IEMOCAP~\cite{busso2008iemocap}     & 542     & 82.00 & 542.00 & 272.03 \\
& LibriSpeech~\cite{panayotov2015librispeech} & 2,248   & 16.02 & 147.59 & 82.82 \\
& Spotify Podcasts~\cite{clifton-etal-2020-100000} & 103920 & 0.06 & 18206.44 & 2002.99 \\
\midrule
Response QA (IE) & VoxPopuli~\cite{wang2021voxpopuli} & 13,913 & 2.12 & 294.80 & 91.38 \\
         & MELD~\cite{poria2018meld} & 1,660 & 1.57 & 108.01 & 31.83 \\
         & IEMOCAP~\cite{busso2008iemocap} & 177 & 82.00 & 542.00 & 272.52 \\
         & MultiDialog~\cite{mesaros2018multi} & 13,505 & 1.95 & 499.33 & 135.40 \\
         & DailyTalk~\cite{lee2023dailytalk} & 4,516 & 5.72 & 103.66 & 30.91 \\
         & Switchboard~\cite{godfrey1992switchboard} & 862 & 22.81 & 148.96 & 88.75 \\

\bottomrule
\end{tabular}
\label{tab:longaudio_data_1}
\end{table*}

\begin{table*}[h]
\centering
\caption{\small Detailed skill-wise and dataset-wise statistics of LongAudio-XL.}
\begin{tabular}{lccccc}
\toprule
\textbf{QA Type} & \textbf{Dataset} & \textbf{\#Instances} & \textbf{Min Dur. (s)} & \textbf{Max Dur.(s)} & \textbf{Avg. Dur.(s)}\\
\midrule
Causal QA (IE) & VoxPopuli~\cite{wang2021voxpopuli} & 12,264 & 4.10 & 240.44 & 92.88 \\
              & LibriSpeech~\cite{panayotov2015librispeech} & 1,166 & 16.02 & 147.59 & 82.04 \\
              & MELD~\cite{poria2018meld} & 2,957 & 1.27 & 108.01 & 31.74 \\
              & IEMOCAP~\cite{busso2008iemocap} & 298 & 82.00 & 542.00 & 273.10 \\
              & EuroParl~\cite{koehn2005europarl} & 7,457 & 7.97 & 176.14 & 70.24 \\
              & Fisher~\cite{cieri2004fisher} & 19,335 & 37.17 & 240.00 & 135.87 \\
              & Switchboard~\cite{godfrey1992switchboard} & 1,352 & 22.81 & 148.96 & 87.40 \\
              & MultiDialog~\cite{mesaros2018multi} & 20,811 & 3.17 & 499.33 & 135.62 \\
              & DailyTalk~\cite{lee2023dailytalk} & 7,368 & 8.05 & 103.66 & 31.15 \\
              & VoxCeleb2~\cite{chung2018voxceleb2} & 6,171 & 8.06 & 1193.60 & 71.08 \\
\bottomrule
\end{tabular}
\label{tab:longaudio_data_2}
\end{table*}

\section{Prompts}
\label{sec.prompts}

We provide all prompting templates used across our datasets and QA types in Figures~\ref{fig:connecting_prompt}, \ref{fig:detail_prompt}, \ref{fig:maintopic_prompt}, \ref{fig:order_prompt}, \ref{fig:reason_behind_prompt}, \ref{fig:summary_prompt}, \ref{fig:emo_flip_prompt}, \ref{fig:emotional_reasono_prompt}, \ref{fig:emotional_sarcasm_prompt}, \ref{fig:emotional_state_prompt}, \ref{fig:respond_prompt}, \ref{fig:speech_sound_prompt}, \ref{fig:msd_caption}, \ref{fig:msd_knowledge}, \ref{fig:msd_reasoning}, \ref{fig:music4all_caption_prompt}, \ref{fig:music4all_knowledge}, \ref{fig:music4all_openqa_prompt}, \ref{fig:music4all_reasoning_qa_prompt}, \ref{fig:yes_no_qa}, \ref{fig:chat_sound}, \ref{fig:music_chat}, \ref{fig:cot_ambient}, \ref{fig:cot_emotion_shift}, \ref{fig:cot_music_2}, \ref{fig:cot_music}, \ref{fig:cot_sound}, \ref{fig:missed_prompt}, \ref{fig:music_texture}, \ref{fig:melody}, \ref{fig:rhythm_tempo}, \ref{fig:harmony_chord}, \ref{fig:sound_1_prompt}, \ref{fig:sound_2_prompt}, \ref{fig:sound_3_prompt}, and \ref{fig:transcript_prompt}.

\section{AF3 Training Datasets}
\label{sec.af3_training_data}

Table~\ref{tab:sft_datasets_app} summarizes all datasets used to train AF3, including total hours, number of audio-QA pairs, and the number of epochs (passes over the dataset) used at each training stage. Similar to ~\citep{kong2025audioflamingo2}, we convert all foundational datasets (captioning, classification, etc.) into QA formats, using the same set of prompts for each task mentioned in ~\citep{kong2025audioflamingo2}.

\begin{table*}[h]
    \centering
    \caption{\small List of fine pre-training and fine-tuning datasets together with their training composition.}
    \begin{tabular}{lccccccc}
    \toprule
Dataset & Hours  & Num. Pairs & St. 1 & St. 2& St. 3 & St. 3.5 & St. 4\\ \midrule
        AudioSkills-XL (Uurs) & - & 9700K & - & 2.0 & 2.0& -\\
        LongAudioXL (Ours)& - & 1000K & $1.0$ & $1.0$& $1.0$& $1.0$& -\\
        AF-Think (Ours)& - & 250K& $1.0$ & $1.0$& $1.0$& $2.0$& -\\
        AF-Chat (Ours)& - & 75K& - & - & - & - & $1.0$\\
        CompA-R~\cite{ghoshcompa} & 159 hrs & 350k & - & 2.0& 2.0& - & -\\
        MusicBench~\cite{melechovsky2023mustango} & 115.5 hrs & 686k & - & 1.0& 1.0& -& -\\
        Mu-LLAMA~\cite{liu2024music} & 62.9 hrs & 70k & 1.0 & 2.0 & 2.0& -& -\\
        Salmonn AQA~\cite{tang2023salmonn} & 800 hrs & 270k & - & 1.0& 1.0& -& -\\
        ClothoAQA~\cite{lipping2022clotho} & 7.4 hrs & 9.7K & - & $8.0$ & $8.0$ & -& -\\
        OpenAQA~\cite{gong2023ltu} & 693.2 hrs & 1959.8K & - & $1.0$ & $1.0$& -& -\\
        Clotho-v2~\cite{drossos2020clotho}  & 24.0 hrs & 19.2K & 1.0 & $2.0$ & $2.0$ & -& -\\
        MACS~\cite{morato2021diversity} & 10.9 hrs & 17.3K & - & $1.0$ & $1.0$& -& -\\
        FSD50k~\cite{fonseca2021fsd50k} & 80.8 hrs & 41.0K & 1.0 & $1.0$ & $1.0$& -& -\\
        CochlScene~\cite{jeong2022cochlscene} & 169.0 hrs & 60.9K & - & $1.0$ & $1.0$ & -& -\\
        NonSpeech 7k~\cite{rashid2023nonspeech7k} & 6.2 hrs & 6.3K & - & $4.0$& $4.0$ & -& - \\
        Chime-home~\cite{foster2015chime} & 5.0 hrs & 4.5K & - & $1.0$ & $1.0$& -& -\\
        Sonyc-UST~\cite{cartwright2020sonyc} & 34.9 hrs & 27.9K & - & $1.0$ & $1.0$& -& -\\
        Emov-DB~\cite{moon2022language} & 7.8 hrs & 6.8K & - & $1.0$ & $1.0$& -& -\\
        JL-Corpus~\cite{james2018open} & 1.4 hrs & 2.4K & - & $6.0$ & $6.0$& -& -\\
        Tess & 1.6 hrs & 2.8K & - & $2.0$ & $2.0$& -\\
        OMGEmotion~\cite{barros2018omg} & 3.0 hrs & 1.7K & - & $3.0$ & $3.0$& -& -\\
        MusicAVQA\textsubscript{audio-only}~\cite{li2022learning} & 77.1 hrs & 5.7K & - & $6.0$ & $6.0$& -& -\\
        MusicQA~\cite{ouyang2025mqad} & 62.9 hrs & 70K & - & $1.0$ & $1.0$& -& -\\
        LP-MusicCaps\textsubscript{MSD}~\cite{doh2023lp} & 5805.7 hrs & 1331.8K & 1.0 & $1.0$ & $1.0$& -& -\\
        LP-MusicCaps\textsubscript{MTT}~\cite{doh2023lp} & 126.4 hrs & 46.9K & 1.0 & $1.0$ & $1.0$ & -& -\\
        LP-MusicCaps\textsubscript{MC}~\cite{doh2023lp} & 7.4 hrs & 7.9K & 1.0 & $2.0$ & $2.0$& - & -\\
        MusicCaps~\cite{agostinelli2023musiclm} & 7.4 hrs & 2.6K & 1.0 & $6.0$ & $6.0$& -& -\\
        NSynth~\cite{engel2017neural} & 321.3 hrs & 289.2K & - & $8.0$ & $8.0$& -& -\\
        MusDB-HQ~\cite{rafii2017musdb18} & 29.1 hrs & 10.2K & - & $2.0$ & $2.0$& -& -\\
        FMA~\cite{defferrard2016fma} & 860.7 hrs & 104.2K & - & $1.0$ & $1.0$& -& -\\
        Laion630k\textsubscript{BBCSoundEffects}~\cite{wu2023large} & 456.9 hrs & 15.1K & $1.0$ & - & $1.0$& -& -\\
        Laion630k\textsubscript{Freesound}~\cite{wu2023large} & 2494.8 hrs & 306.5K & $1.0$ & -& $1.0$& -& -\\
        SoundDescs~\cite{koepke2022audio} & 749.7 hrs & 23.1K & $1.0$ & -& $1.0$& -& -\\
        WavCaps~\cite{mei2024wavcaps} & 3793.3 hrs & 402.6 K & $1.0$ & -& $1.0$& -& -\\
        AudioSet~\cite{gemmeke2017audio} & 2617.8 hrs & 950.8K & 1.0 & -& $1.0$& -& -\\
        WavText5K~\cite{deshmukh2022audio} & 23.8 hrs & 4.3K & $1.0$ & -& $1.0$& -& -\\
        MSP-Podcast~\cite{martinez2020msp} & 73.9 hrs & 45.1K & $1.0$ & $1.0$& $1.0$& -& -\\
        MELD~\cite{poria2018meld} & 8.7 hrs & 32.9K & $1.0$ & $1.0$ & $1.0$& -& -\\
        MusicAVQA\textsubscript{audio-visual}~\cite{li2022learning} & 142.4 hrs & 17.9K & $1.0$ & $6.0$& $6.0$& -& -\\
        Music4All Captions (ours)& 910.5 hrs & 55.6K & $1.0$ & -& $1.0$& -& -\\
        MSD Captions (ours)& 15449.9 hrs & 55.6K & $1.0$ & -& $1.0$& -& -\\
        Speech-in-Sound Captions (ours)& 6227.6 hrs & 1999959 & $1.0$ & -& $1.0$& -& -\\
        LibriSpeech~\cite{panayotov2015librispeech} & 960 hrs & 281.2K & $1.0$ & $1.0$& $1.0$& -& -\\
        Switchboard~\cite{godfrey1992switchboard} & 109.9 hrs & 76.6K & $1.0$ & $1.0$& $1.0$& -& -\\
        GigaSpeech (L)~\cite{chen2021gigaspeech}& 2499.8 hrs & 2266.3K & $1.0$ & $1.0$& $1.0$& -& -\\
        Common Voice 15~\cite{commonvoice:2020} & 1752.1 hrs & 1109.6K & $1.0$ & $1.0$& $1.0$& -& -\\
        VoxPopuli (en)~\cite{wang2021voxpopuli} & 501.8 hrs & 177K & $1.0$ & $1.0$& $1.0$& - & -\\
        TEDLIUM (en)~\cite{hernandez2018ted} & 472.3 hrs & 68K & $1.0$ & $1.0$& $1.0$& -& -\\
        SPGISpeech~\cite{o2021spgispeech} & 4999.8 hrs & 1966.1K & $1.0$ & $1.0$& $1.0$& -& -\\
        VoiceAssistant400K~\citep{xie2024mini} & 684 hrs & 470K & - & - & - & - & $1.0$\\
    \bottomrule
    \end{tabular}
    
    \label{tab:sft_datasets_app}
\end{table*}

\section{AF3 Training Details}
\label{sec.af3_training_details}
In this section, we present the training settings of our models across all 5 stages, each with specific configurations. Details are in \cref{tab:hyperparams}.

\begin{table}[!ht]
\centering
\begin{tabular}{lccccc}
\toprule
\textbf{Settings} & \textbf{Stage1} & \textbf{Stage2} & \textbf{Stage3} & \textbf{Stage3.5} & \textbf{Stage4}\\
\midrule
per device batch size & 64 & 16 & 4 & 4 & 2\\
learning rate & 1e-3 & 2e-5 & 2e-5 & 5e-5 & 5e-5\\
learning schedule & \multicolumn{5}{c}{Cosine decay} \\
warm up ratio & \multicolumn{5}{c}{0.03}\\
weight decay & \multicolumn{5}{c}{0.0}\\
epoch & 1 & 1 & 1 & 2& 2\\
bf16 & \checkmark & \checkmark & \checkmark & \checkmark & \checkmark\\
grad accumulate & 1 & 2 & 4 & 4 & 8\\
DeepSpeed stage & \multicolumn{5}{c}{Zero3}\\

GPUs & \multicolumn{5}{c}{128$\times$A100} \\
\bottomrule
\end{tabular}
\caption{\small Training settings across stages.}
\label{tab:hyperparams}

\end{table}

\section{Streaming TTS System Architecture and Training Details}
\label{sec.af3_voice}

To enable voice output capabilities within our system, we incorporate a text-to-speech (TTS) module that operates on subword text tokens. For efficient and simplified streaming speech synthesis, our TTS module employs a decoder-only architecture.

\begin{figure}[t]
    \centering

    \begin{subfigure}[b]{0.48\textwidth}
        \centering
        \includegraphics[width=\textwidth]{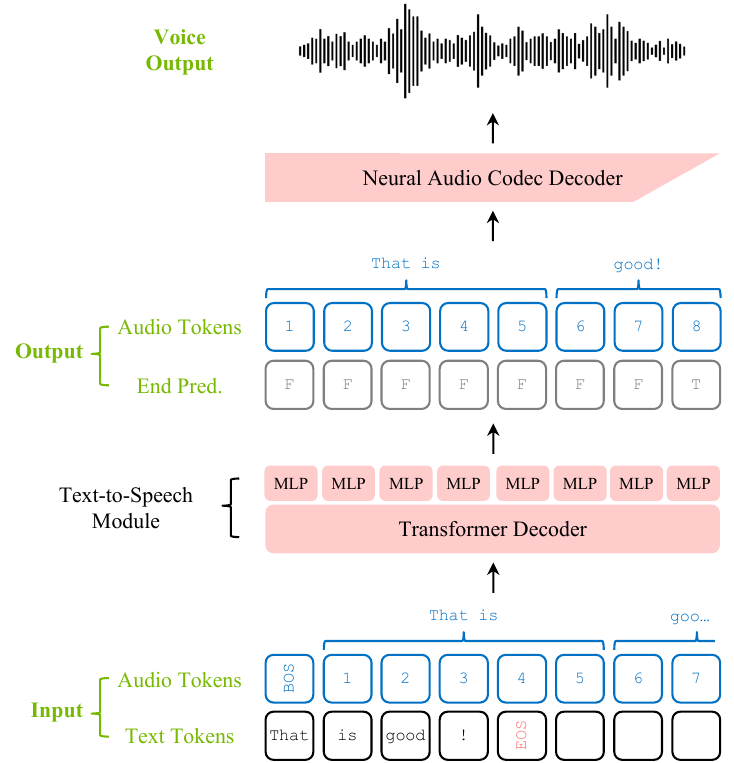}
        \caption{Streaming TTS system architecture.}
        \label{fig:af3_voice_diagram}
    \end{subfigure}
    \hfill 
    \begin{subfigure}[b]{0.48\textwidth}
        \centering
        \includegraphics[width=\textwidth]{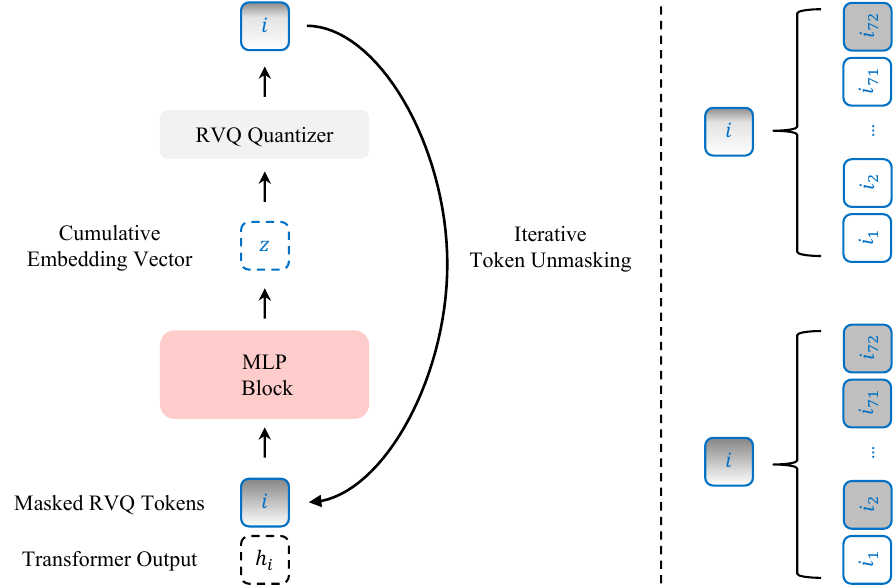}
        \caption{Iterative unmasking of RVQ audio tokens.}
        \label{fig:af3_voice_mlp}
    \end{subfigure}

    \caption{\small Streaming TTS is enabled by autoregressive audio token generation coupled with a neural audio codec decoder. (a) The streaming TTS system predicts audio tokens conditioned on incoming subword text tokens (e.g., from the main AF3 model) and the history of previously generated audio tokens; these audio tokens are then decoded into voice output by the neural audio codec. (b) The iterative audio token unmasking process relies on an MLP block. This block takes partially masked RVQ tokens and transformer decoder output as input, predicts a cumulative embedding vector, which is subsequently quantized into progressively more unmasked RVQ tokens.}
    \label{fig:af3_voice}
    
\end{figure}

As illustrated in Fig.~\ref{fig:af3_voice}, the TTS module predicts the subsequent audio token conditioned on incoming subword text tokens from the main AF3 model and the history of previously generated audio tokens. These audio tokens are then decoded into voice output by the neural audio codec. This design simplifies the speech generation pipeline and minimizes latency, which are critical for real-time speech streaming.

\subsection{Neural Audio Codec}
\label{sec.af3_voice_neural_audio_codec}
We utilize a fully causal convolutional neural audio codec  for efficient streaming audio decoding, following \cite{kim2024efficient, siuzdakvocos}.

{\noindent \textbf{Encoder.}} Input audio is first resampled to 44.1~kHz. It is then converted into Short-Time Fourier Transform~(STFT) parameters using a hop size of 8 and a window size of 32. This STFT representation is processed by an initial 1x1 convolutional layer to produce 384-dimensional hidden embeddings. Following this, the signal undergoes three downsampling stages. Each stage consists of three causal 1D-ConvNeXt blocks~\cite{liu2022convnet, siuzdakvocos} followed by a strided convolutional layer for downsampling. These strided convolutional layers use a stride and kernel size of 8. Each such layer doubles the hidden dimension, except for the final one, which produces a 512-dimensional output. The encoded output sequence is 4096 times shorter than the raw waveform, corresponding to approximately 10.8 frames per second.

{\noindent \textbf{Quantization.}} The encoded output is quantized into audio tokens using Residual Vector Quantization (RVQ)~\cite{lee2022autoregressive, kumarhigh}. The number of RVQ levels is set to 72.

{\noindent \textbf{Decoder.}} The decoder mirrors the encoder's architecture symmetrically, employing 1D transposed convolutional layers for upsampling and causal 1D-ConvNeXt blocks. The final convolutional layer reconstructs the STFT parameters, which are then transformed back into a raw audio waveform via an inverse STFT~(iSTFT) similar to Vocos~\cite{siuzdakvocos}.

{\noindent \textbf{Training.}} The codec is trained using a combination of adversarial training and a mel-spectrogram reconstruction loss, following methodologies from DAC~\cite{kumarhigh}.

\subsection{Text-to-Speech (TTS) Module}
\label{sec.af3_voice_tts_module}

{\noindent \textbf{Architecture and Operation.}} The TTS module's transformer decoder processes a sequence formed by concatenating subword text tokens (from the main AF3 model) and previously generated audio tokens. The resulting hidden states from the transformer serve as conditional input to a multi-layer perceptron~(MLP) block. This MLP block then iteratively predicts progressively higher levels of the RVQ tokens, a technique inspired by~\cite{kim2024efficient}. In practice, we employ 4 iteration steps during inference. A key aspect contributing to the system's simplicity and low latency is that the model is designed to generate an audio token whenever a text token is emitted by the AF3 model, without requiring explicit alignments between text and speech.

{\noindent \textbf{Training and Configuration.}} During training, the transformer decoder utilizes teacher-forcing with ground-truth audio tokens. The MLP block is trained to estimate the parameters of a mixture-of-gaussians distribution where the number of mixtures is 1024. The objective is to maximize the log-likelihood of predicting the cumulative RVQ token embedding, following~\cite{kim2024efficient}.
The decoder-only transformer has a configuration similar to DiT-XL~\cite{peebles2023scalable, lee2025dittotts}. The MLP block consists of 3 layers, totaling 644 M parameters.

\subsection{Training Data and Processing}
\label{sec.af3_voice_training_data}

The models are trained on a comprehensive dataset comprising LibriTTS-R~\cite{koizumi2023libritts}, LibriHeavy~\cite{kang2024libriheavy}, the English portion of Multilingual LibriSpeech~\cite{pratap2020mls}, and proprietary datasets, totaling approximately 100,000 hours of audio. To support long-form speech generation, audio segments from the same speaker are randomly concatenated to create training samples with durations ranging from 1 to 120 seconds.

\section{Qualitative Examples}
\label{sec.qual}

\begin{figure}[h]
    \centering
    \includegraphics[width=\linewidth]{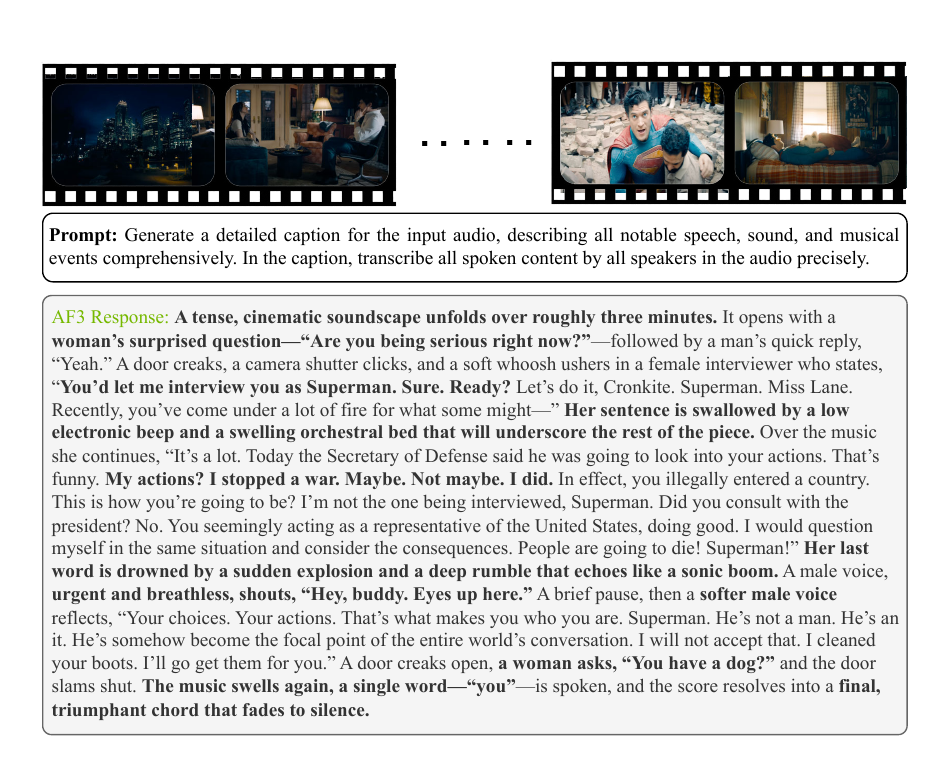}
    \caption{\small Demonstration of AF3's capabilities on an audio captioning task. We prompt AF3 with an unseen audio clip—extracted from the Superman 2025 trailer (\url{https://www.youtube.com/watch?v=2woCZg5QdVE})—captured in the wild. The model accurately identifies and describes background sounds, spoken content, speaker turns, and transcriptions, demonstrating strong audio understanding. Beyond this example, AF3 supports significantly more complex reasoning tasks. We invite readers to explore these capabilities via our public demo: \url{https://huggingface.co/spaces/nvidia/audio-flamingo-3}.}
    \label{fig:superman}
\end{figure}

\begin{figure}
    \centering
    \includegraphics[width=\linewidth]{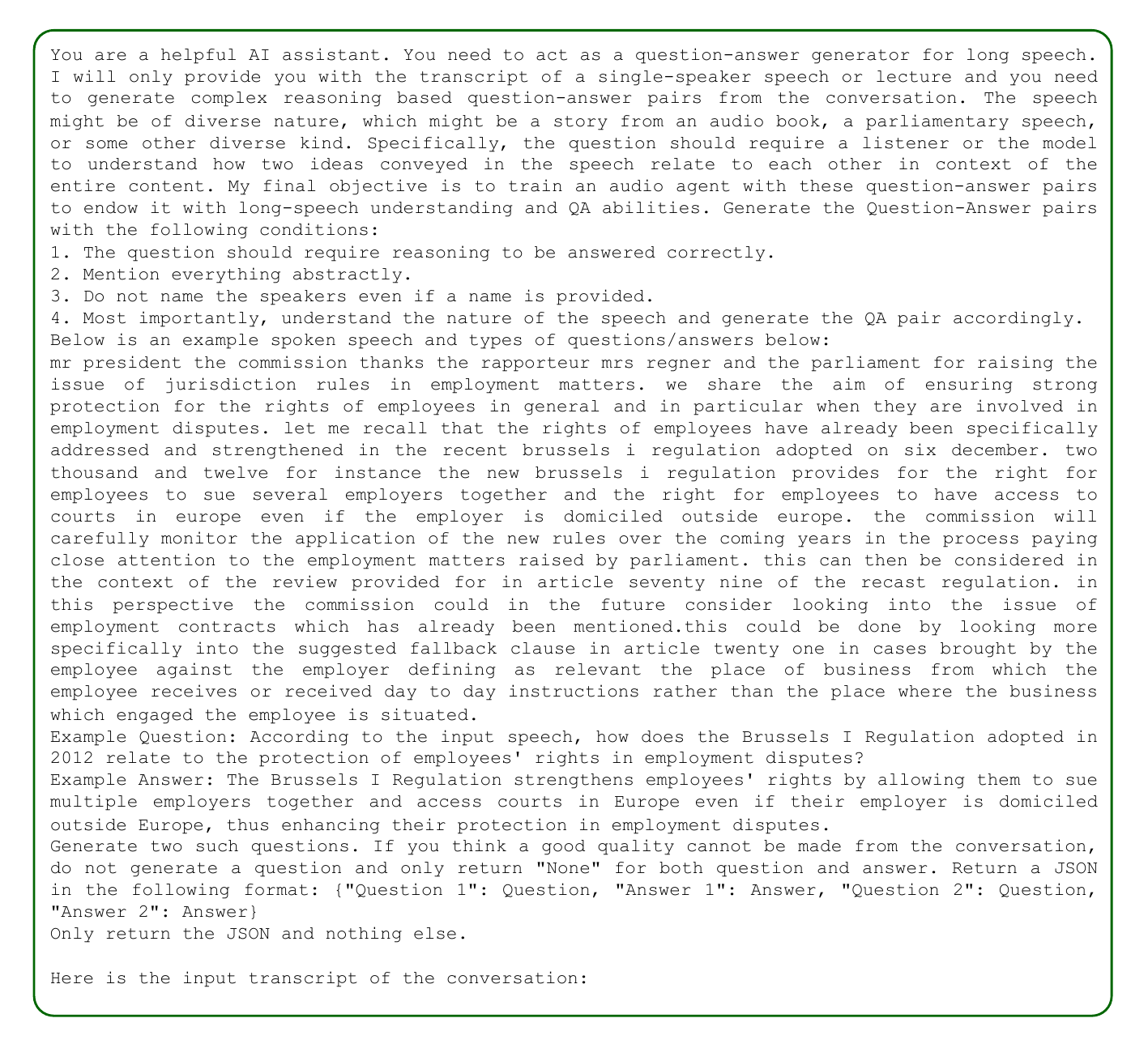}
    \caption{\small Prompt used for generating \textbf{Topic Relationship QA} for LongAudioXL.}
    \label{fig:connecting_prompt}
\end{figure}

\begin{figure}
    \centering
    \includegraphics[width=\linewidth]{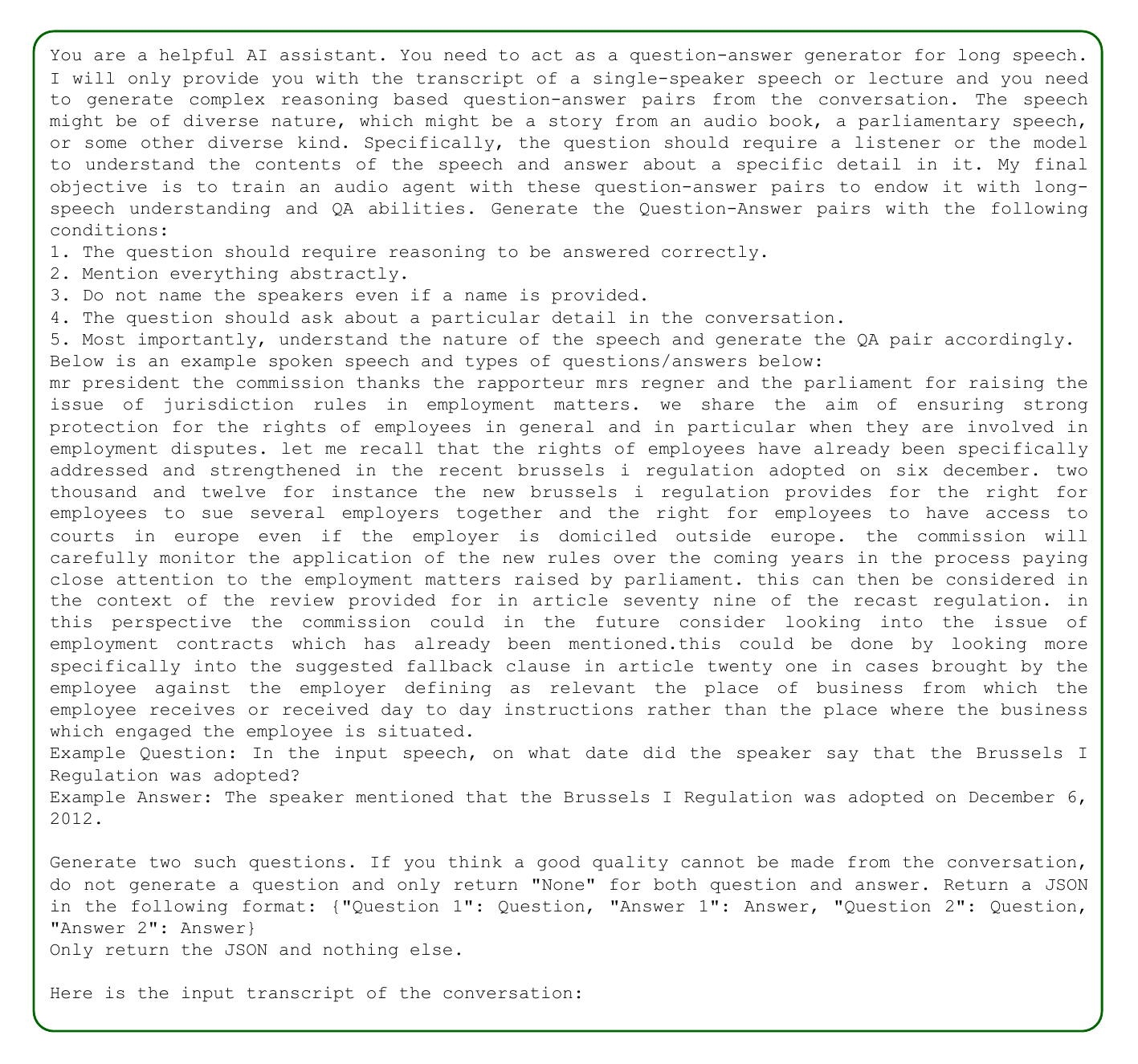}
    \caption{\small Prompt used for generating \textbf{Needle QA} (Information Extraction type) for LongAudioXL.}
    \label{fig:detail_prompt}
\end{figure}

\begin{figure}
    \centering
    \includegraphics[width=\linewidth]{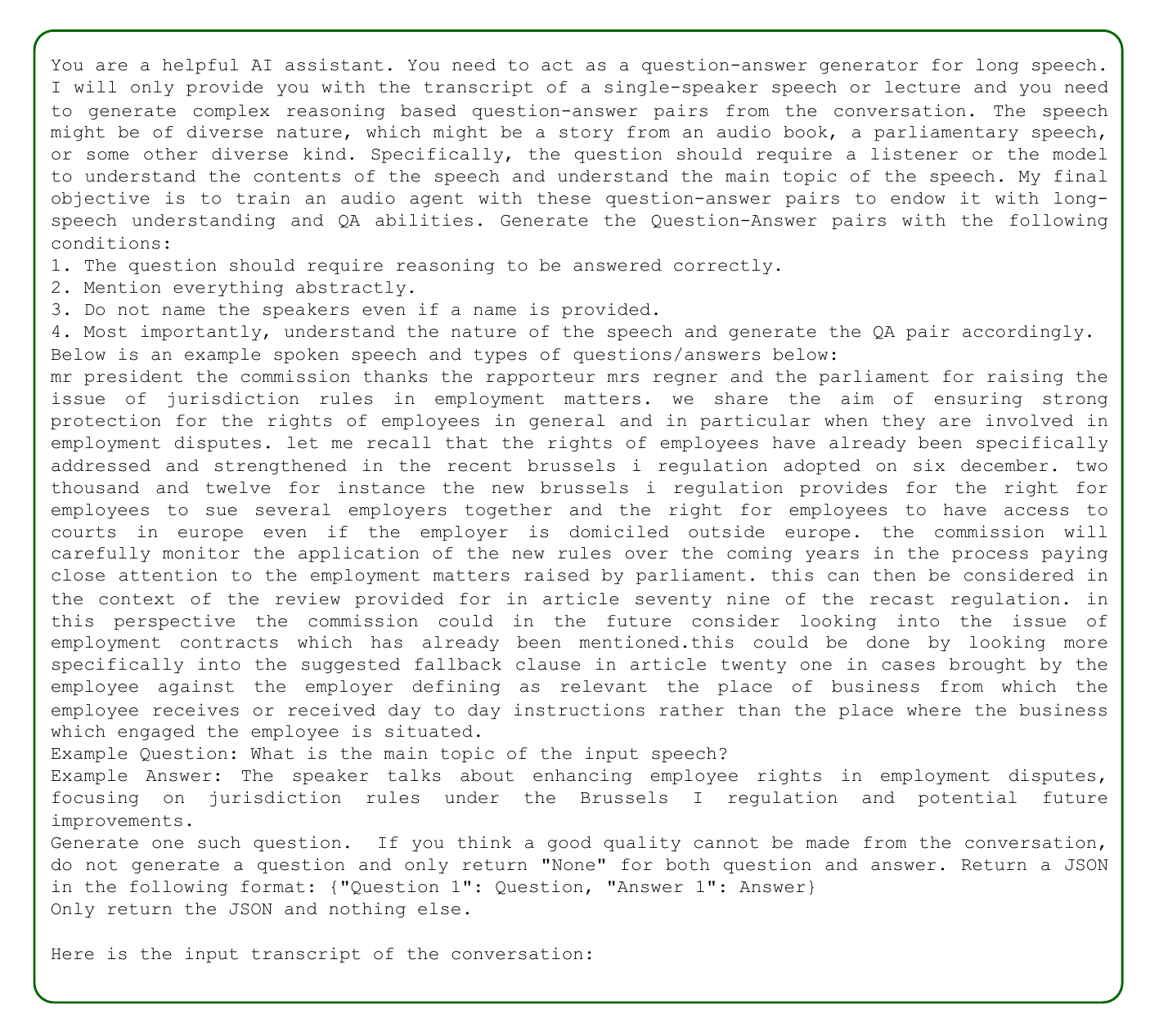}
    \caption{\small Prompt used for generating \textbf{Topic QA} (Information Extraction type) for LongAudioXL.}
    \label{fig:maintopic_prompt}
\end{figure}

\begin{figure}
    \centering
    \includegraphics[width=\linewidth]{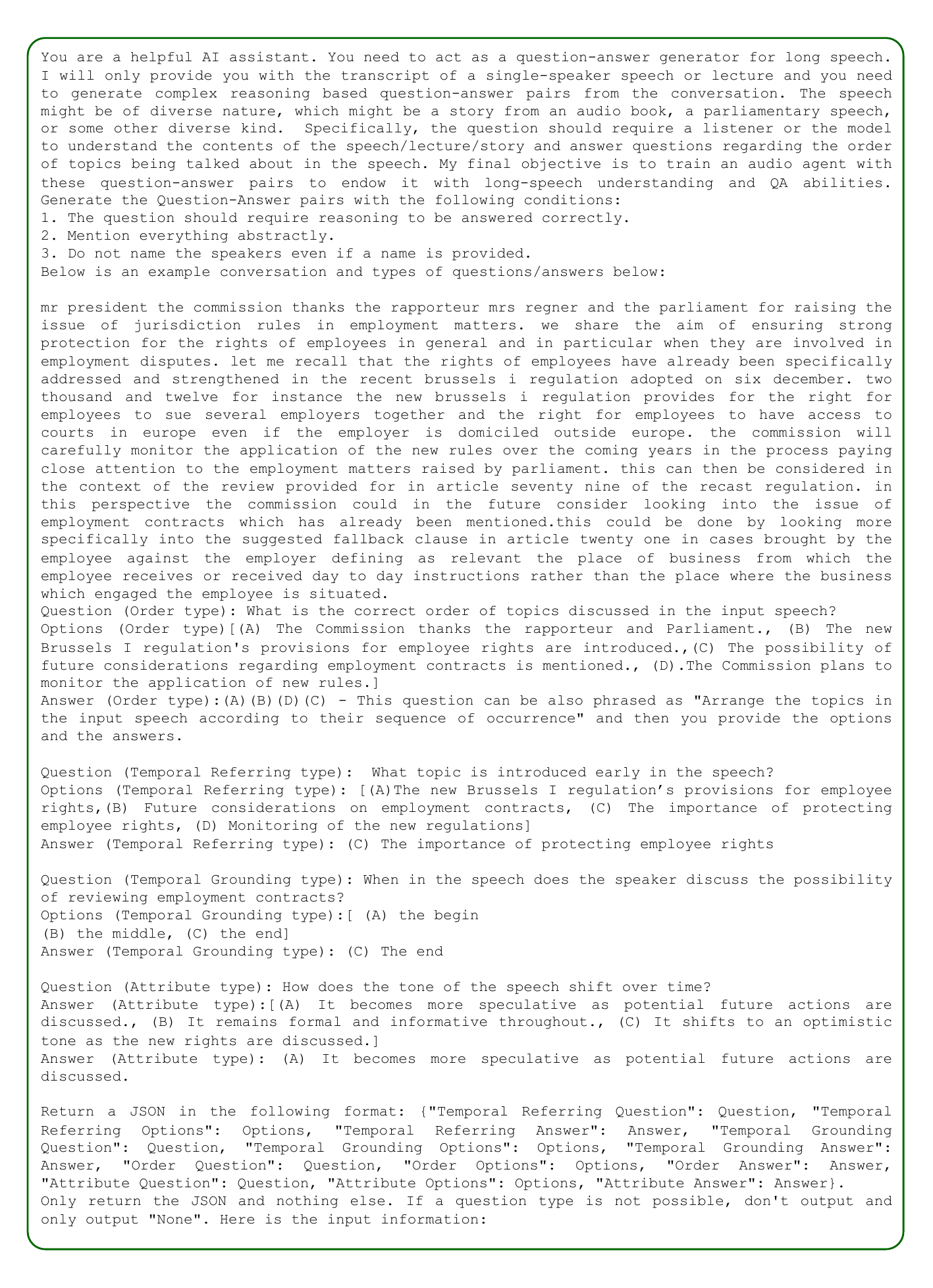}
    \caption{\small Prompt used for generating \textbf{Order QA} for LongAudioXL.}
    \label{fig:order_prompt}
\end{figure}

\begin{figure}
    \centering
    \includegraphics[width=\linewidth]{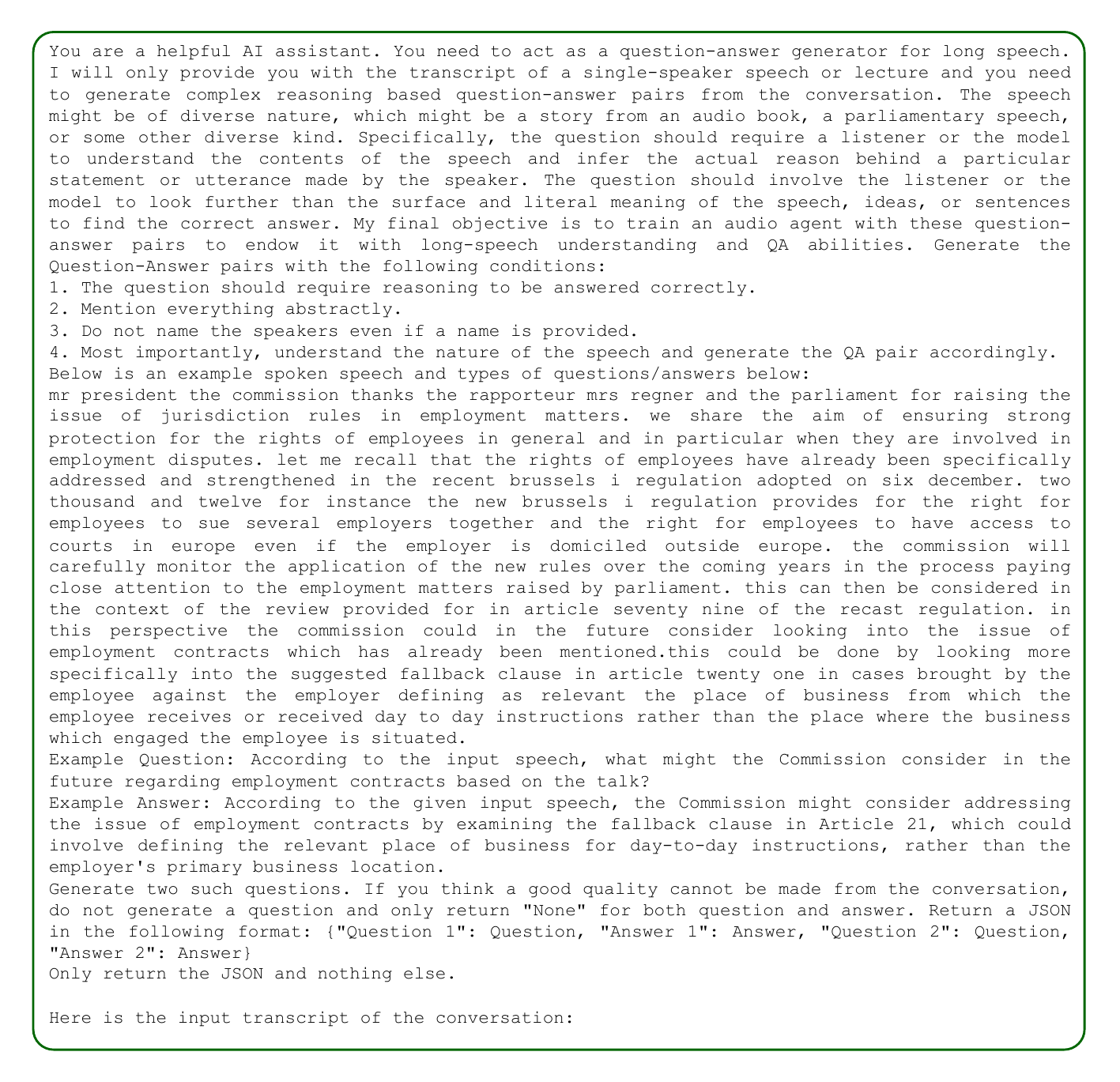}
    \caption{\small Prompt used for generating \textbf{Causal QA} for LongAudioXL.}
    \label{fig:reason_behind_prompt}
\end{figure}

\begin{figure}
    \centering
    \includegraphics[width=\linewidth]{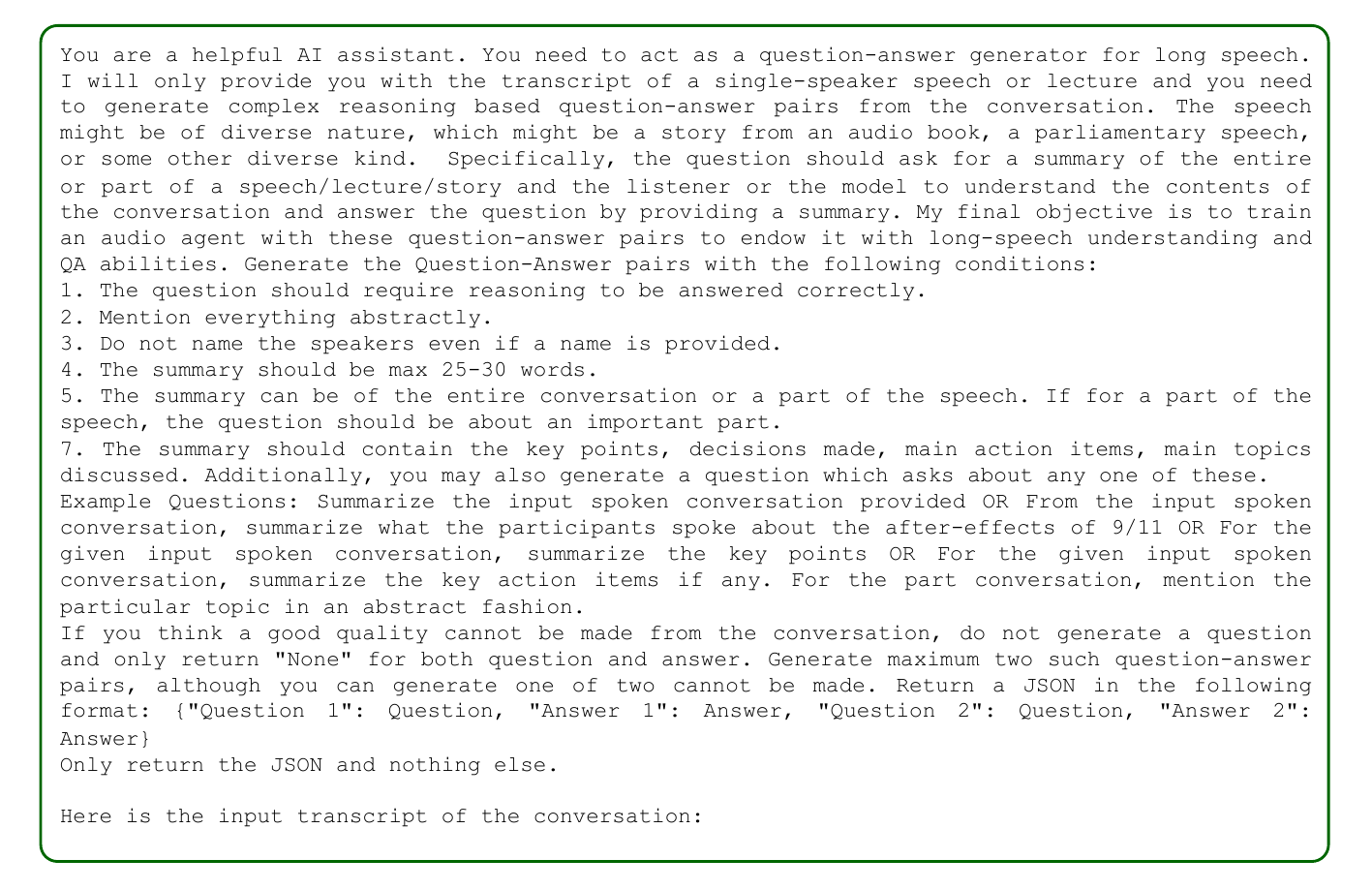}
    \caption{\small Prompt used for generating \textbf{Summarization QA} (Summary QA) for LongAudioXL.}
    \label{fig:summary_prompt}
\end{figure}

\begin{figure}
    \centering
    \includegraphics[width=\linewidth]{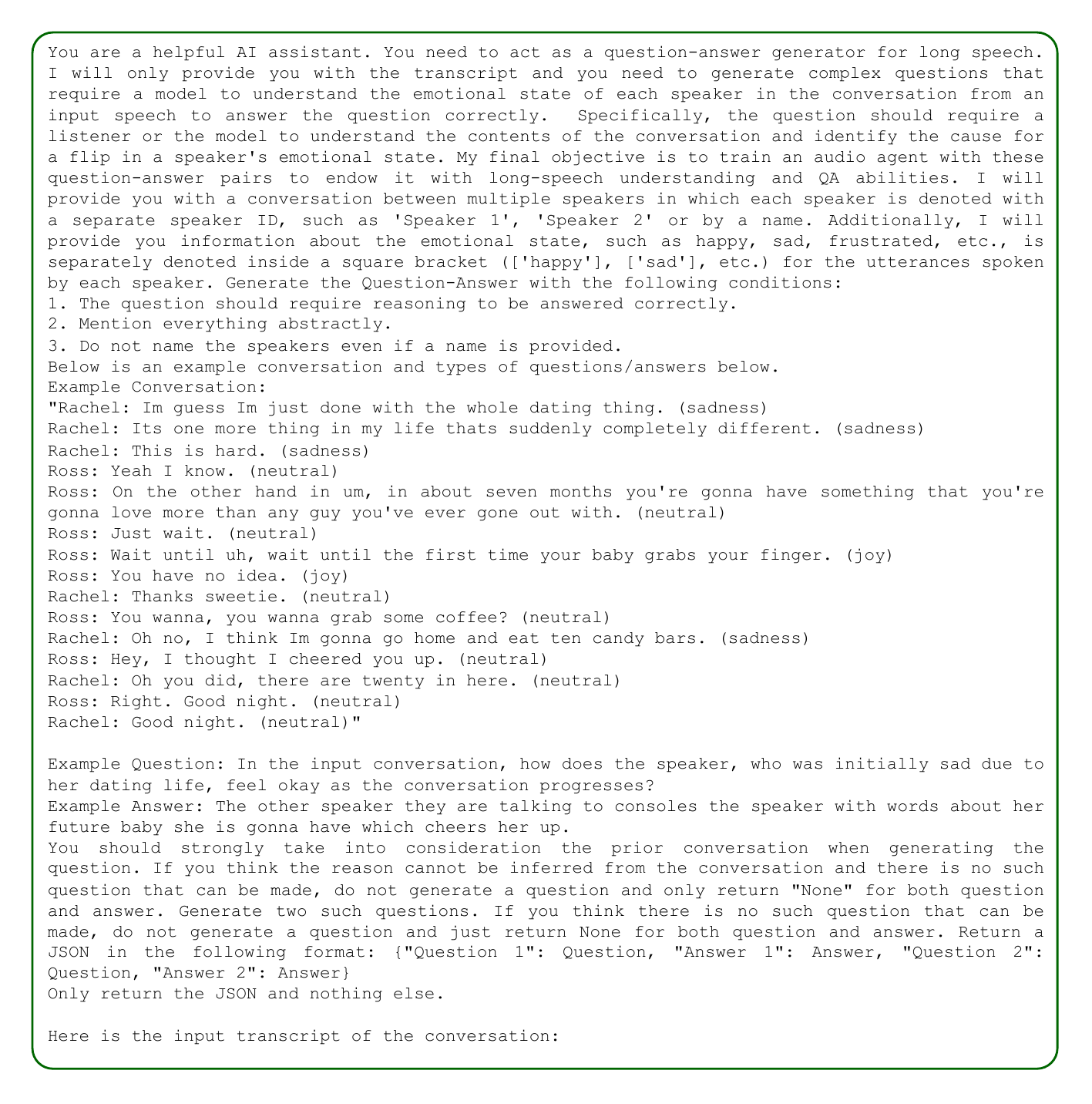}
    \caption{\small Prompt used for generating \textbf{Emotion Flip QA} (Emotional State Reasoning type) for LongAudioXL.}
    \label{fig:emo_flip_prompt}
\end{figure}

\begin{figure}
    \centering
    \includegraphics[width=\linewidth]{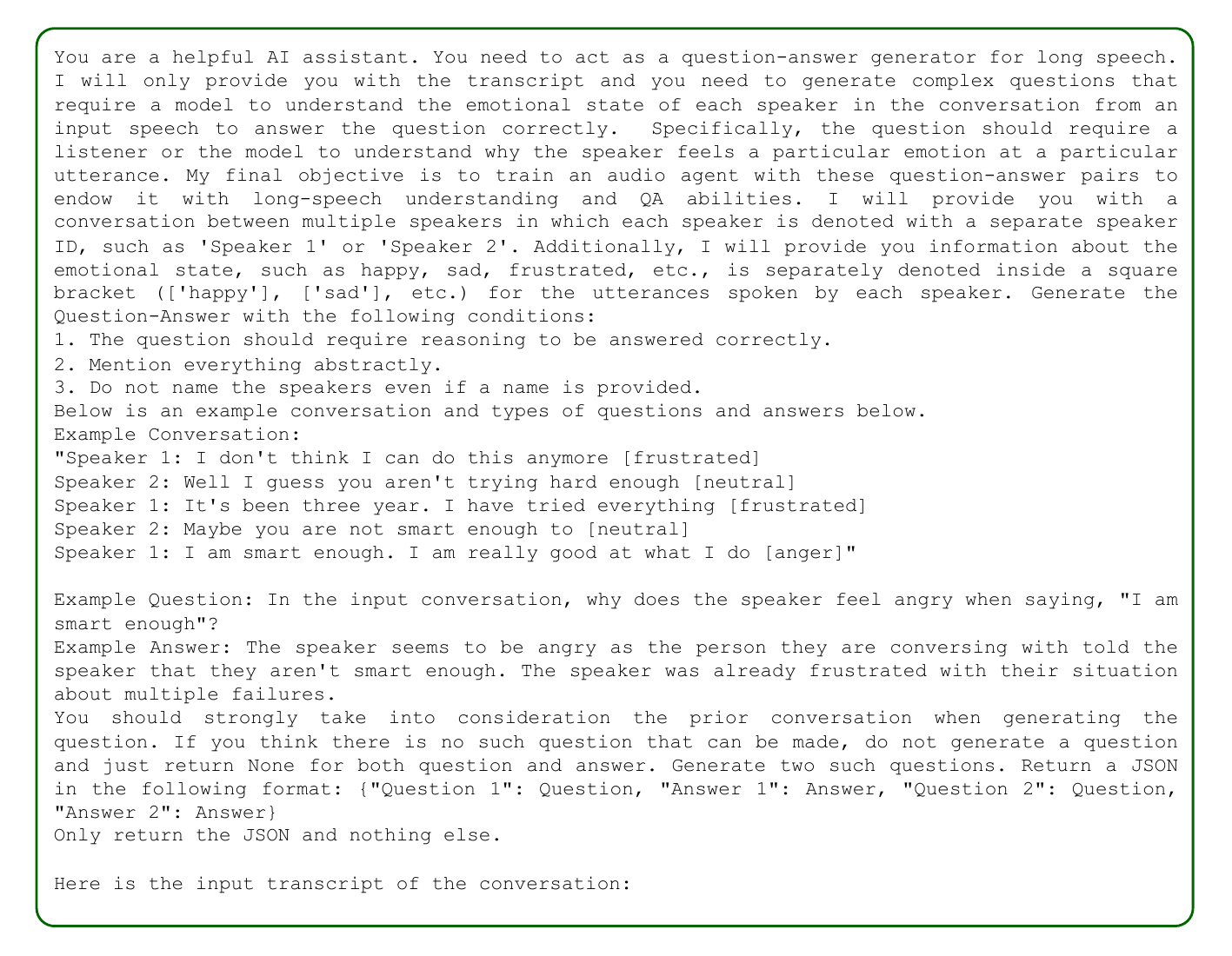}
    \caption{Prompt used for generating \textbf{Causal Reasoning} (Emotional State Reasoning type) for LongAudioXL.}
    \label{fig:emotional_reasono_prompt}
\end{figure}

\begin{figure}
    \centering
    \includegraphics[width=\linewidth]{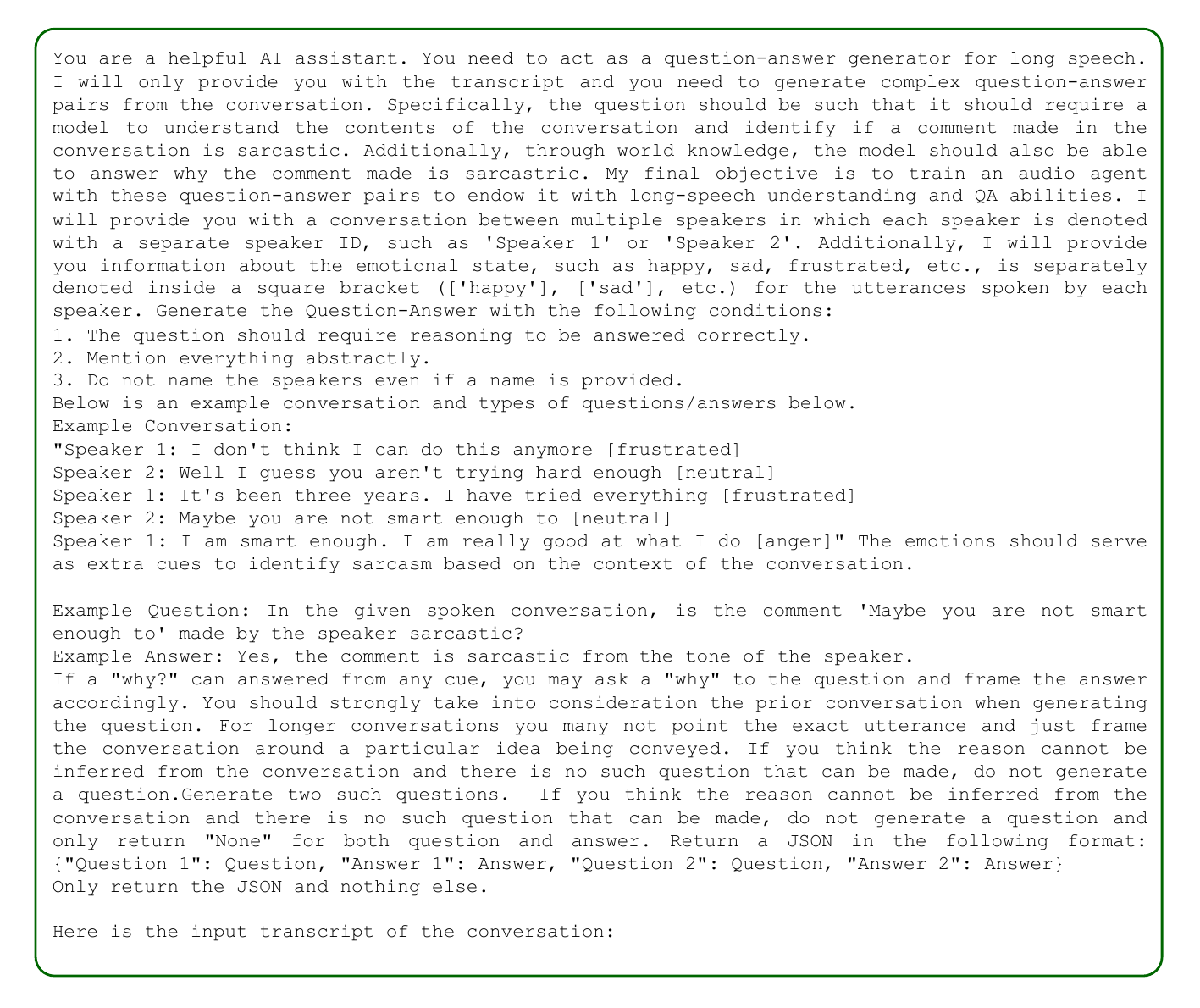}
    \caption{\small Prompt used for generating \textbf{Sarcasm Identification QA} for LongAudioXL.}
    \label{fig:emotional_sarcasm_prompt}
\end{figure}

\begin{figure}
    \centering
    \includegraphics[width=\linewidth]{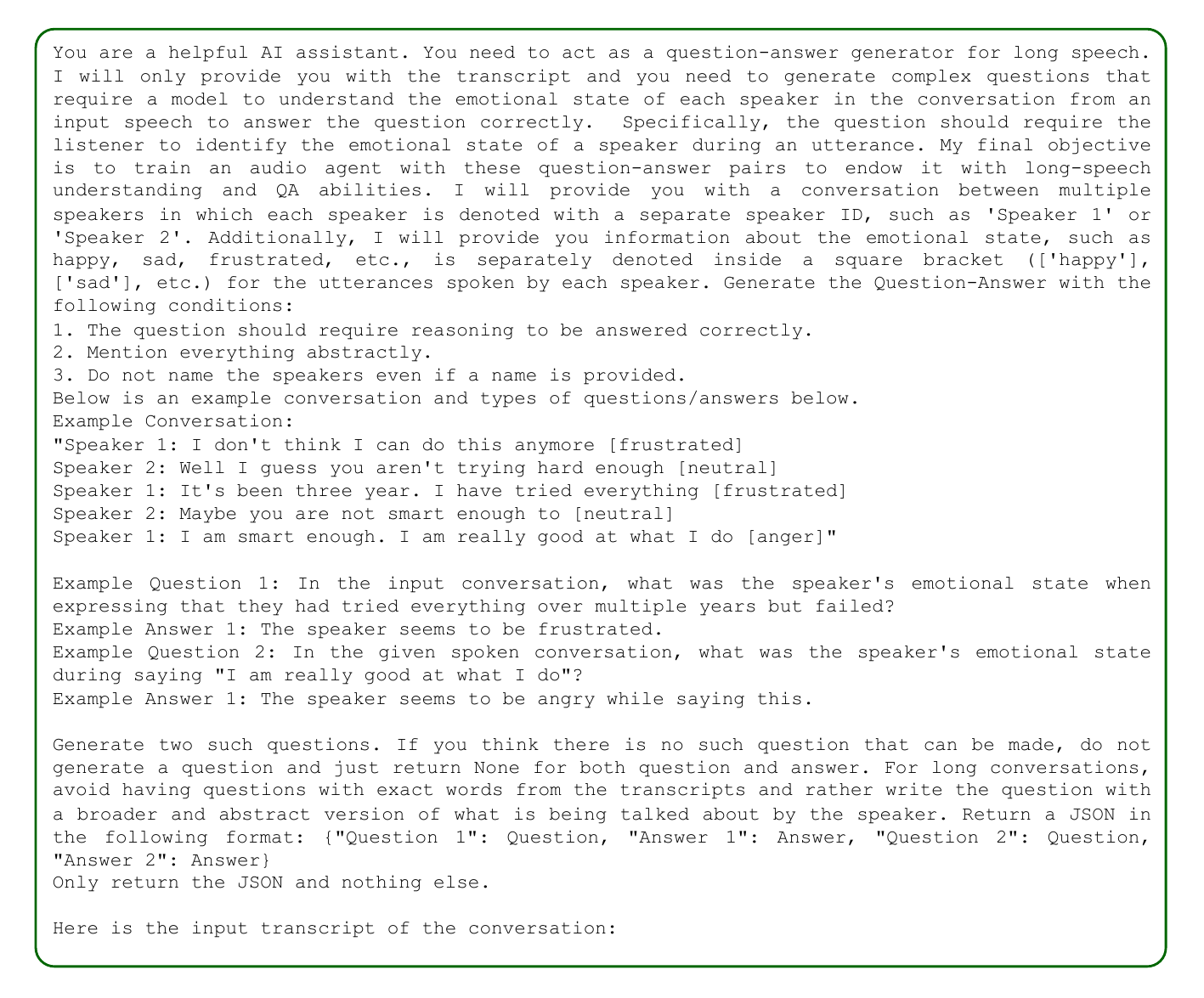}
    \caption{\small Prompt used for generating \textbf{Identification QA} (Emotional State Reasoning type) for LongAudioXL.}
    \label{fig:emotional_state_prompt}
\end{figure}

\begin{figure}
    \centering
    \includegraphics[width=\linewidth]{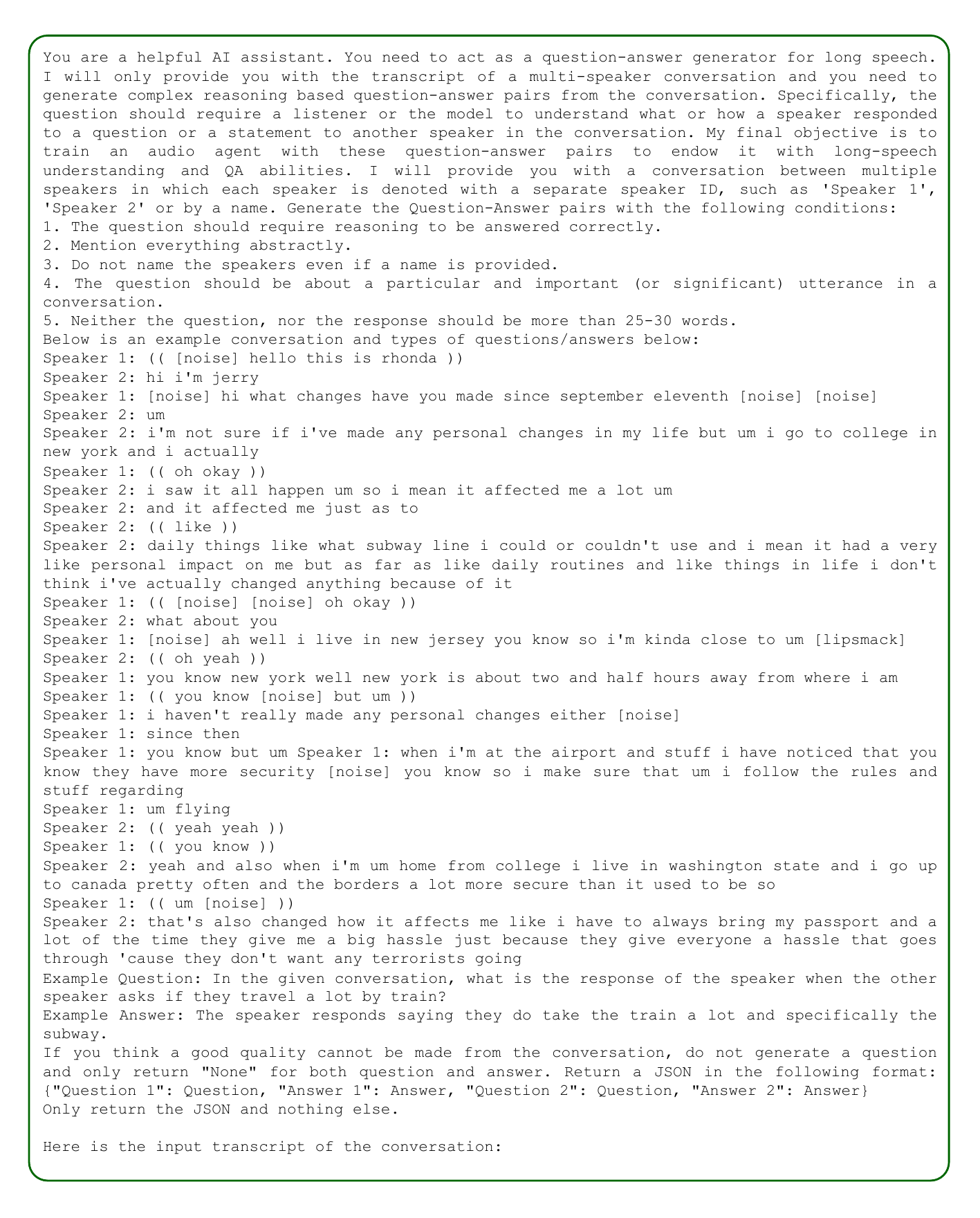}
    \caption{\small Prompt used for generating \textbf{ResponseQA} (Information Extraction type) for LongAudioXL.}
    \label{fig:respond_prompt}
\end{figure}

\begin{figure}
    \centering
    \includegraphics[width=\linewidth]{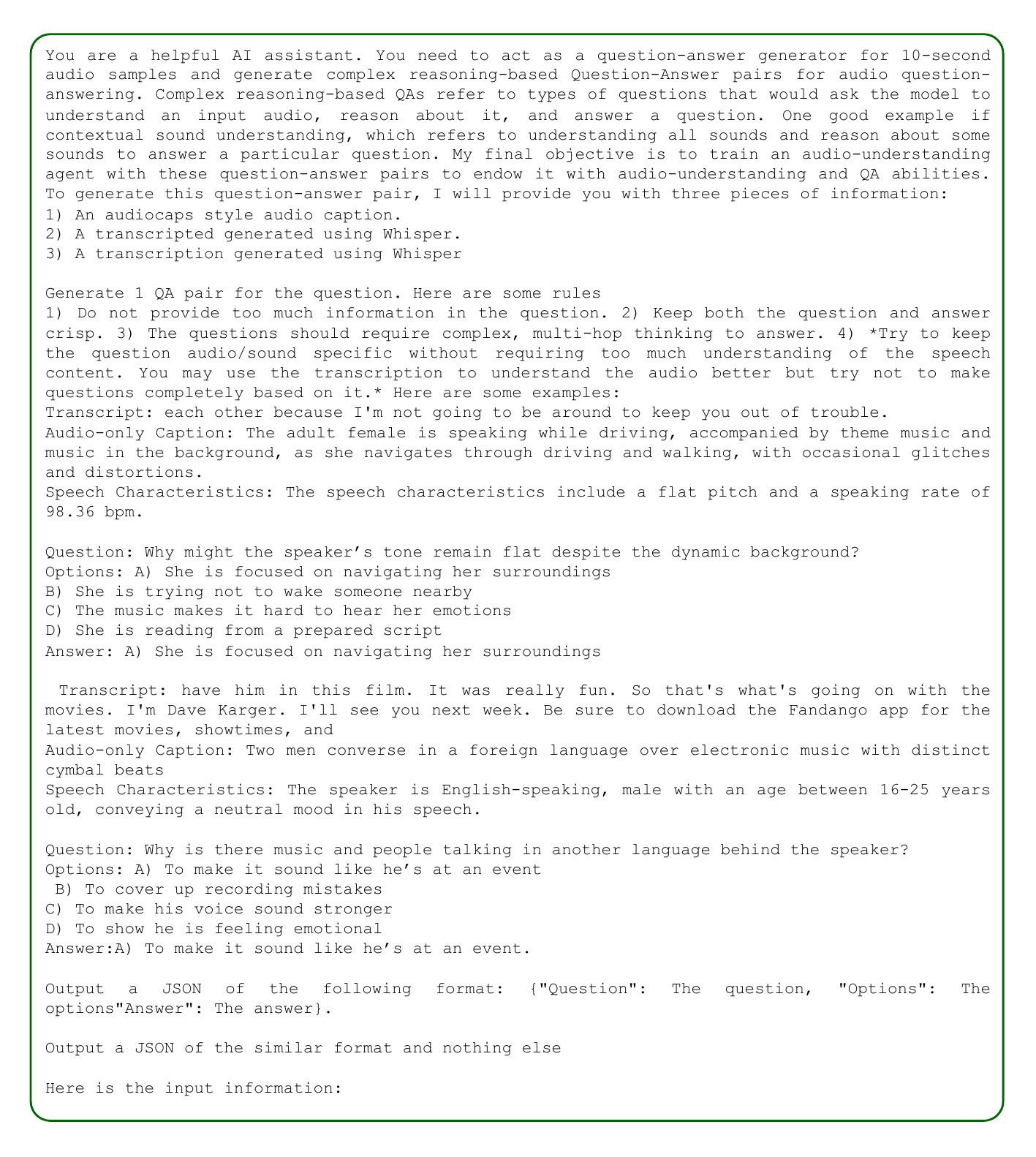}
    \caption{\small Prompt used for generating \textbf{Speech-in-Sound QA} for AudioSkills-XL.}
    \label{fig:speech_sound_prompt}
\end{figure}

\begin{figure}[h]
    \centering
    \includegraphics[width=\linewidth]{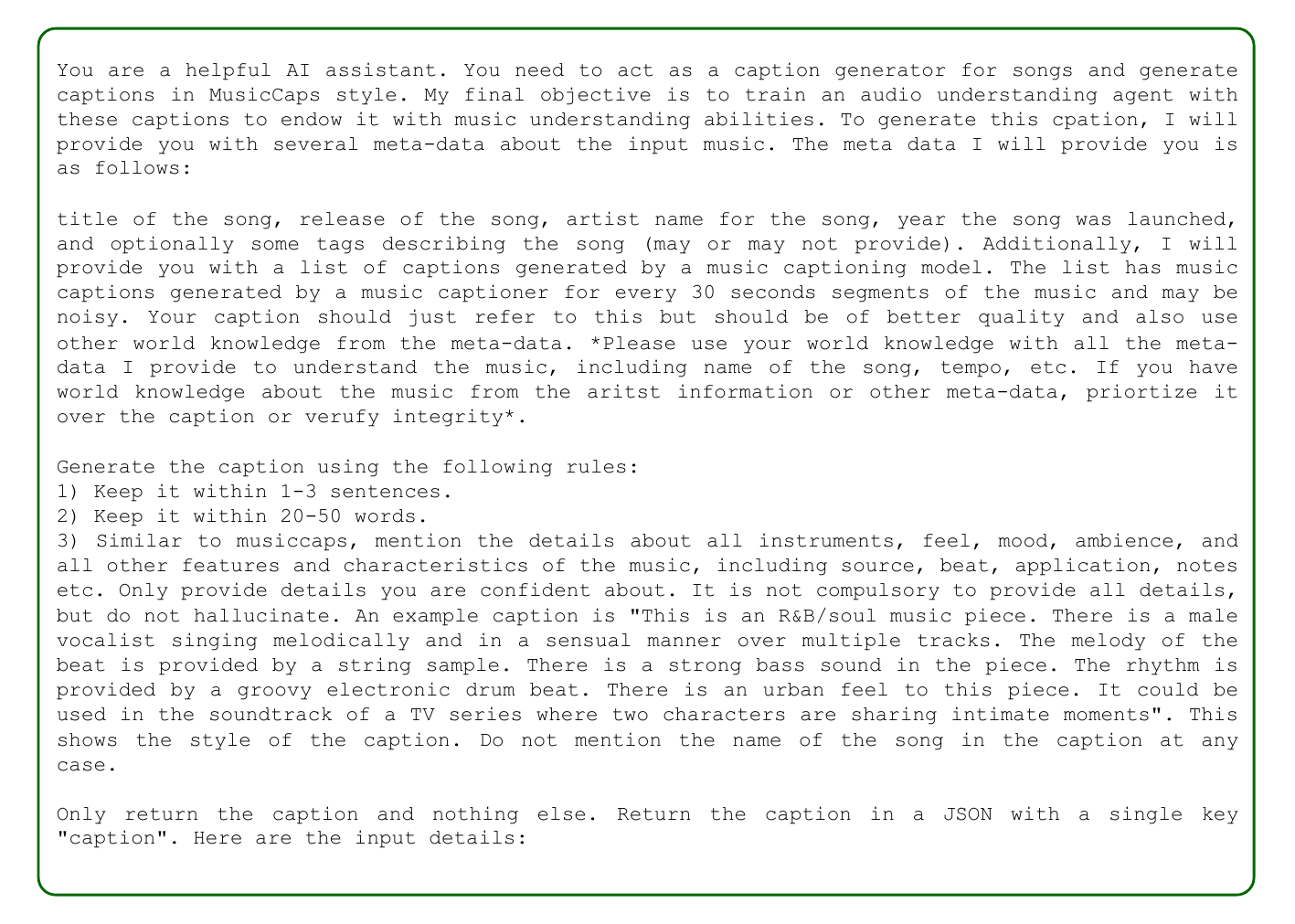}
    \caption{\small Prompt used for generating \textbf{captions for Million Songs Dataset}. Noisy captions for the prompt are generated using AF2.}
    \label{fig:msd_caption}
\end{figure}

\begin{figure}[h]
    \centering
    \includegraphics[width=\linewidth]{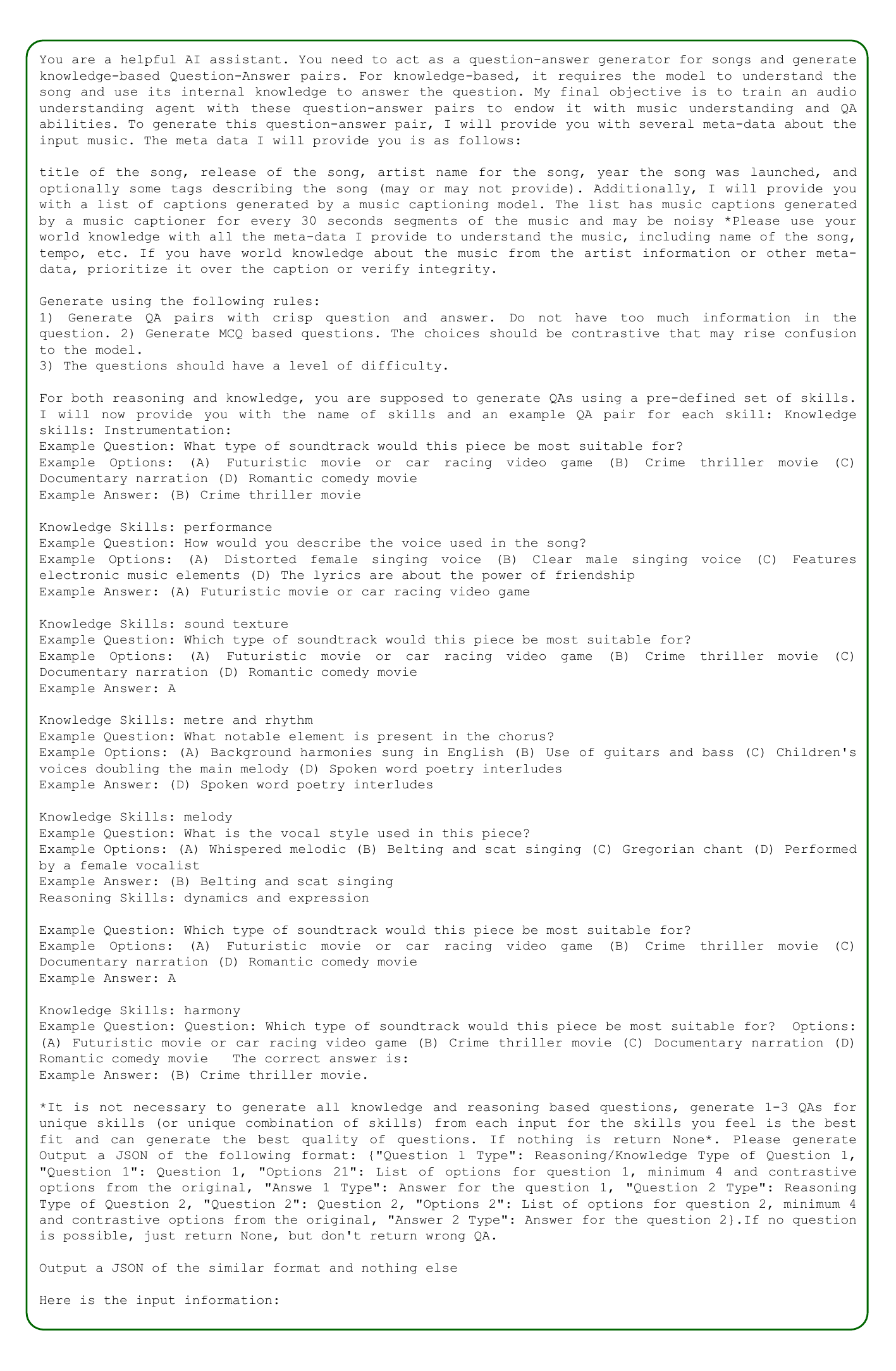}
    \caption{\small Prompt used for generating \textbf{Music Knowledge QA} from Million Songs Dataset for AudioSkills-XL. Noisy captions for the prompt are generated using AF2.}
    \label{fig:msd_knowledge}
\end{figure}

\begin{figure}[h]
    \centering
    \includegraphics[width=\linewidth]{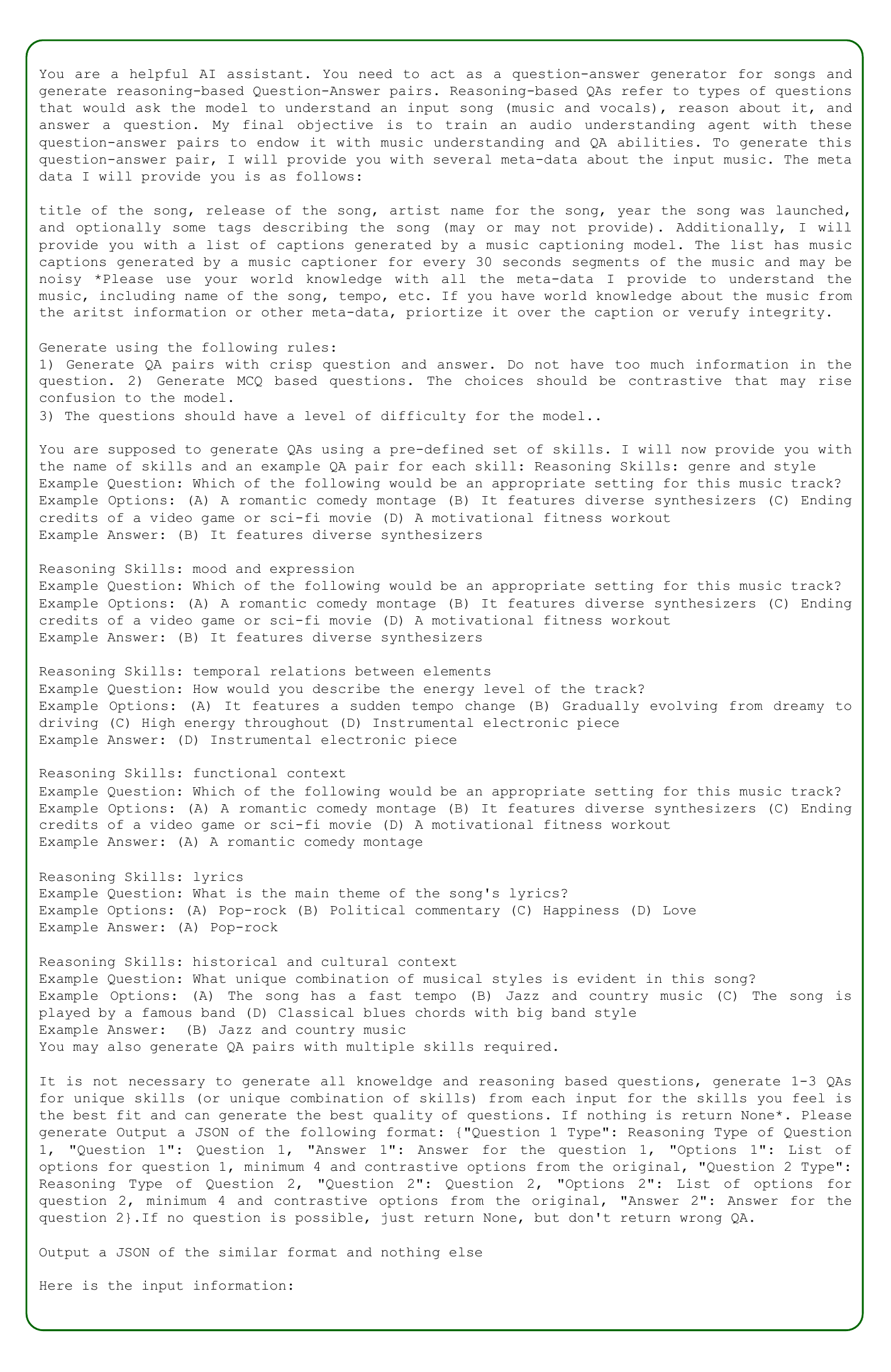}
    \caption{\small Prompt used for generating \textbf{Music Reasoning QA} from Million Songs Dataset for AudioSkills-XL. Noisy captions for the prompt are generated using AF2.}
    \label{fig:msd_reasoning}
\end{figure}

\begin{figure}[h]
    \centering
    \includegraphics[width=\linewidth]{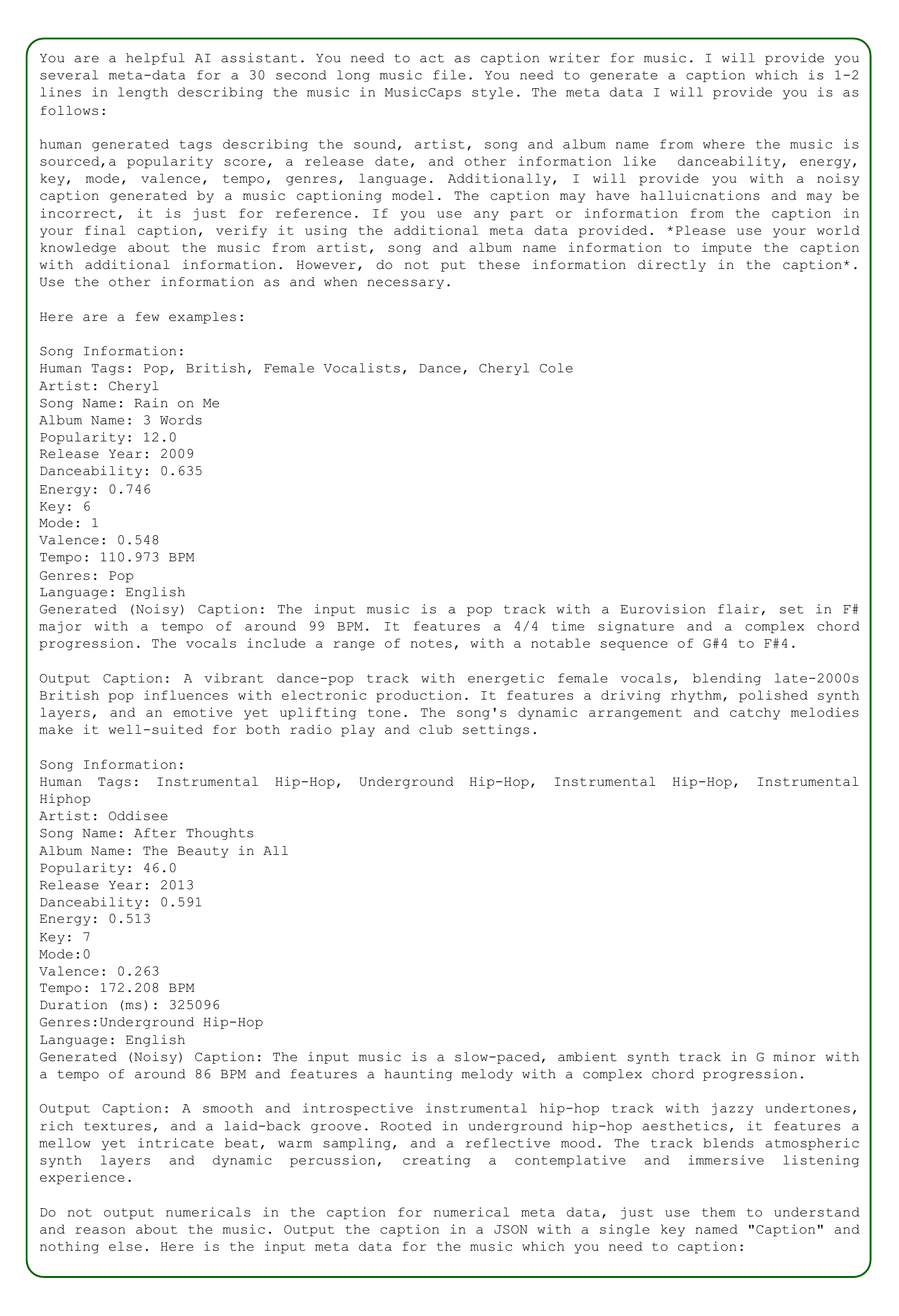}
    \caption{\small Prompt used for generating \textbf{captions for Music4All}. Noisy captions for the prompt are generated using AF2.}
    \label{fig:music4all_caption_prompt}
\end{figure}

\begin{figure}[h]
    \centering
    \includegraphics[width=\linewidth, trim=0 1cm 0 0]{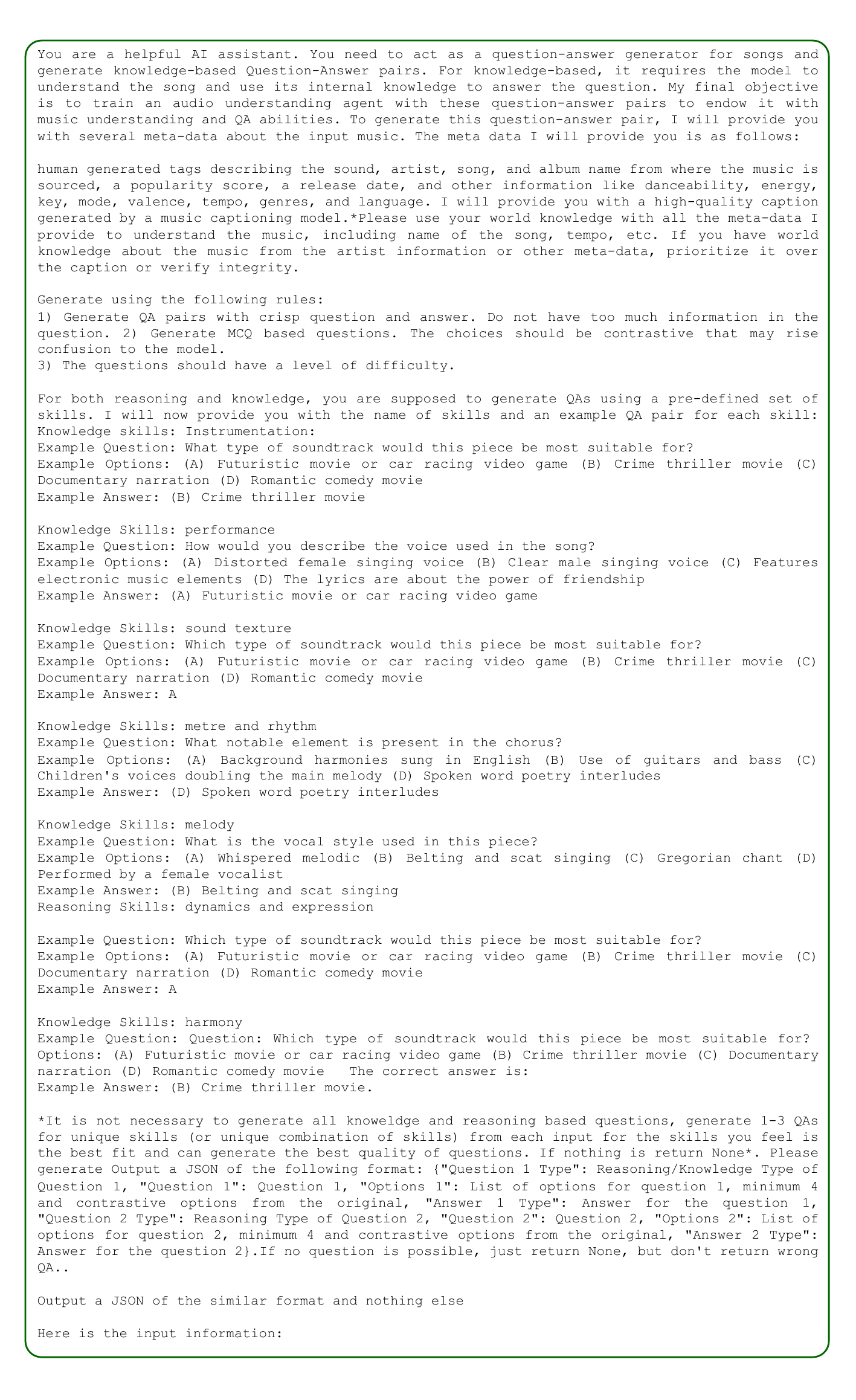}
    \caption{Prompt used for generating \textbf{Music Knowledge QA} from Music4All for AudioSkills-XL. Noisy captions for the prompt are generated using AF2.}
    \label{fig:music4all_knowledge}
\end{figure}

\begin{figure}
    \centering
    \includegraphics[width=\linewidth, trim=0 1cm 0 0]{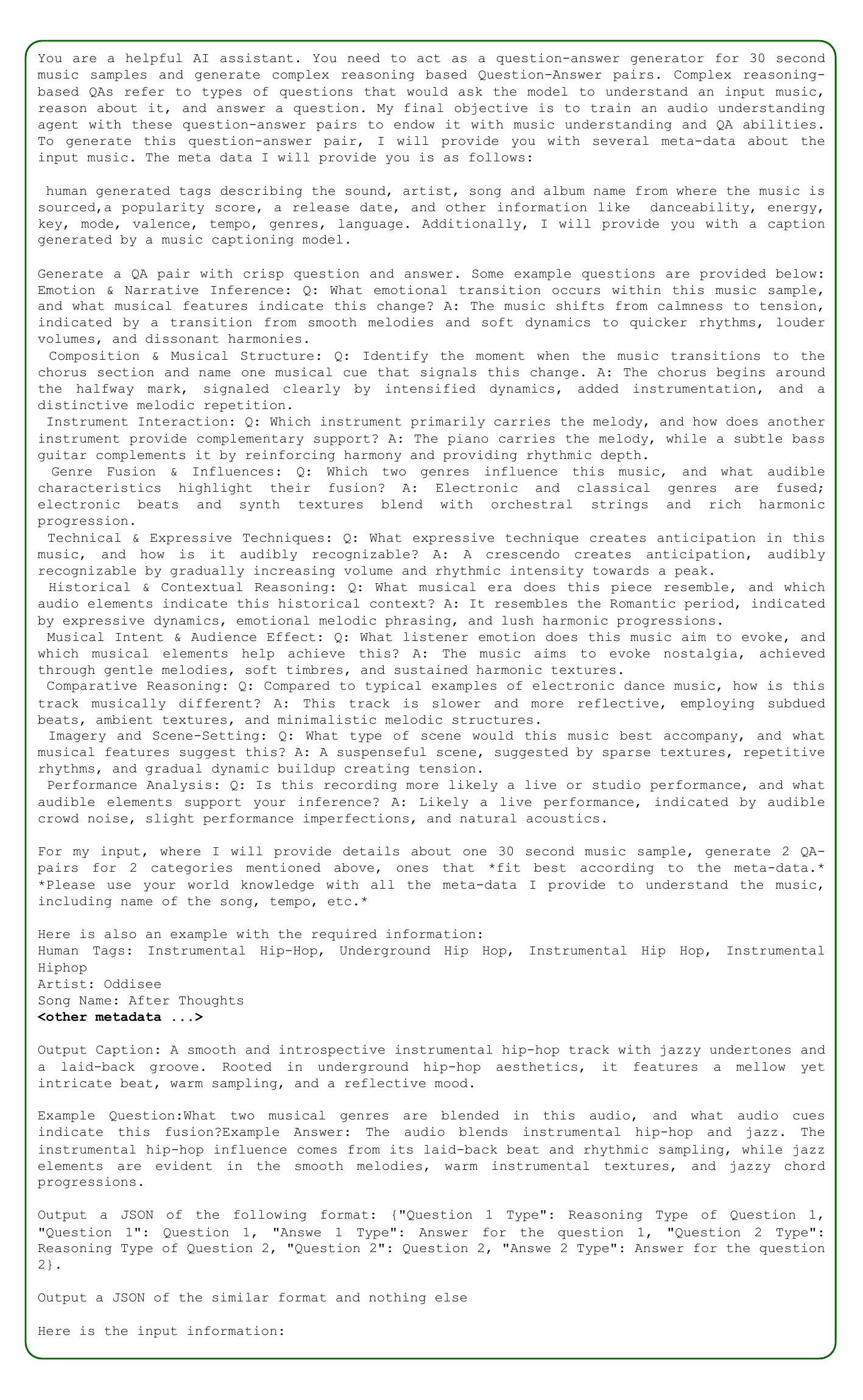}
    \caption{Prompt used for generating \textbf{General Open-Ended Complex Reasoning QA} for Music Reasoning QA from Music4All for AudioSkills-XL.}
    \label{fig:music4all_openqa_prompt}
\end{figure}

\begin{figure}
    \centering
    \includegraphics[width=\linewidth]{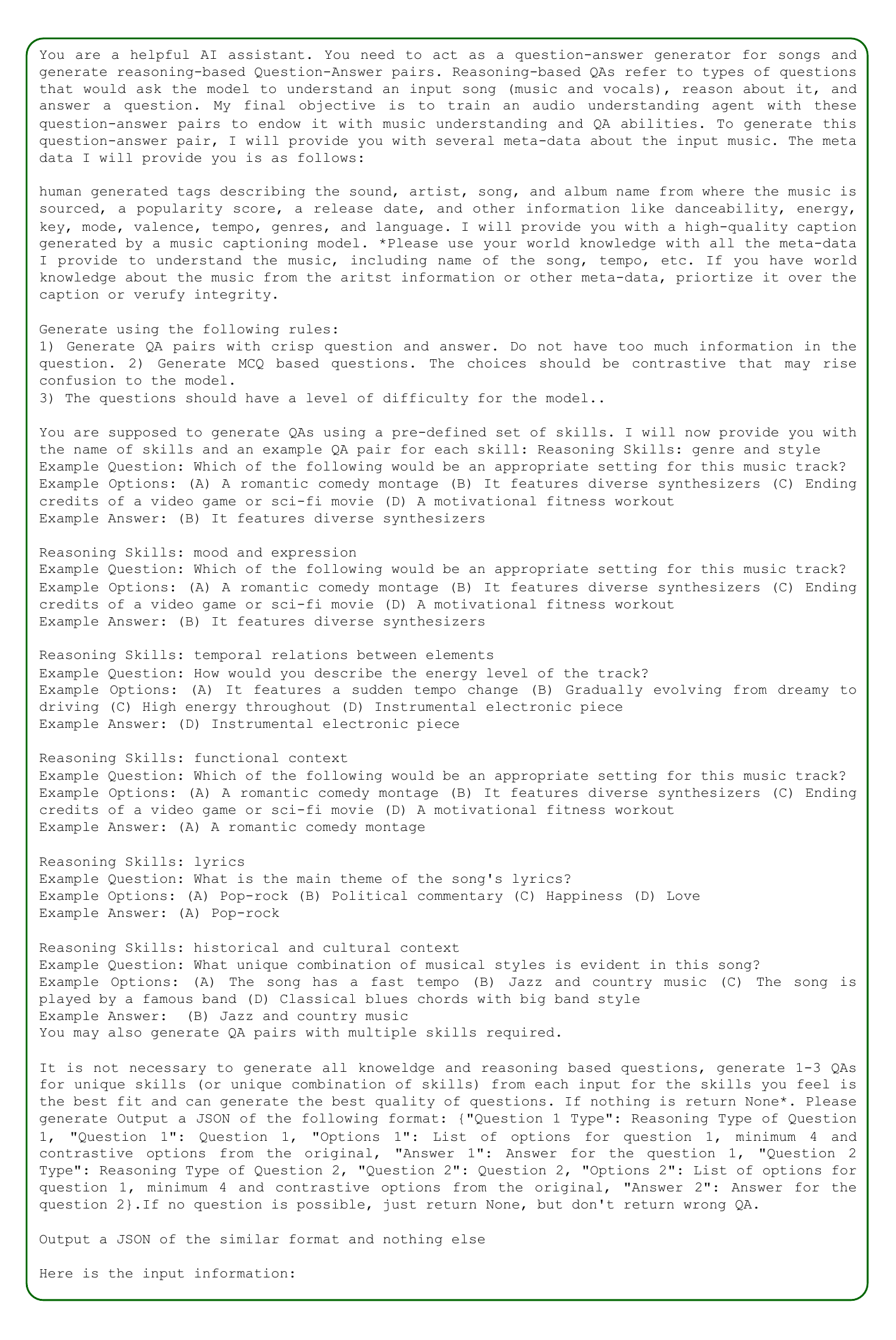}
    \caption{\small Prompt used for generating \textbf{Music Reasoning QA} from Music4All for AudioSkills-XL. Noisy captions for the prompt are generated using AF2.}
    \label{fig:music4all_reasoning_qa_prompt}
\end{figure}

\begin{figure}
    \centering
    \includegraphics[width=\linewidth]{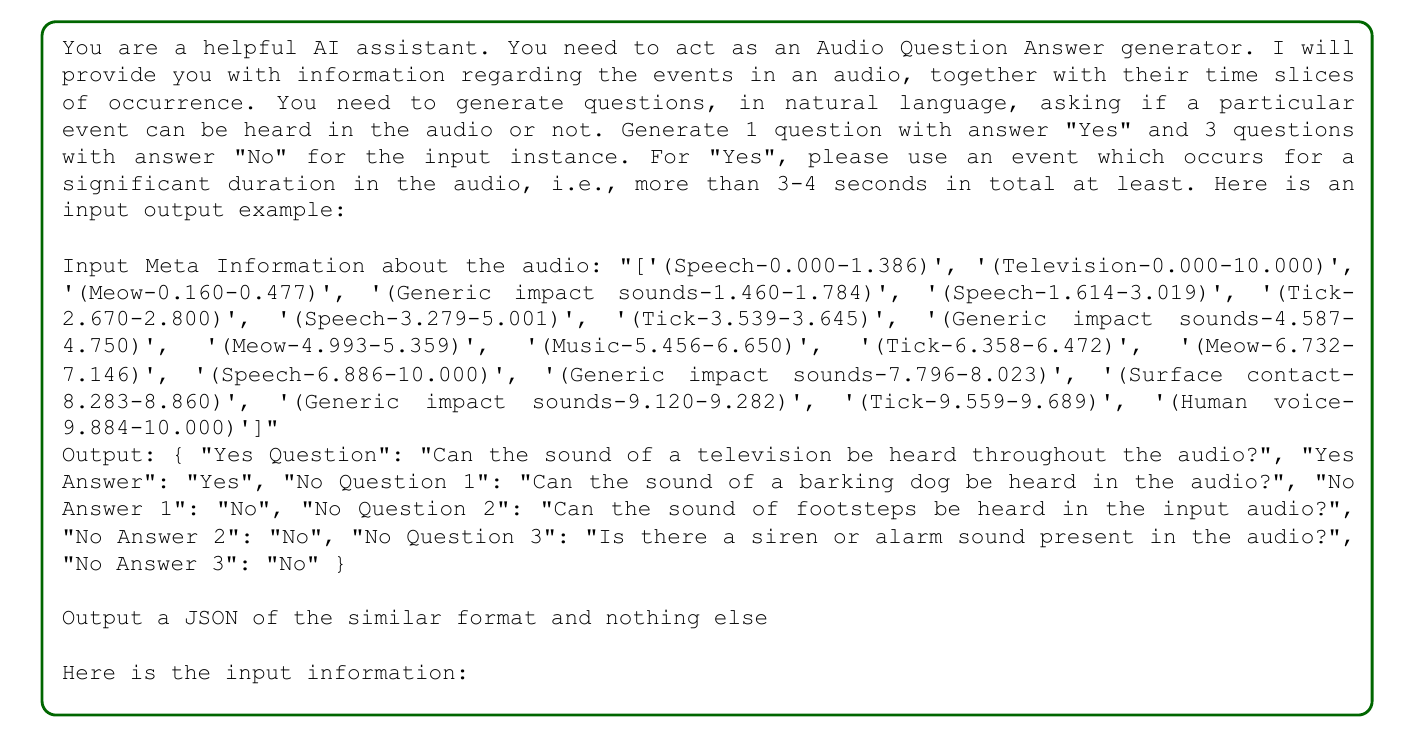}
    \caption{\small Prompt used for generating \textbf{Yes-No QA} (part of General Reasoning+ours (only sound) from AudioSet-SL) for AudioSkills-XL.}
    \label{fig:yes_no_qa}
\end{figure}

\begin{figure}
    \centering
    \includegraphics[width=\linewidth]{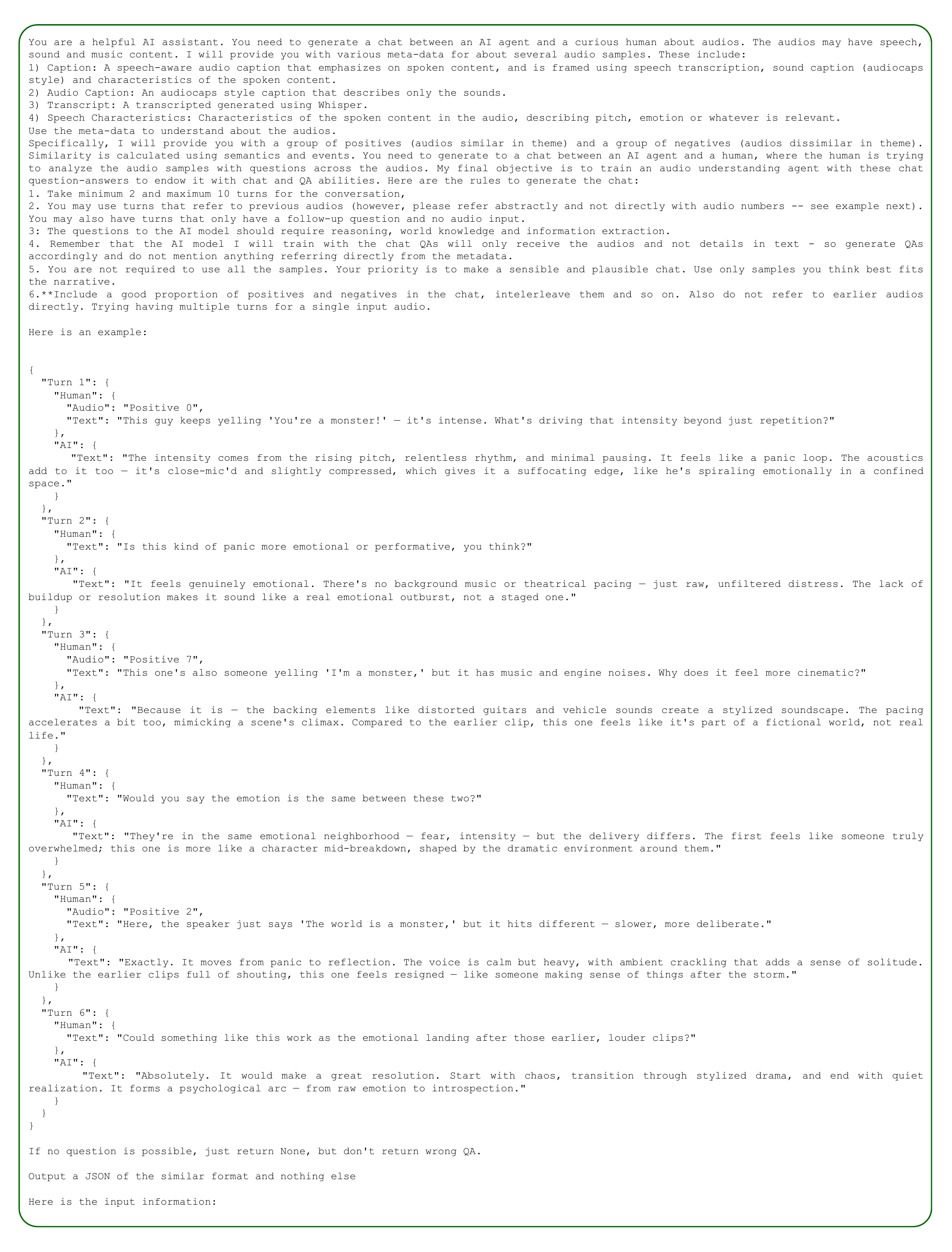}
    \caption{\small Prompt used for generating \textbf{multi-turn, multi-audio chat/dialogues (speech and sounds)} for AF-Chat.}
    \label{fig:chat_sound}
\end{figure}

\begin{figure}
    \centering
    \includegraphics[width=\linewidth]{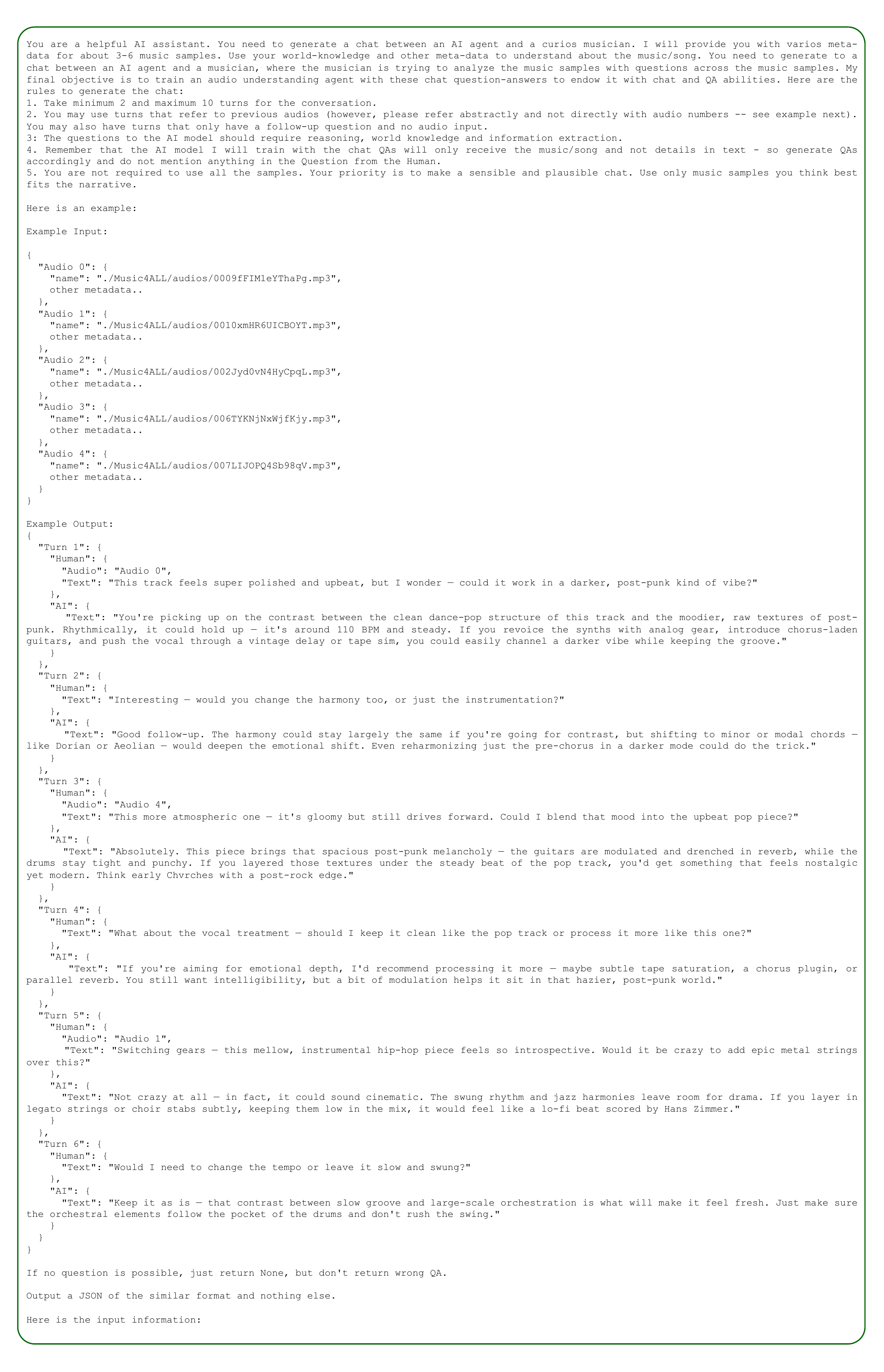}
    \caption{\small Prompt used for generating \textbf{multi-turn, multi-audio chat/dialogues (music)} for AF-Chat.}
    \label{fig:music_chat}
\end{figure}

\begin{figure}
    \centering
    \includegraphics[width=\linewidth]{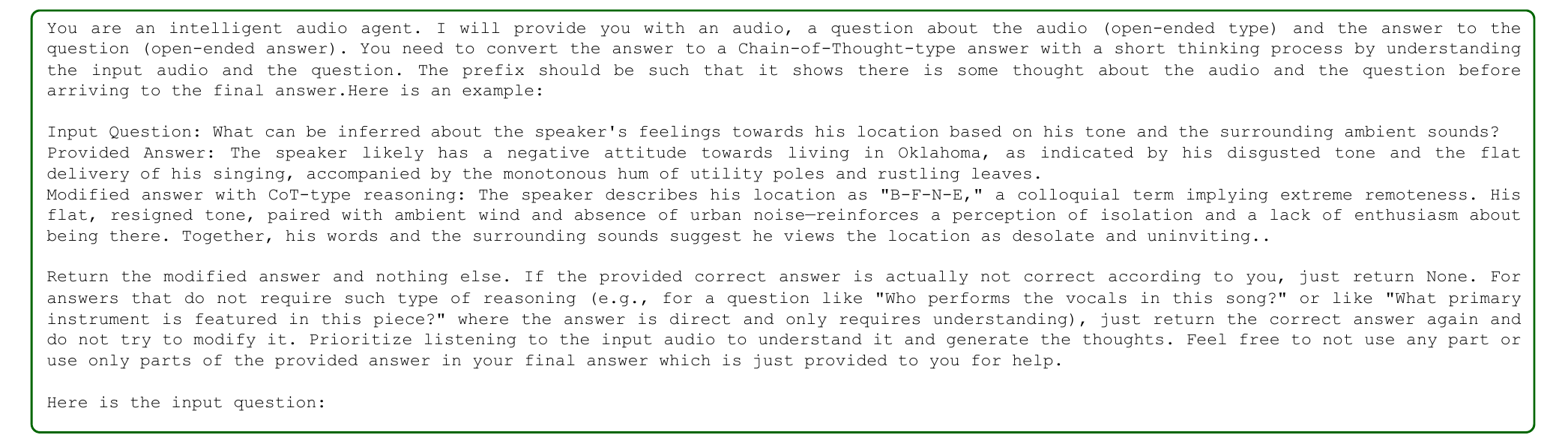}
    \caption{\small Prompt 1 used for generating CoT-style reasoning focused on speech and ambient sounds (input instances sampled from Speech-in-Sound Caps, which is curated using YouTube8M) for AF-Think.}
    \label{fig:cot_ambient}
\end{figure}

\begin{figure}
    \centering
    \includegraphics[width=\linewidth]{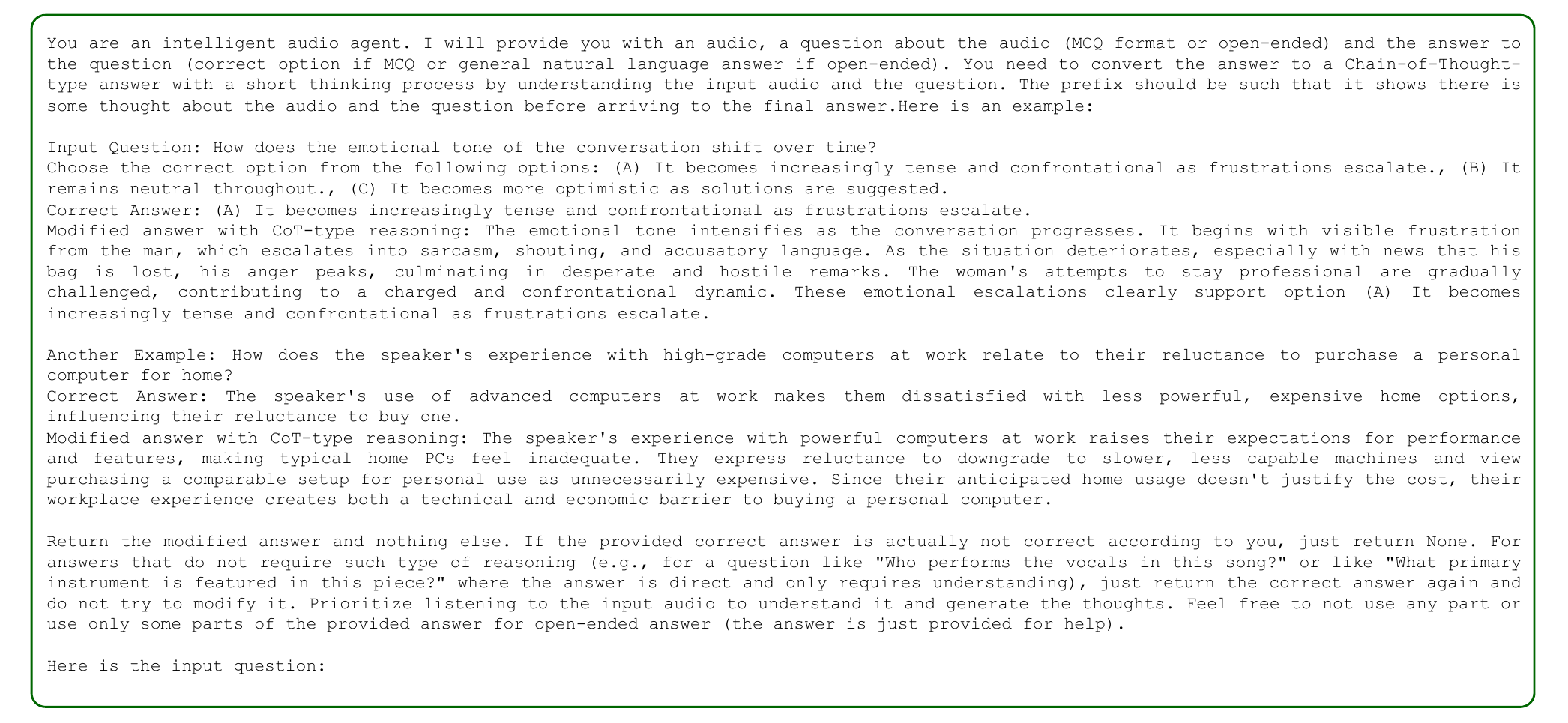}
    \caption{\small Prompt 2 used for generating CoT-style reasoning focused on SpeechQAs (input instances randomly sampled from LongAudio-XL speech subset) for AF-Think.}
    \label{fig:cot_emotion_shift}
\end{figure}

\begin{figure}
    \centering
    \includegraphics[width=\linewidth]{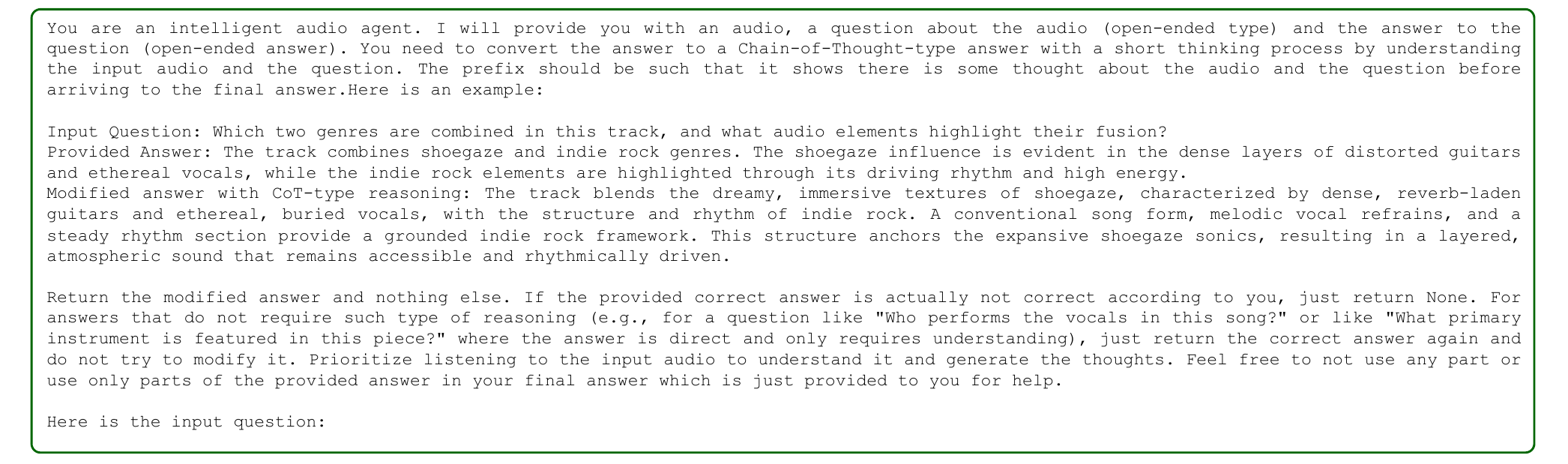}
    \caption{Prompt 3 used for generating CoT-style reasoning focused on music (input instances sampled from our Music Knowledge and Reasoning subset of AudioSkills-XL) for AF-Think. This focuses on open-ended QA.}
    \label{fig:cot_music_2}
\end{figure}

\begin{figure}
    \centering
    \includegraphics[width=\linewidth]{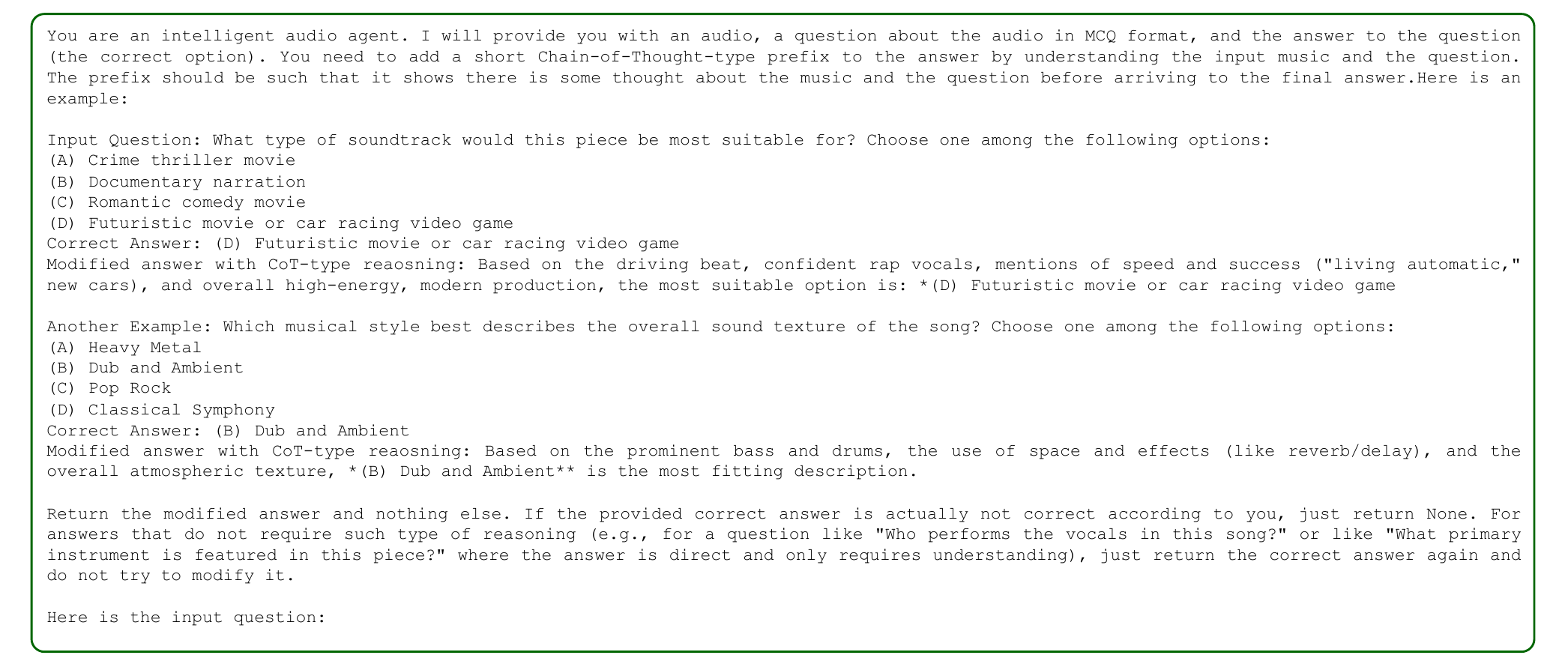}
    \caption{\small Prompt 4 used for generating CoT-style reasoning focused on music (input instances sampled from our Music Knowledge and Reasoning subset of AudioSkills-XL) for AF-Think. This focuses on MCQ-based QA.}
    \label{fig:cot_music}
\end{figure}

\begin{figure}
    \centering
    \includegraphics[width=\linewidth]{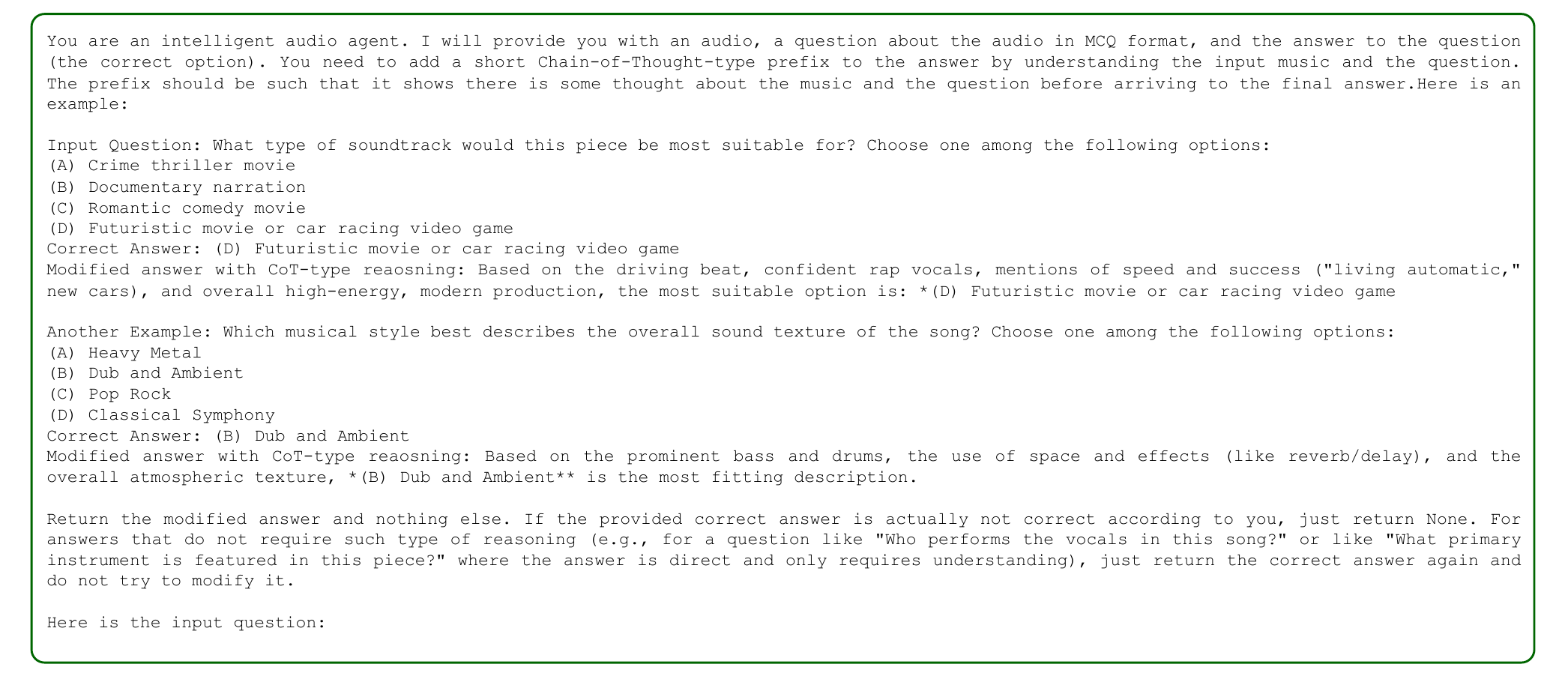}
    \caption{\small Prompt 5 used for generating CoT-style reasoning focused on ambient sounds only (input instances sampled from our Sound Reasoning subset of AudioSkills-XL, which is curated from YouTube8M) for AF-Think. This focuses on MCQ-based QA.}
    \label{fig:cot_sound}
\end{figure}

\begin{figure}
    \centering
    \includegraphics[width=\linewidth]{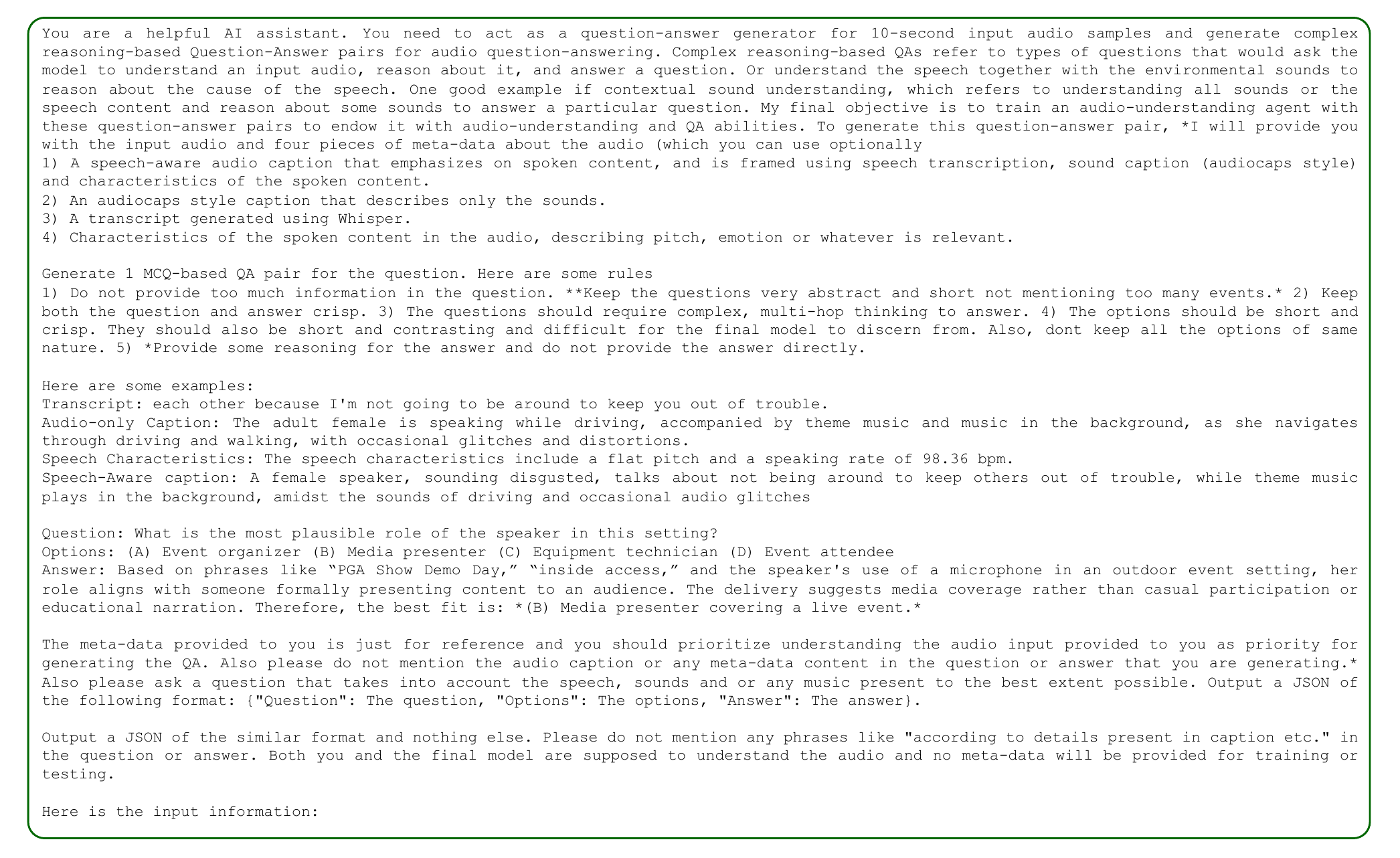}
    \caption{\small Prompt 6 used for generating CoT-style reasoning focused on speech and ambient sounds (input instances sampled from Speech-in-Sound Caps, which is curated using YouTube8M) for AF-Think. This focuses on MCQ-based QA.}
    \label{fig:missed_prompt}
\end{figure}

\begin{figure}
    \centering
    \includegraphics[width=\linewidth]{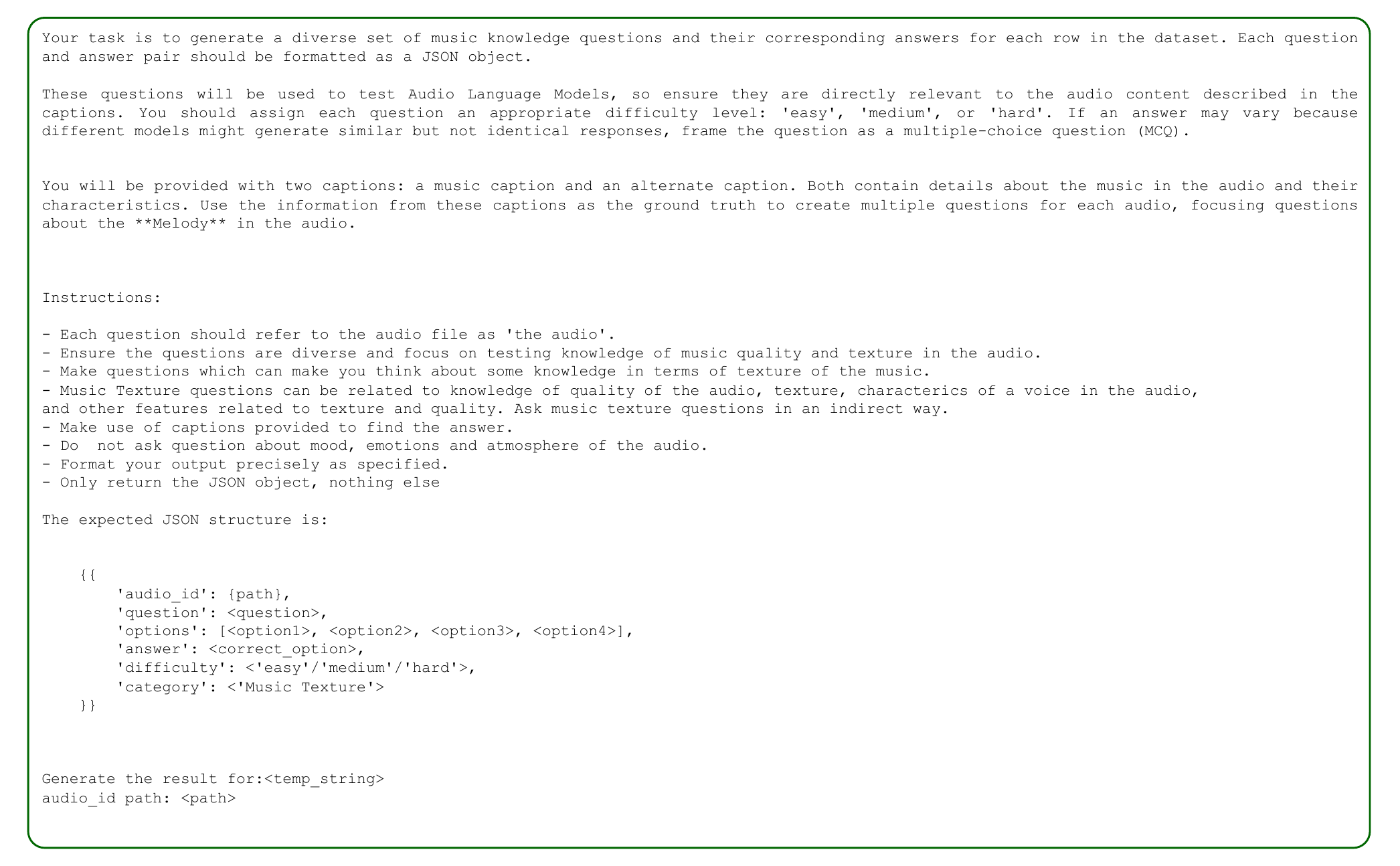}
    \caption{\small Prompt 1 used for \textbf{Music Reasoning} for AudioSkills-XL. The QAs are focused on music texture reasoning.}
    \label{fig:music_texture}
\end{figure}

\begin{figure}
    \centering
    \includegraphics[width=\linewidth]{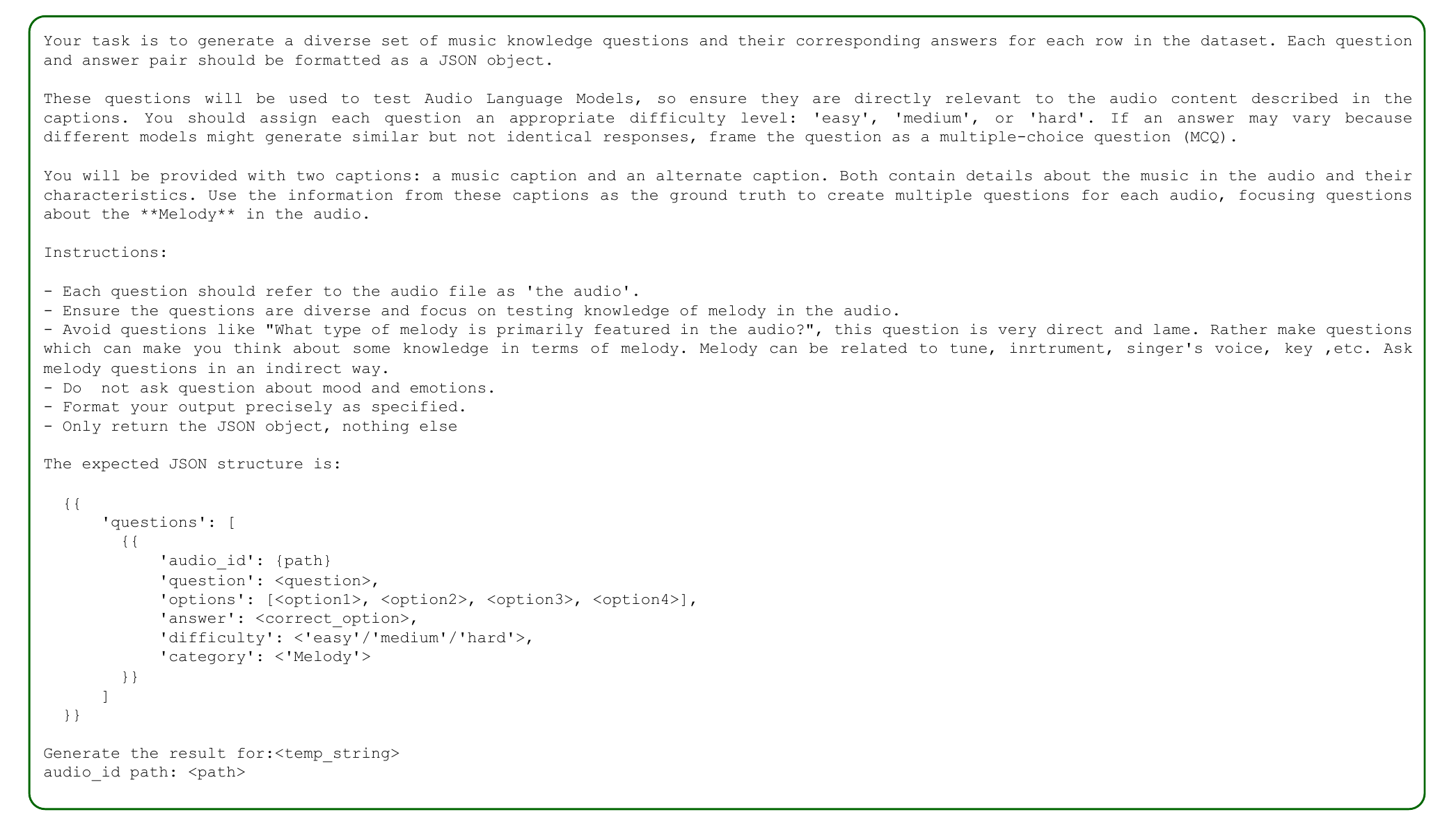}
    \caption{\small Prompt 2 used for \textbf{Music Reasoning} for AudioSkills-XL. The QAs are focused on melody reasoning.}
    \label{fig:melody}
\end{figure}

\begin{figure}
    \centering
    \includegraphics[width=\linewidth]{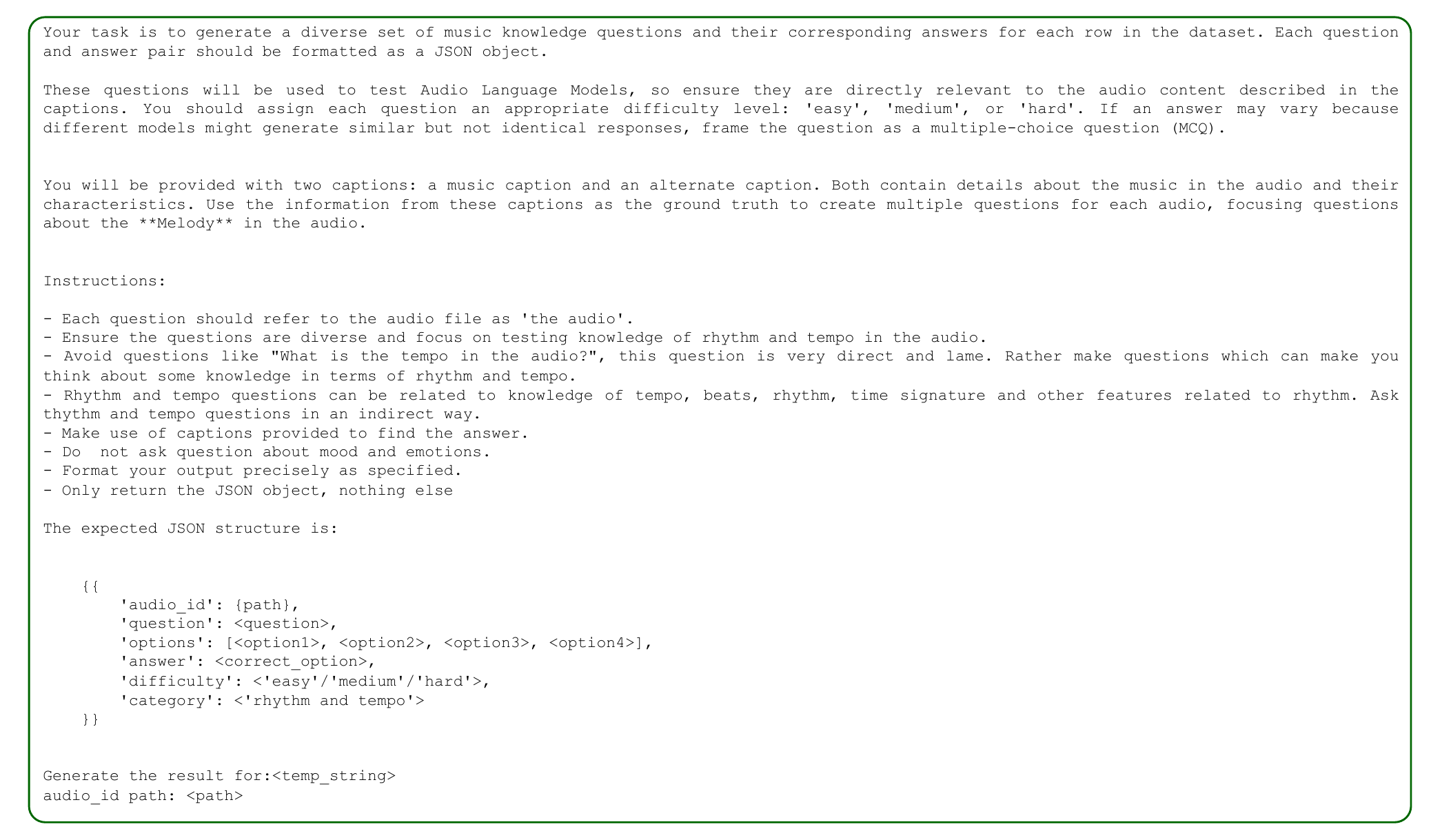}
    \caption{\small Prompt 3 used for \textbf{Music Reasoning} for AudioSkills-XL. The QAs are focused on rhythm and tempo reasoning.}
    \label{fig:rhythm_tempo}
\end{figure}

\begin{figure}
    \centering
    \includegraphics[width=\linewidth]{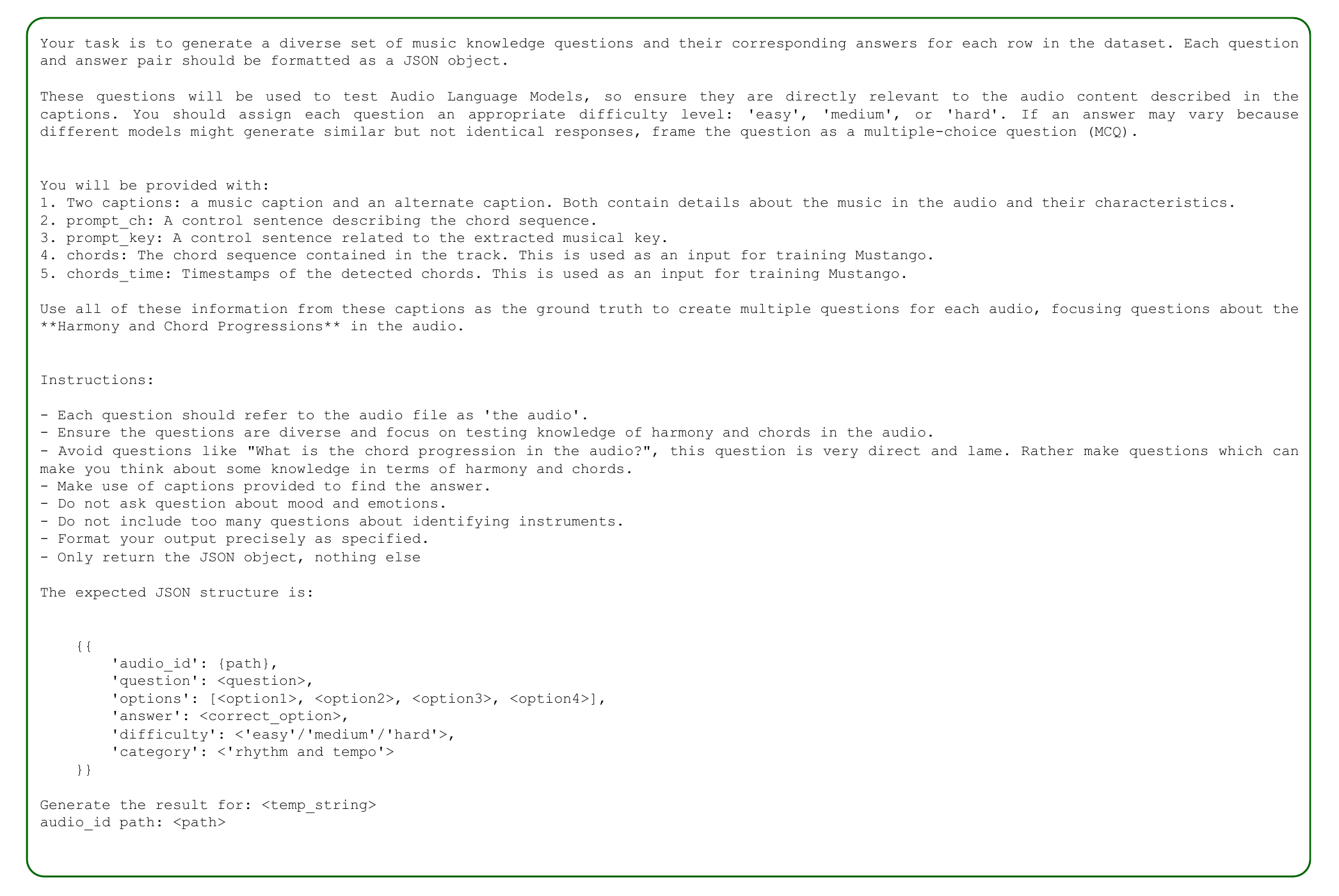}
    \caption{\small Prompt 4 used for \textbf{Music Reasoning} for AudioSkills-XL. The QAs are focused on harmony and chord reasoning.}
    \label{fig:harmony_chord}
\end{figure}

\begin{figure}
    \centering
    \includegraphics[width=\linewidth]{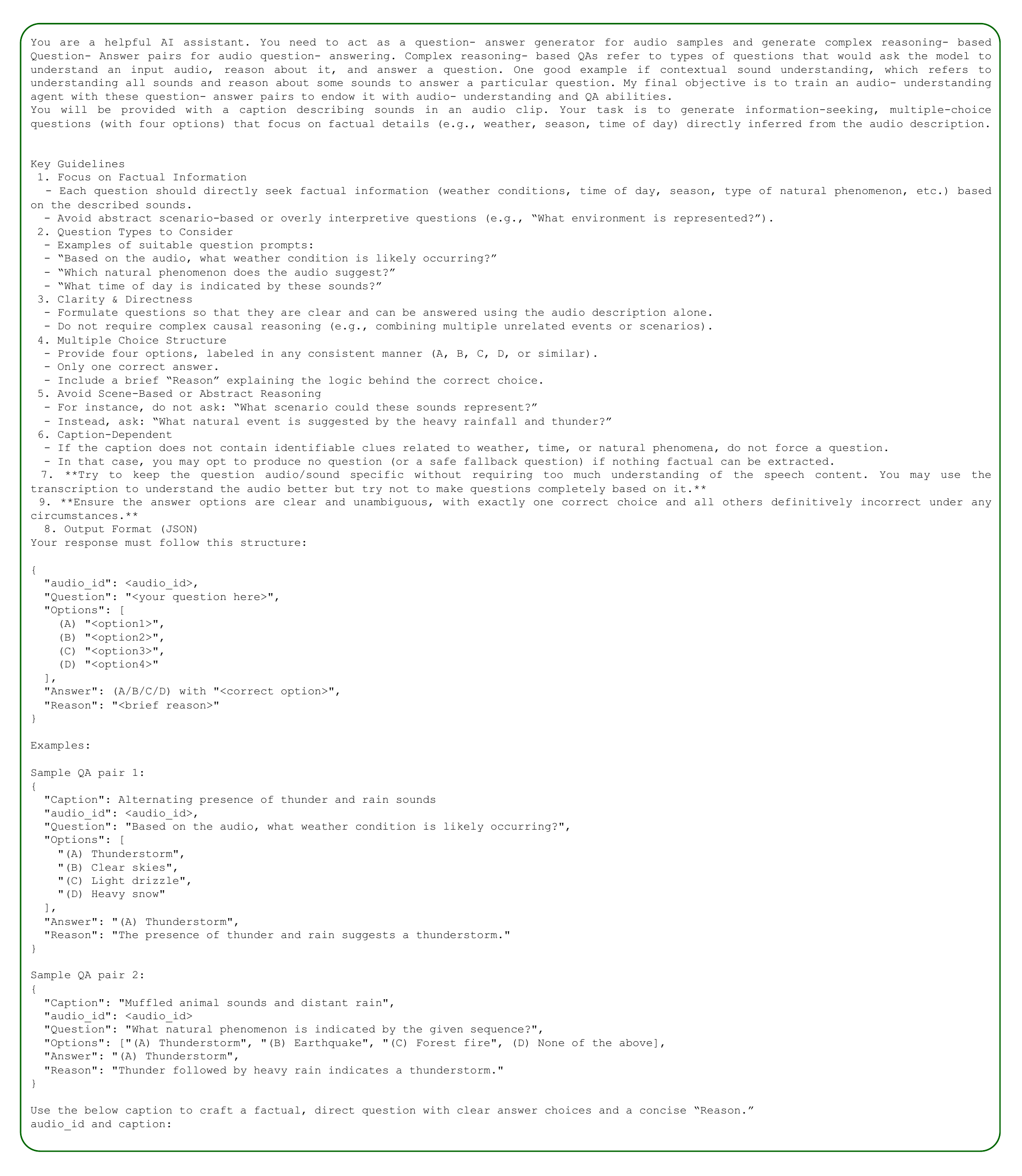}
    \caption{\small Prompt 1 used for \textbf{Sound Reasoning} for AudioSkills-XL. The QAs are focused on - eco-acoustic sound reasoning.}
    \label{fig:sound_1_prompt}
\end{figure}

\begin{figure}
    \centering
    \includegraphics[width=\linewidth]{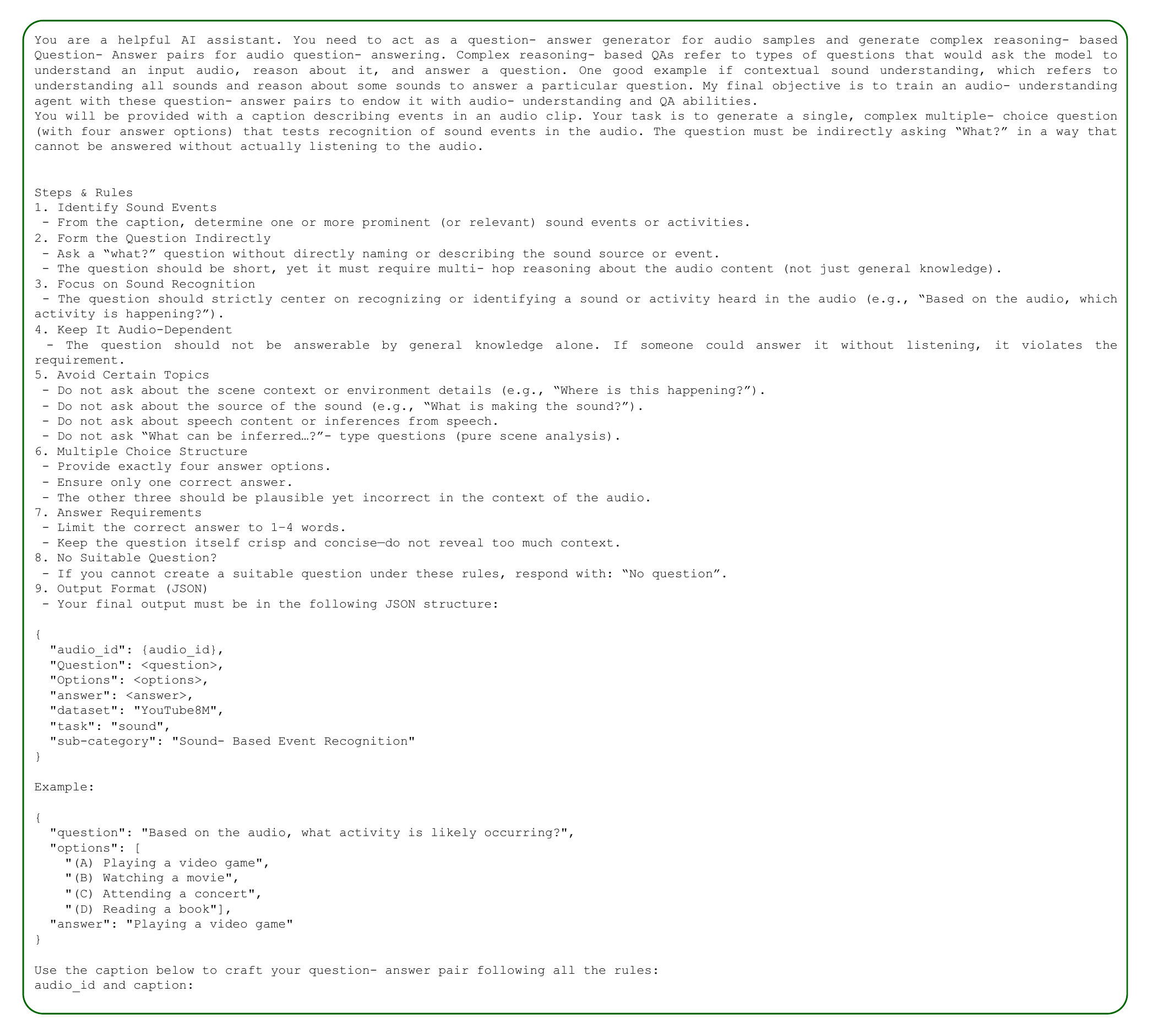}
    \caption{\small Prompt 1 used for \textbf{Sound Reasoning} for AudioSkills-XL. The QAs are focused on -Acoustic Scene Reasoning.}
    \label{fig:sound_2_prompt}
\end{figure}

\begin{figure}
    \centering
    \includegraphics[width=\linewidth]{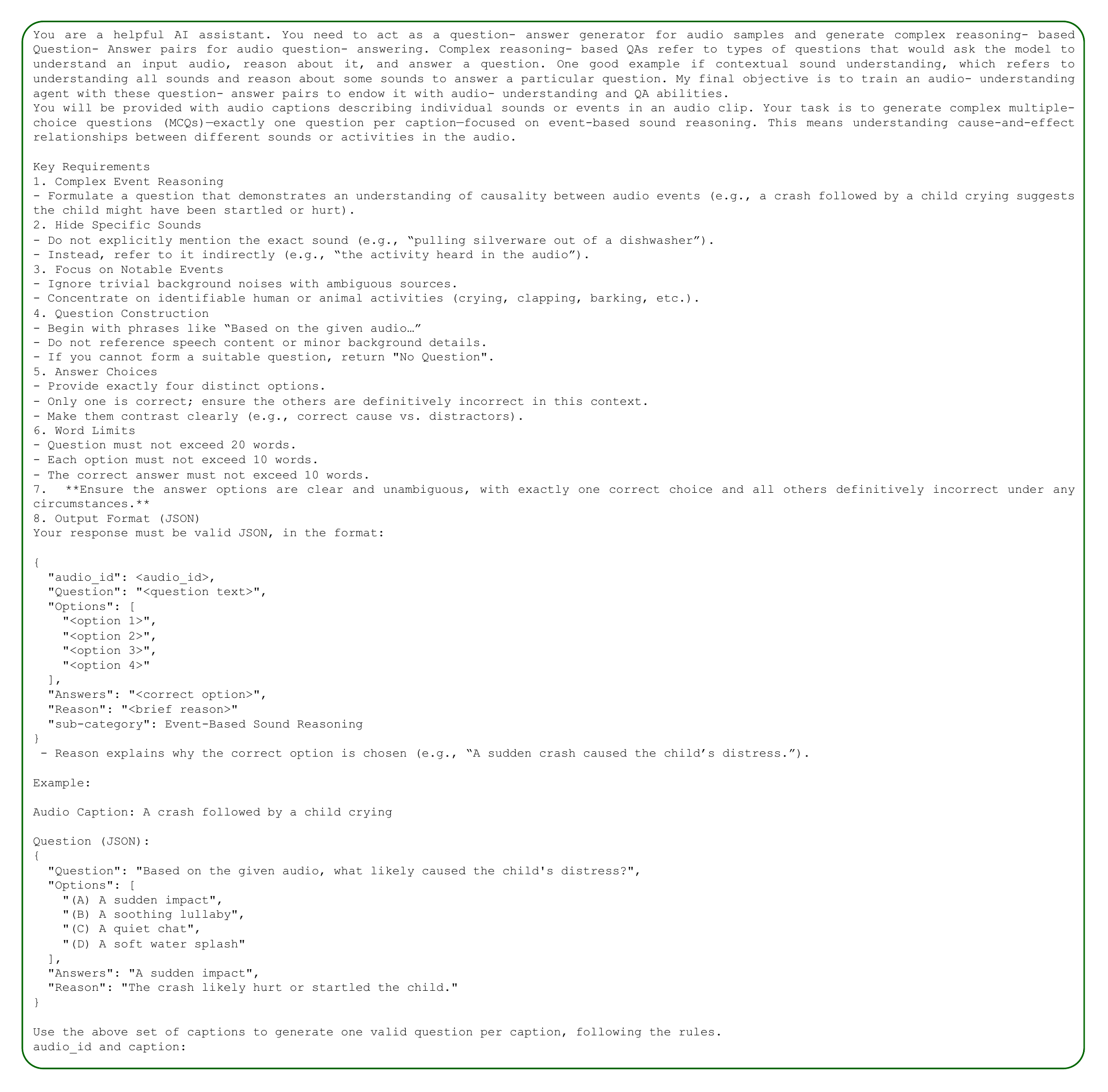}
    \caption{\small Prompt 1 used for \textbf{Sound Reasoning} for AudioSkills-XL. The QAs are focused on -Sound-Based Event Reasoning.}
    \label{fig:sound_3_prompt}
\end{figure}

\begin{figure}
    \centering
    \includegraphics[width=\linewidth]{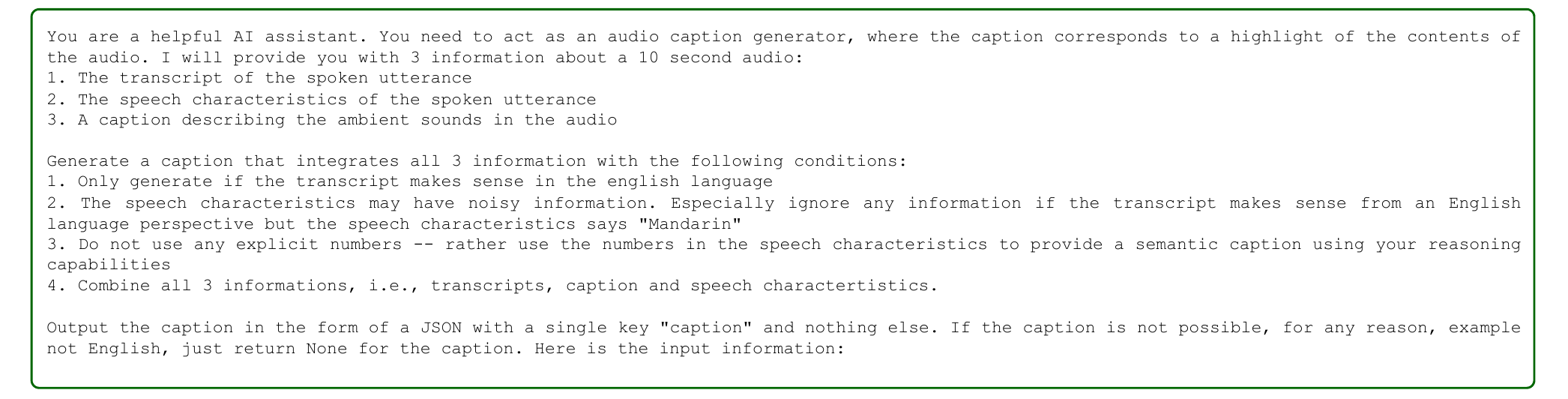}
    \caption{\small Prompt used for generating \textbf{Speech-in-Sound captions} used in pre-training and further used in generating other QAs.}
    \label{fig:transcript_prompt}
\end{figure}